\documentclass[aps,prb,notitlepage,twocolumn,superscriptaddress,showpacs]{revtex4-1}
\usepackage{amsfonts}
\usepackage{graphicx, bm, amsmath, amsfonts,amssymb}
\usepackage{tabularx}
\usepackage{caption}
\usepackage{subcaption}
\graphicspath{{figures/}}

\begin{document}

\title{Accurate simulation of q-state clock model}
\author{Guanrong Li}
 \affiliation{Department of Physics, The Chinese University of Hong Kong, Shatin, New Territories, Hong Kong, China}

\author{Kwok Ho Pai}
 \affiliation{Department of Physics, The Chinese University of Hong Kong, Shatin, New Territories, Hong Kong, China}

\author{Zheng-Cheng Gu}
\email{zcgu@phy.cuhk.edu.hk}
 \affiliation{Department of Physics, The Chinese University of Hong Kong, Shatin, New Territories, Hong Kong, China}

\begin{abstract}
We accurately simulate the phase diagram and critical behavior of the
$q$-state clock model on the square lattice by using the
state-of-the-art loop optimization for tensor network renormalzation(loop-TNR)
algorithm. The two phase transition points for $q \geq 5$ are determined
with very high accuracy. Furthermore, by computing the
conformal scaling dimensions, we are able to
accurately determine the compactification radius $R$ of the compactified boson theories at both phase
transition points. In particular, the compactification radius $R$ at high-temperature critical point is precisely the same as the predicted $R$ for Berezinskii-Kosterlitz-Thouless (BKT) transition. Moreover, we find that the fixed point tensors at high-temperature critical point also converge(up to numerical errors) to the same one for large enough $q$ and the corresponding operator product expansion(OPE) coefficient of the compactified boson theory can also be read out directly from the fixed point tensor.
\end{abstract}

\maketitle

\affiliation{Department of Physics, The Chinese University of Hong Kong,
Shatin, New Territories, Hong Kong, China}

\affiliation{Department of Physics, The Chinese University of Hong Kong,
Shatin, New Territories, Hong Kong, China}

\affiliation{Department of Physics, The Chinese University of Hong Kong,
Shatin, New Territories, Hong Kong, China}

\section{Introduction}

Berezinskii-Kosterlitz-Thouless (BKT)\cite{BKT1,BKT2,BKT3} transition was originally proposed in classical XY model with a continuum $U(1)$ symmetry. It is well known that spontaneous breaking of continuum symmetry is not allowed in 2D classical systems and the BKT transition provides us the first example beyond Landau's symmetry breaking paradigm. On the contrary, spontaneous breaking of discrete symmetry is generally allowed for 2D classical systems and BKT transition is usually not expected for these systems. In recent years, people find very strong numerical evidence that BKT transition actually also happens in systems with discrete symmetry, e.g., the $q$-state clock model.
It has been pointed out that for $q\geq 5$, the $q$-state clock
model typically has two critical points\cite{JosSe}. At high-temperature critical
point, the system undergoes a BKT transition, while at low-temperature critical
point, the long-range order would emerge and the usual symmetry breaking transition happens. Theoretically, it was well known that $q$-state model with $q\geq 5$ is effectively described by $Z_q$ deformed sine-Gordon model\cite%
{Wiegm,Matsu}, and
the renormalization analysis also suggests that the model will undergo two phase transitions as the temperature decreases.
Between the two phase transition points, the effective field theory
reduces to a simple compactified boson theory with emergent $U(1)$ symmetry.
 Previously, a lot of studies have been focused on how to
determine the two critical temperatures\cite{Toboc,Baek,Baeks,Kuman,Chate,Krcma,ChenJ,HKim,ChenY,SHong}, but how to accurately extract the exact conformal data at critical points is
still a very challenging problem.\cite{Chall,Zhang,Yamag,Tomit,Hwang,Brito,Bori1,Bori2,Raym,Baek1,Chatt,Suru,Li}


Tensor renormalization group(TRG) algorithm\cite{Levin,Gu} is a powerful tool to study the phase diagram of 2D classical statistical models and 1+1D quantum models.
By investigating the properties of the corresponding fixed point tensor, many important properties of the phase diagram can be read out directly\cite{Gu}. In recent years, the so-called loop optimization for tensor network renormalzation(loop-TNR)\cite{Yang} method was proposed as a real space renormalization algorithm to accurately study critical properties of 2D classical statistical models and 1+1D quantum models.
Comparing with singular value decomposition based methods, e.g., TRG and higher order TRG(HOTRG)\cite{Xie,HOTRG}, the loop-TNR algorithm
has extremely high accuracy and makes it possible for us to read out all the conformal data for critical systems, such as scaling dimensions, operator product expansion(OPE) coefficient for primary fields from the corresponding fixed point tensor.


In this paper, we use loop-TNR algorithm to study the phase transition properties
of the $q$-state clock model. We find very
strong numerical evidence that the physics of self-dual critical points for $%
q<5$ model matches very well with the previous proposal from conformal field(CFT)
theory and other numerical results. For $q\geq 5$ model, the middle phase between the
symmetry-breaking phase transition point and BKT critical point is described by
the compactified boson theory with central charge $c=1$. By computing the scaling dimensions of the two phase transition points as well as the so-called self-dual points, we are able to determine the compactification radius $R$ of the corresponding compactified boson theory with very high accuracy. We find that the obtained compactification radius $R$ perfectly agree with the field theory predictions. Furthermore, we also find that for big enough $q$, the corresponding fixed point tensors at high-temperature critical point $T_{c2}$ converge to the same one(up to numerical errors) describing BKT transition with an emergent $U(1)$ symmetry, and the corresponding OPE coefficient of the compactified boson theory can also be read out directly.

We stress that our method not only gives accurate critical temperature, but also produces accurate conformal data, especially for the cases with $q=5$ and $q=6$, which are very hard to be simulated by density matrix renormalization group(DMRG)/matirx product state(MPS) based methods\cite{Chate,Li} as well as Monte Carlo simulation\cite{Toboc,Tomit,Hwang,Baeks,Bori1,Bori2,Kuman,Chatt,HKim,Suru} due to the presence of marginal irrelevant terms\cite{marginal}. Our numerical results also suggest that 2D CFT could be reformulated as an infinite dimensional fixed point tensor which encodes the complete conformal data, such as scaling dimensions and OPE coefficients. This might provide us an algebraic way to reformulate and classify all 2D CFT.


\section{$q<5$ models}

The $q$-state clock model is describe by the Hamiltonian%
\begin{equation}
H=-J\underset{\left\langle ij\right\rangle }{\sum }\cos \left( \theta
_{i}-\theta _{j}\right) ,
\end{equation}%
where $\theta _{i}=2\pi n_{i}/q,$ and $n_{i}\in \left\{ 1,2,...q\right\} .$
We note that for $q=2$ and $q=3$ the model is equivalent to classical Ising model and
3-states Potts model. The partition function of the $q$-state clock model can be expressed as a trace of local tensors:
\begin{equation}
Z=\text{Tr}\otimes T.
\end{equation}%

\begin{figure}[h]
\centering
\includegraphics[width=8cm]{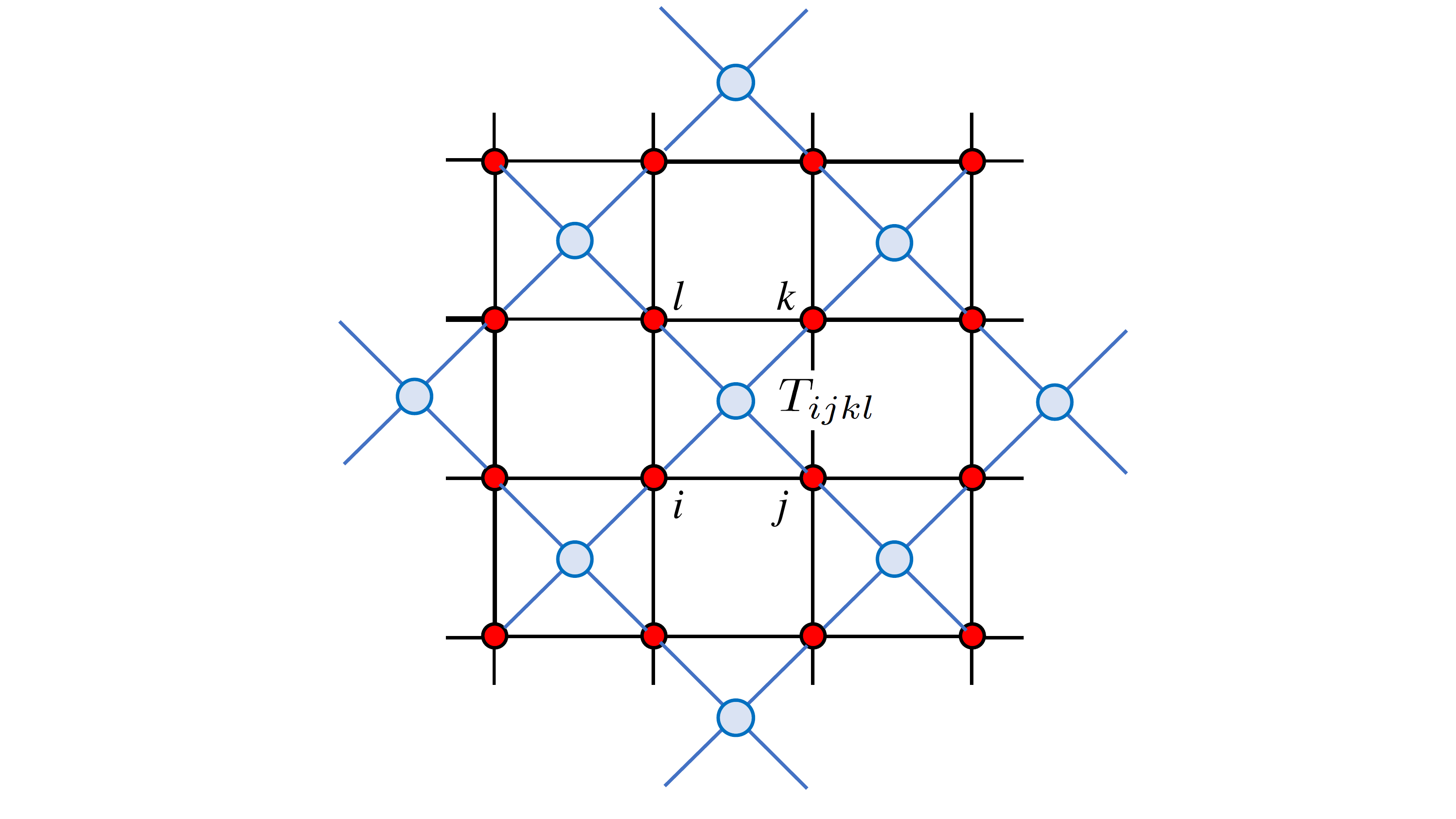}
\caption{Tensor network representation of $q$-state clock model on square lattice.}
\label{F41}
\end{figure}

On square lattice, the partition function of the $q$-state model is expressed by the trace of the following element tensor $T_{ijkl}$(seen in Fig. \ref{F41}):
\begin{eqnarray}
T_{ijkl}
=\exp \beta \left( \cos \theta _{ij}+\cos \theta _{jk}+\cos \theta
_{kl}+\cos \theta _{li}\right),
\label{element}
\end{eqnarray}%
where $\theta _{ij}=2\pi \left( i-j\right) /q$
and $i,j,k,l$ take values $\left\{ 1,2,...q\right\}$.

\begin{figure}[h]
\includegraphics[width=6cm]{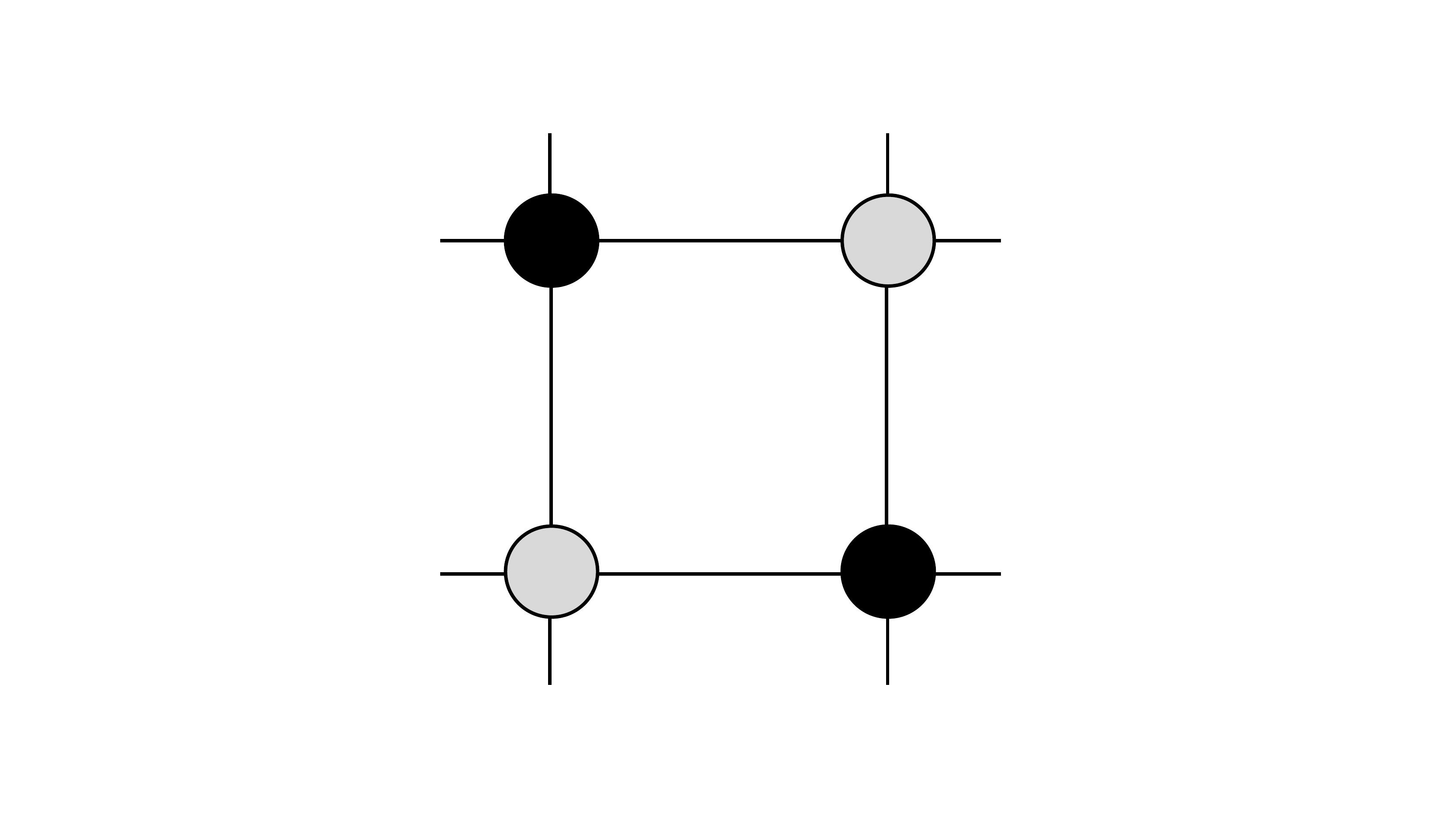}
\caption{We use the 2 by 2 block to represent the fixed point tensor $T$ when
calculating $\protect\chi$, where we group the index $(i_{1},i_{2})$ into a single index $i$ for tensor $T$}
\label{F42}
\end{figure}

\begin{figure}[h]
\includegraphics[width=8cm]{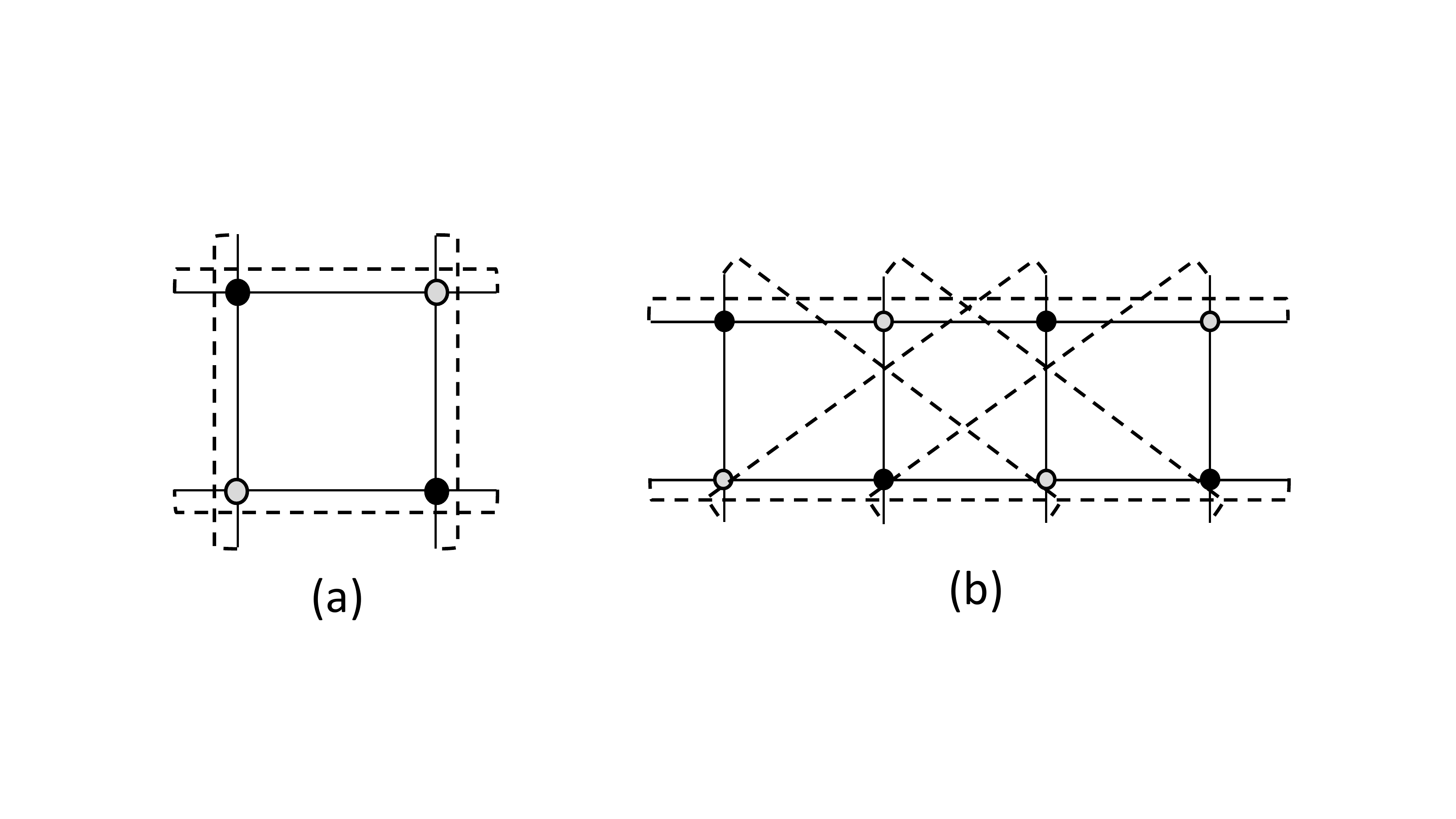}
\caption{Gauge invariant quantity $\protect\chi$ formed by $T$, where the numerator is
represented by the square of part (a), and the denominator is given by part
(b)}
\label{F43}
\end{figure}

For $q<5$, it is well known that the self-dual critical temperature reads\cite{Kramers}:%
\begin{equation}
\beta _{c}=\left\{
\begin{array}{c}
\ln \left( \sqrt{2}+1\right) /2,\text{ }q=2 \\
2\ln \left( \sqrt{3}+1\right) /3,\text{ }q=3 \\
\ln \left( \sqrt{2}+1\right) ,\text{ }q=4%
\end{array}%
\right.
\end{equation}%
We will first benchmark with these exact results to examine the accuracy of our algorithm. Since the $q=2$ case has already been studied before, here we will begin with the $q=3$ and $q=4$ cases.
To find
the critical point, we first calculate the gauge invariant quantity $\chi$
introduced in Ref. \cite{Gu}:
\begin{equation}
\chi =\frac{\left( \underset{ij}{\sum }T_{ijij}\right) ^{2}}{\underset{ijkl}{%
\sum }T_{ijkl}T_{klij}},
\label{chi}
\end{equation}
where we use the 2 by 2 block to represent the fixed point tensor $T$(composed by $T_A$ and $T_B$ on sublattices A and B, respectively) when calculating the gauge invariant quantity $\chi$, as shown in Fig. \ref{F42} and Fig. \ref{F43}.

\begin{figure}[h]
\includegraphics[width=8cm]{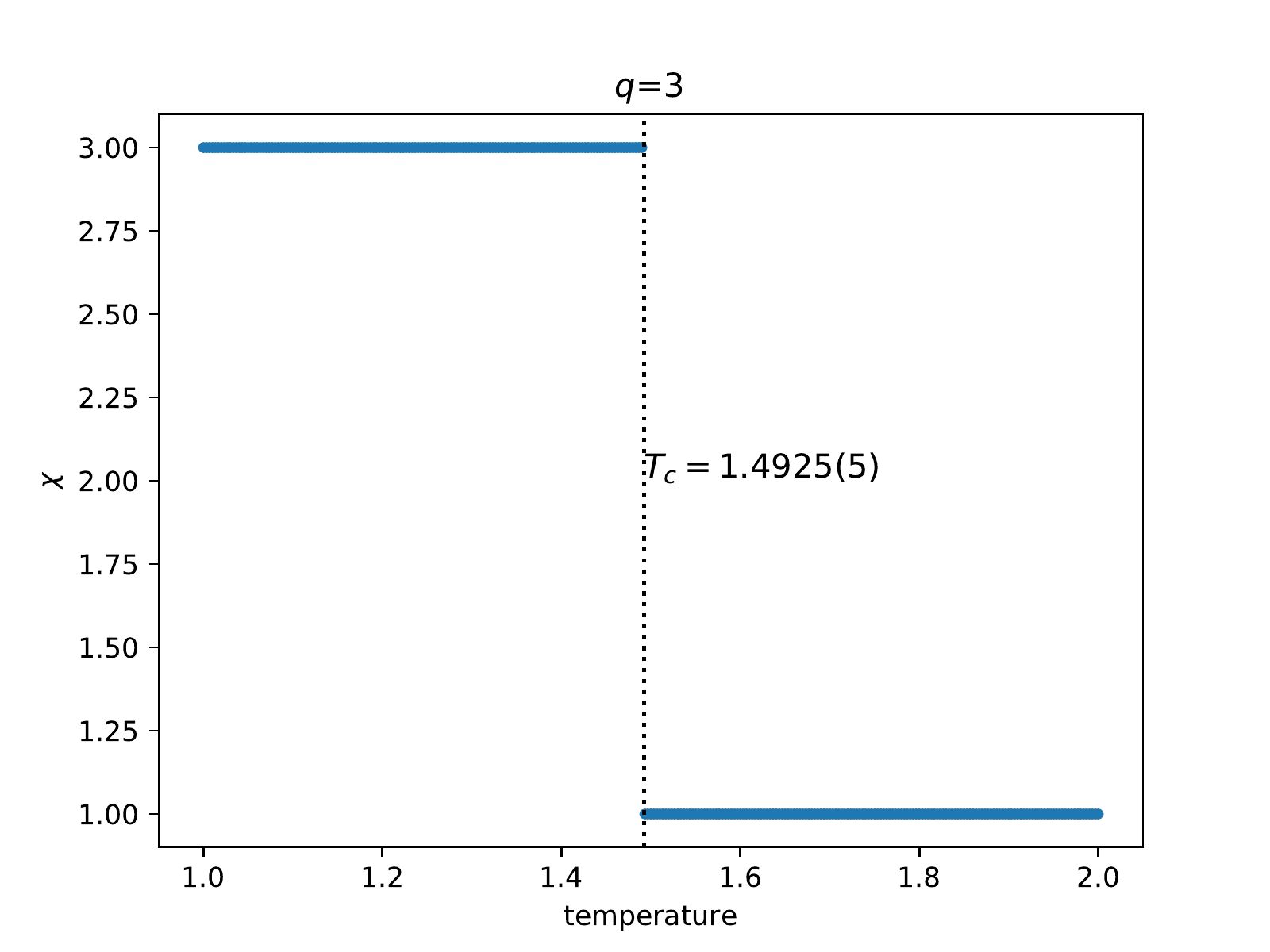}
\caption{The invariant quantity $\chi$ as a function of temperature. We find that the critical temperature $T_{c}$ for $q=3$ model is around 1.4925(5), which is intrinsically close to the prediction of the self-dual analysis. Here we keep $D_{cut}=36$ in the loop-TNR algorithm and system size up to $2^{23}$.}
\label{F1}
\end{figure}

As seen in Fig. \ref{F1}, we see that there is a sudden jump from ordered phase to
disordered phase. This is because the tensors for
ordered and disordered phase would flow to different fixed points. To understand better for the gauge invariant quantity $\chi$, we
introduce matrix $M^{h} $ and $M^{v} $:
\begin{eqnarray}
M_{ij}^{h} &=&\underset{k}{\sum }T_{ikjk}^{\text{fixed-point}}  \notag \\
M_{ij}^{v} &=&\underset{k}{\sum }T_{kikj}^{\text{fixed-point}}.
\end{eqnarray}%
We see that for ordered phase, the eigenvalue $\lambda $ of $ M^{h} $ and $M^{v}$ is:
\begin{equation}
\left\{
\begin{array}{c}
\lambda _{1},\lambda _{2},...\lambda _{q}=1/q \\
\text{others}=0%
\end{array}%
\right. .
\end{equation}%
And in disordered phase, we have $\lambda _{1}=1$, and all the others
approach 0, which shows clearly the symmetry breaking nature of the phase
transition. Here, we have already normalized the fixed point tensor as:
\begin{equation}
\underset{jk}{\sum }T_{jkjk}^{\text{fixed-point}}=1.
\end{equation}%

Next, we compute the central charge and scaling dimensions for $q=3$ model(here we keep $%
D_{cut}=36 $ in our loop-TNR algorithm). We find that the central charge $c=0.80005$, which is intrinsically close to the value predicted by the CFT with $c=4/5$.
We see that both central charge and scaling dimensions are very
stable up to 20 renormalization steps, which corresponds to a total system size $2^{23}$.

\begin{figure}[h]
\includegraphics[width=8cm]{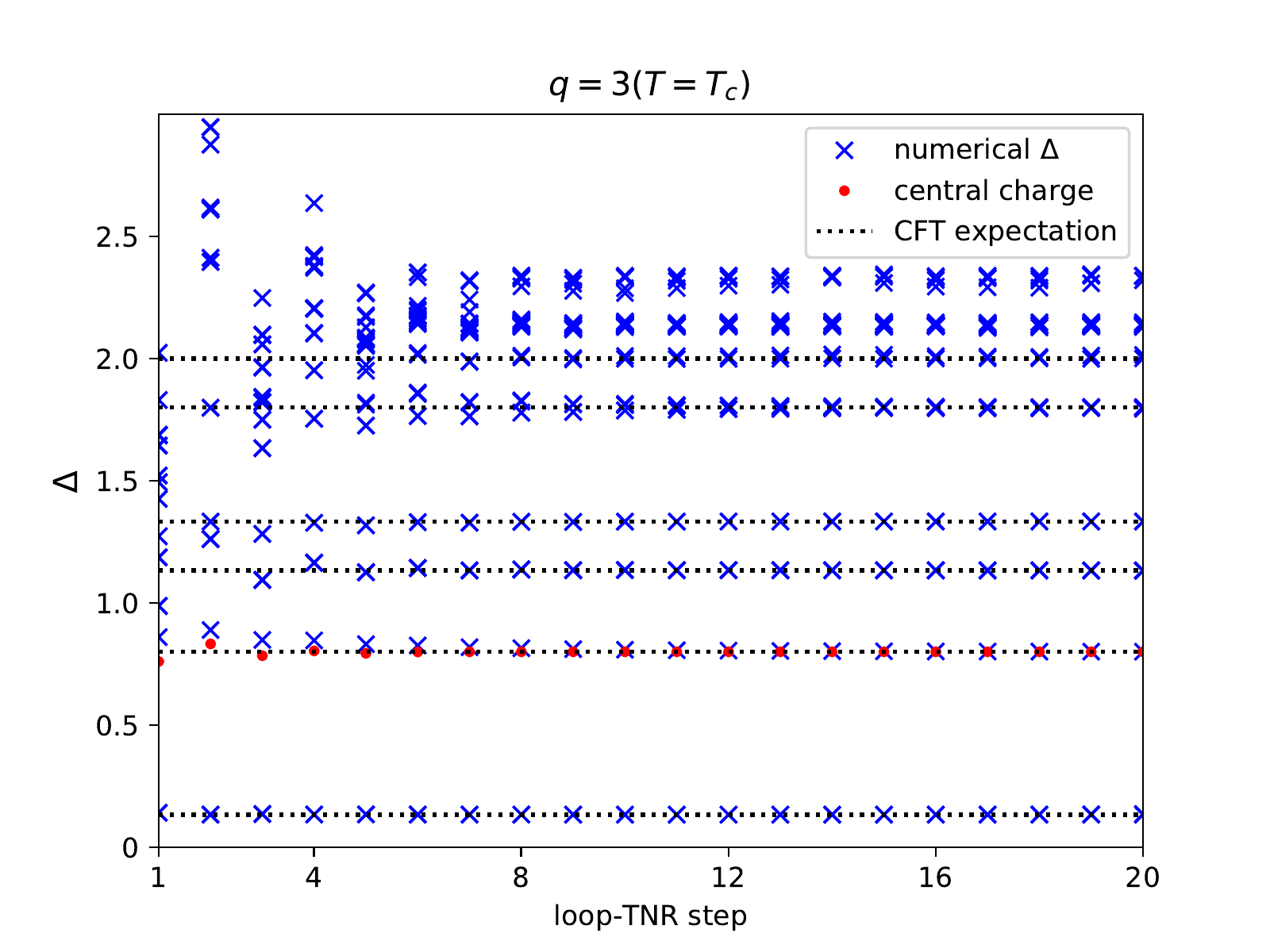}
\caption{The scaling dimensions at the critical point of $q=3$ model with $D_{cut}$=36. We see that the conformal data rapidly converges to CFT predictions during the renormalization process.}
\label{F2}
\end{figure}

\begin{figure}[h]
\includegraphics[width=8cm]{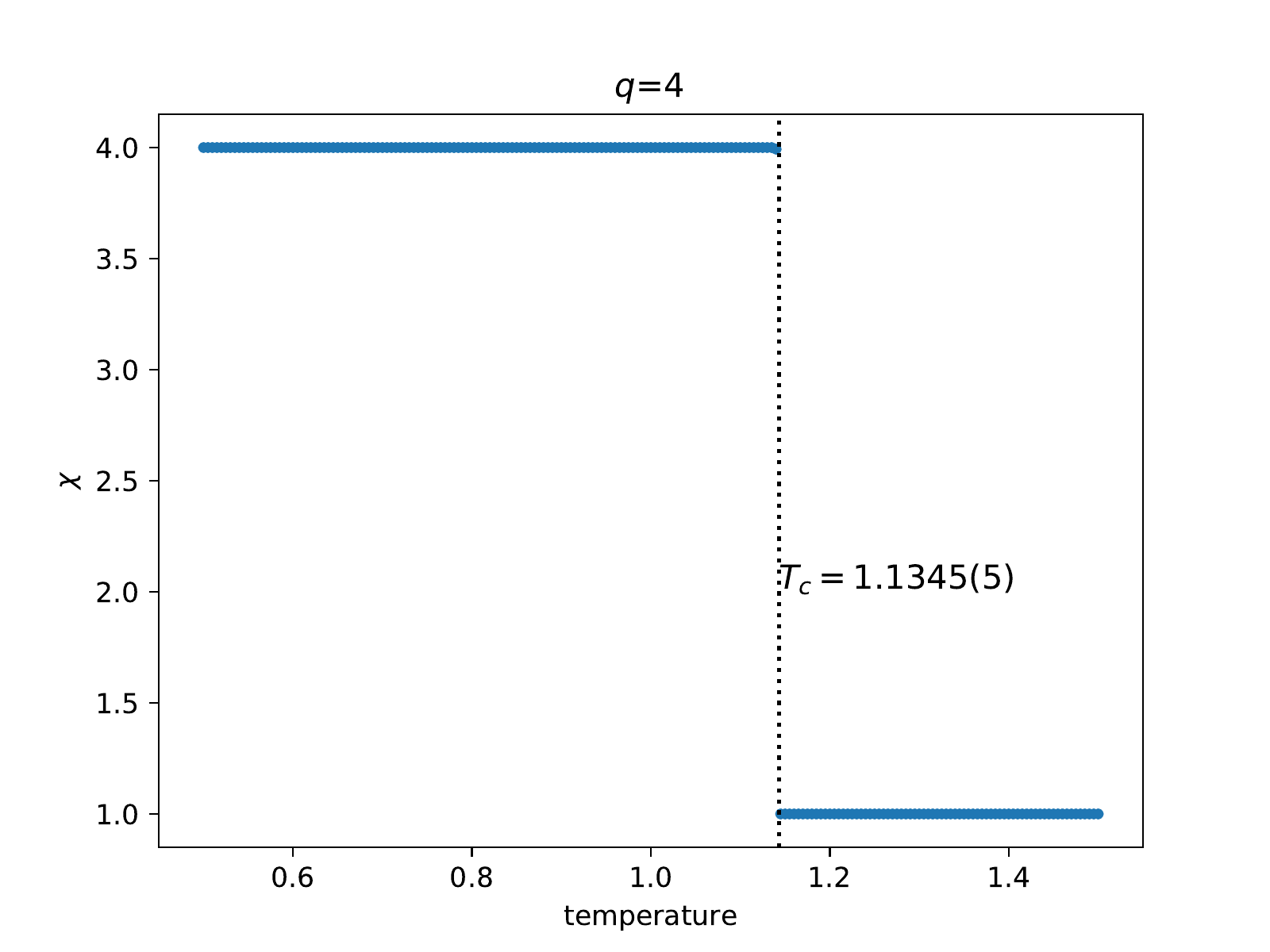}
\caption{The invariant quantity $\chi$ as a function of temperature. We find that the critical temperature $T_{c}$ for $q=4$ model is around 1.1345(5), which is intrinsically close to the prediction of the self-dual analysis\cite{Kramers}. Here we also keep $D_{cut}=36$ in the loop-TNR algorithm and system size up to $2^{23}$.}
\label{F3}
\end{figure}

\begin{figure}[h]
\includegraphics[width=8cm]{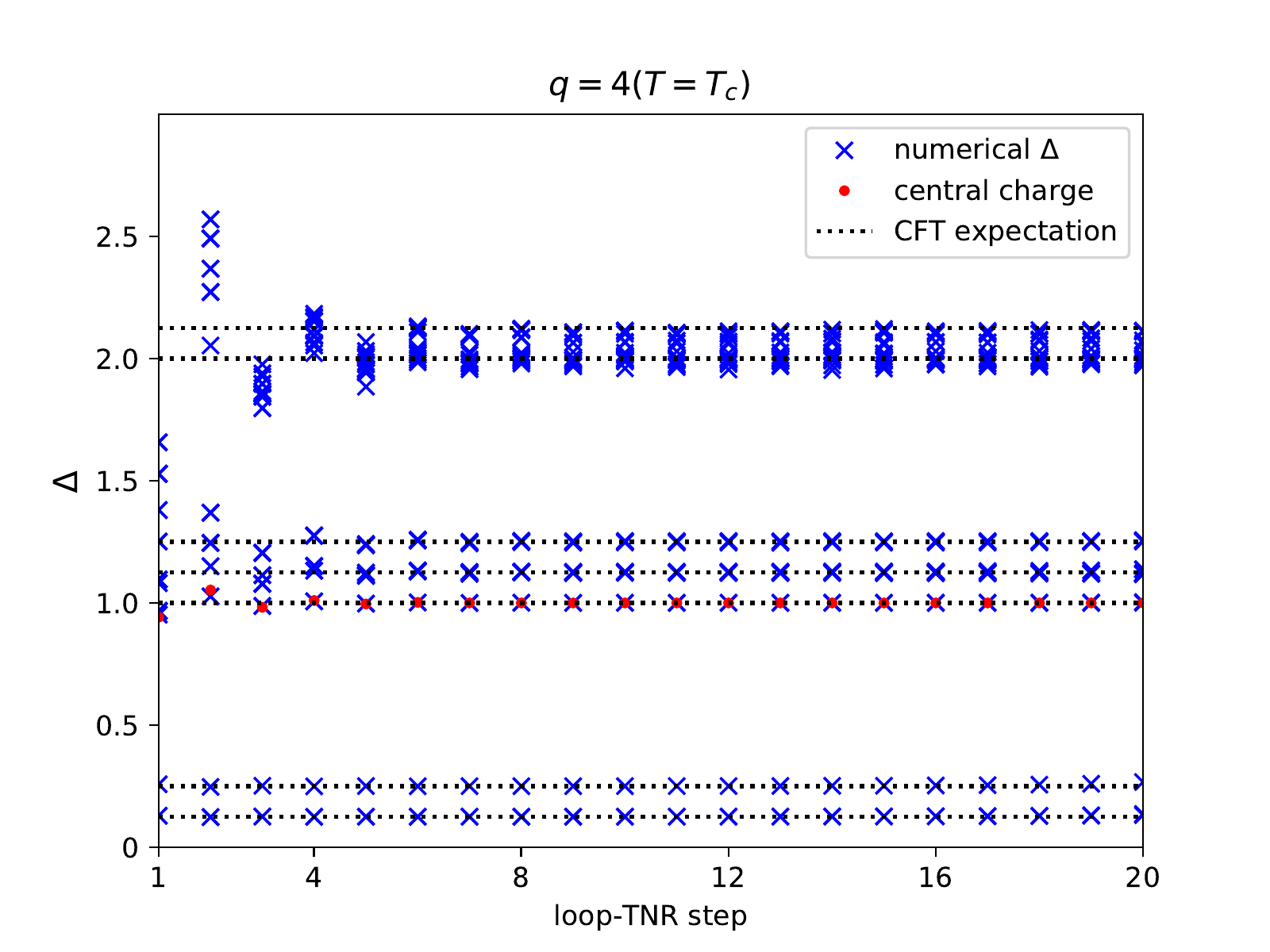}
\caption{The scaling dimensions at the critical point of $q=4$ model with $D_{cut}$=36.}
\label{F4}
\end{figure}

Similarly, we can compute the gauge invariant quantity $\chi$, central charge and scaling dimensions for the $q=4$ model(again, we keep $%
D_{cut}=36 $ in our loop-TNR algorithm), as shown in Fig. \ref{F3} and Fig. \ref{F4}. We find that $c=1.00021$, which is also consistent with previous theoretical predictions with $c=1$.
In fact, the critical point of $q=4$ model can be just regarded as two copies of the Ising CFT. Again, we see that both central charge and scaling dimensions are very
stable up to 20 renormalization steps

\section{$q=5$ and $q=6$ models}

For $q \geq 5$, it is conjectured that the $q$-state clock model is described by $Z_{q}$
-deformed sine-Gordon theory\cite{Wiegm,Matsu}%
\begin{eqnarray}
\label{sine}
S &=&\frac{1}{2\pi K}\int d^{2}\mathbf{r}\left( \mathbf{\nabla }\phi \right)
^{2}+\frac{g_{1}}{2\pi \alpha ^{2}}\int d^{2}\mathbf{r}\cos \left( \sqrt{2}%
\phi \right)  \notag \\
&&+\frac{g_{2}}{2\pi \alpha ^{2}}\int d^{2}\mathbf{r}\cos \left( q\sqrt{2}%
\Theta \right) ,
\end{eqnarray}%
where $\phi ,\Theta $ are compactified as $\phi \equiv \phi +\sqrt{2}\pi$, $\Theta \equiv \Theta +\sqrt{2}\pi ,$ and they satisfy the dual relation $\partial _{x}\phi =-K\partial
_{y}\Theta ,$ $\partial _{y}\phi =K\partial _{x}\Theta .$ The coupling
constants $K,g_{1},g_{2}$ are temperature-dependent, and $\alpha $ is a UV
cutoff.

\begin{figure}[h]
\includegraphics[width=8cm]{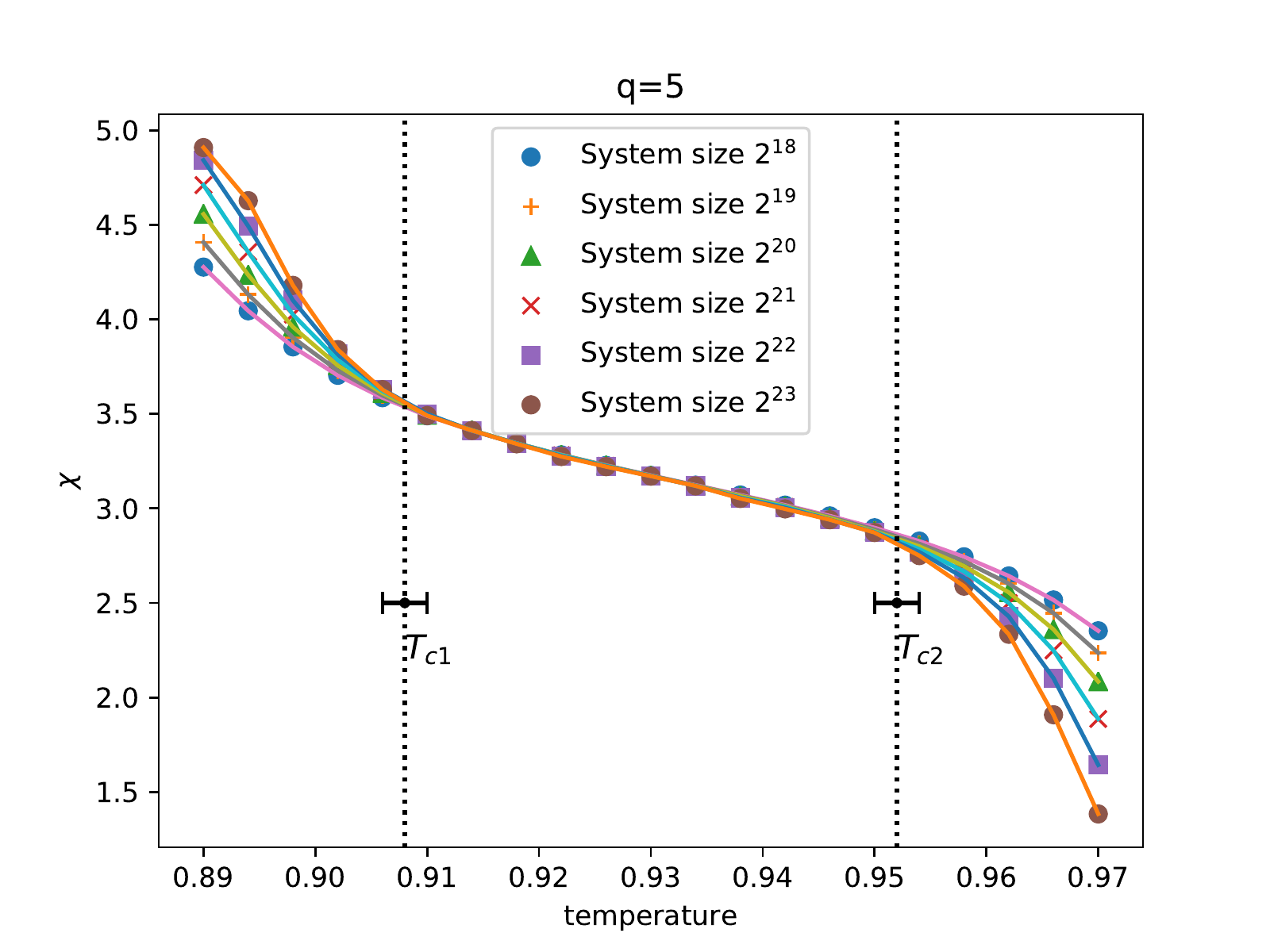}
\caption{Invariant quantity of $q=5$ model. In ordered phase, the gauge invariant quantity
$\chi$  should flow to the fixed point $\chi=5$, while in disordered
phase, $\chi$ should flow to $\chi=1$. In the middle phase, the value of $\chi$
varies with temperature. We can
read out that $T_{c1}=0.908(2)$, and $T_{c2}=0.952(2)$.}
\label{F6}
\end{figure}

\begin{figure}[h]
\begin{subfigure}{.5\textwidth}
\includegraphics[width=8cm]{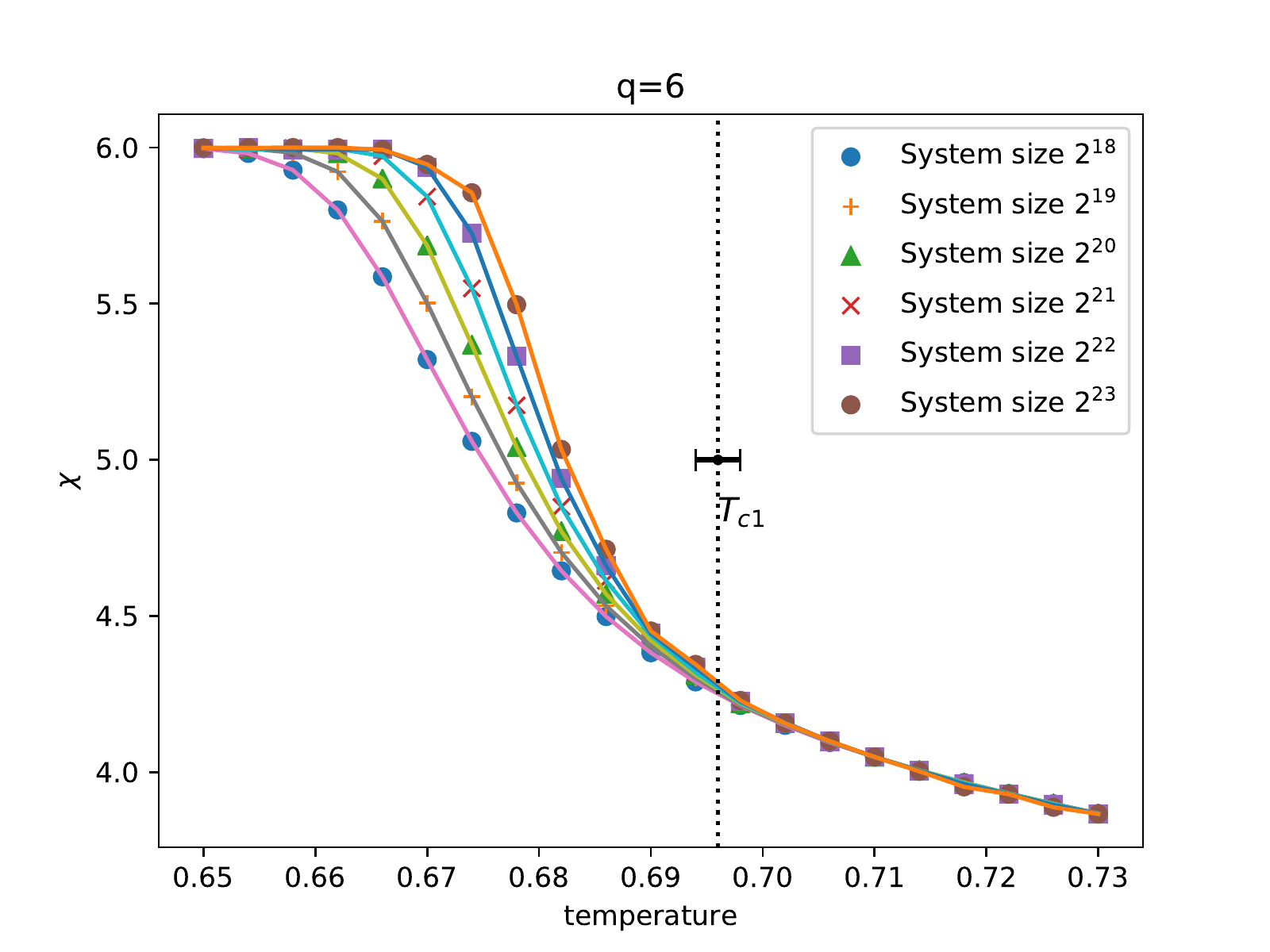}
\end{subfigure}
\par
\begin{subfigure}{.5\textwidth}
\includegraphics[width=8cm]{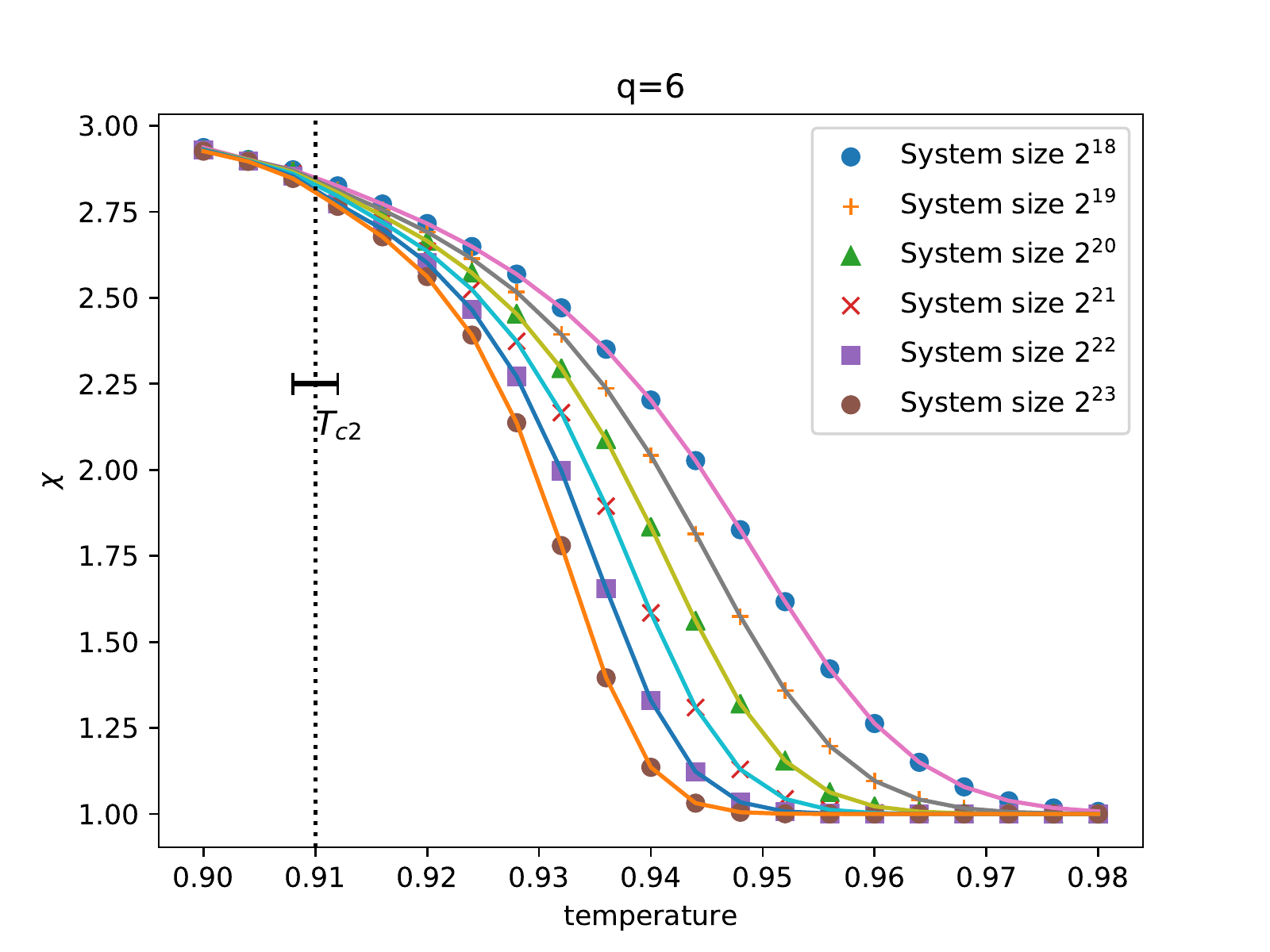}
\end{subfigure}
\caption{Invariant quantity of $q=6$ model around $T_{c1}$ and $T_{c2}$. We can read
that $T_{c1}=0.696(2)$, and $T_{c2}=0.912(2)$, with the same analysis with $q=5$ model.}
\label{F20}
\end{figure}

\begin{figure}[h]
\includegraphics[width=8cm]{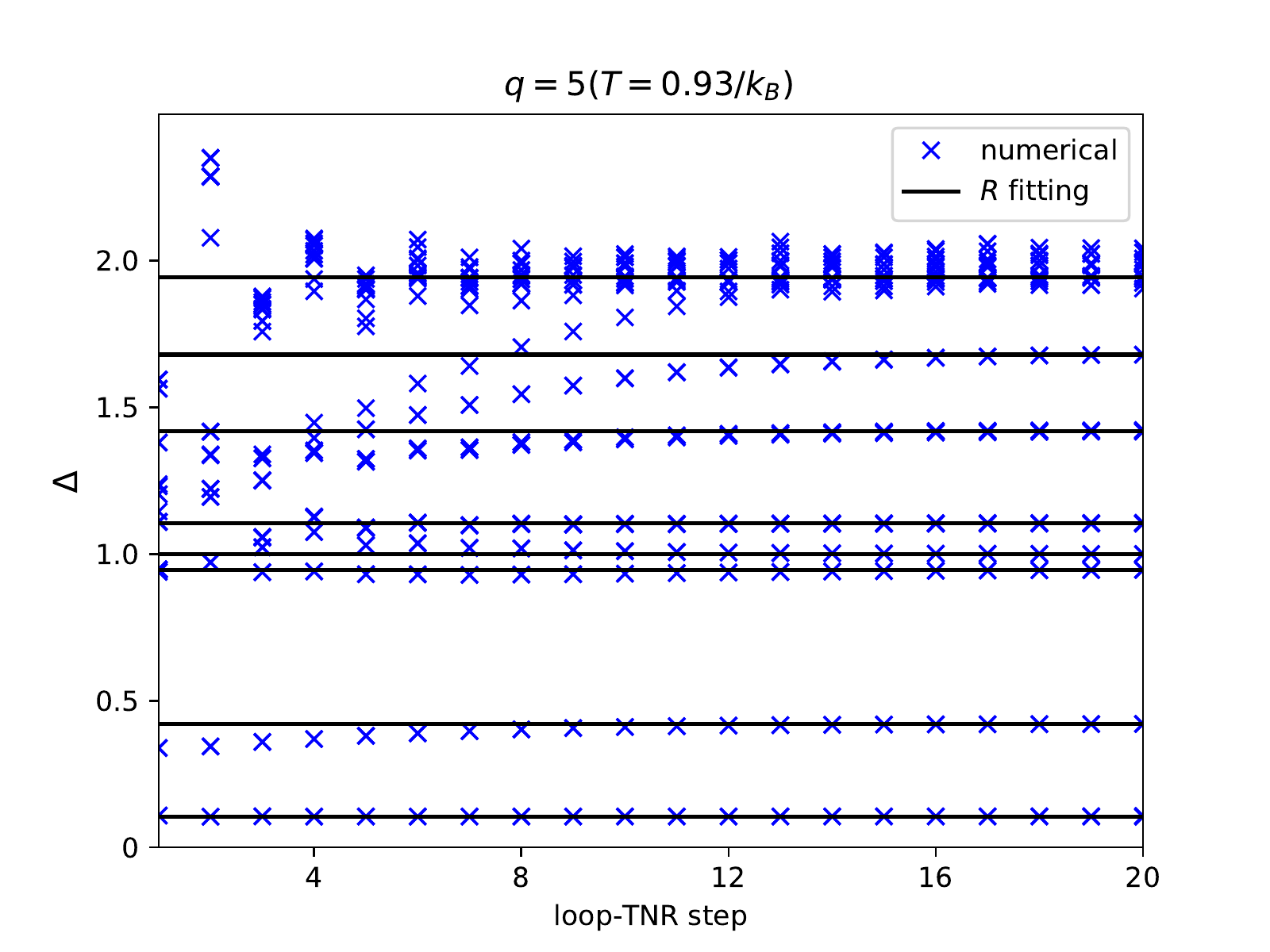}
\caption{An example of scaling dimensions in the critical phase for $q=5$ model.}
\label{F5}
\end{figure}


With decreasing temperature, the above effective theory will describe two phase transitions, which can be understood from the renormalization group flow of the second and third terms. The high-temperature critical point is described by the well known BKT transition while the low-temperature transition is described by the usual symmetry breaking transition.
As the coupling $g_{1}$ and $g_{2}$ become irrelevant
between the two critical point $T_{c1}<T<T_{c2}$,   the effective theory reduces to the compactified boson theory in the middle phase, 
with compatification radius $R=\sqrt{2K}$. In addition, if $g_1=g_2$, Eq. (\ref{sine}) is self-dual.
From the scaling dimension analysis, the compactification radius can be computed exactly for both phase transition points as well as for the self-dual point\cite{Li}. We have:
\begin{eqnarray*}
R_{c2} &=&2\sqrt{2},\text{ BKT transition point} \\
R_{self-dual} &=&\sqrt{2q},\text{ self-dual point} \\
R_{c1} &=&q/\sqrt{2},\text{ symmetry-breaking point}
\end{eqnarray*}%

\begin{figure}[h]
\begin{subfigure}{.5\textwidth}
\includegraphics[width=8cm]{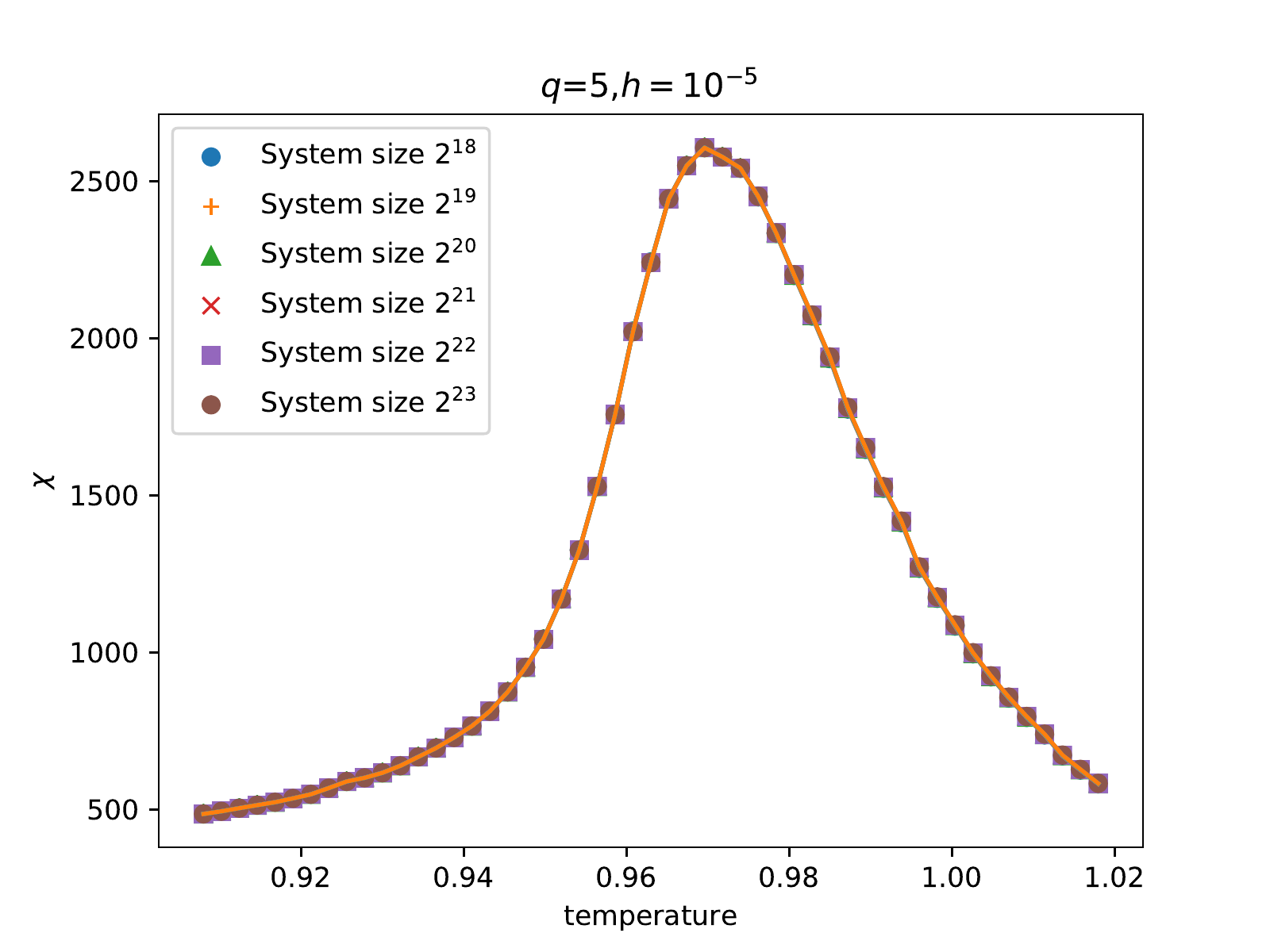}
\end{subfigure}
\par
\begin{subfigure}{.5\textwidth}
\includegraphics[width=8cm]{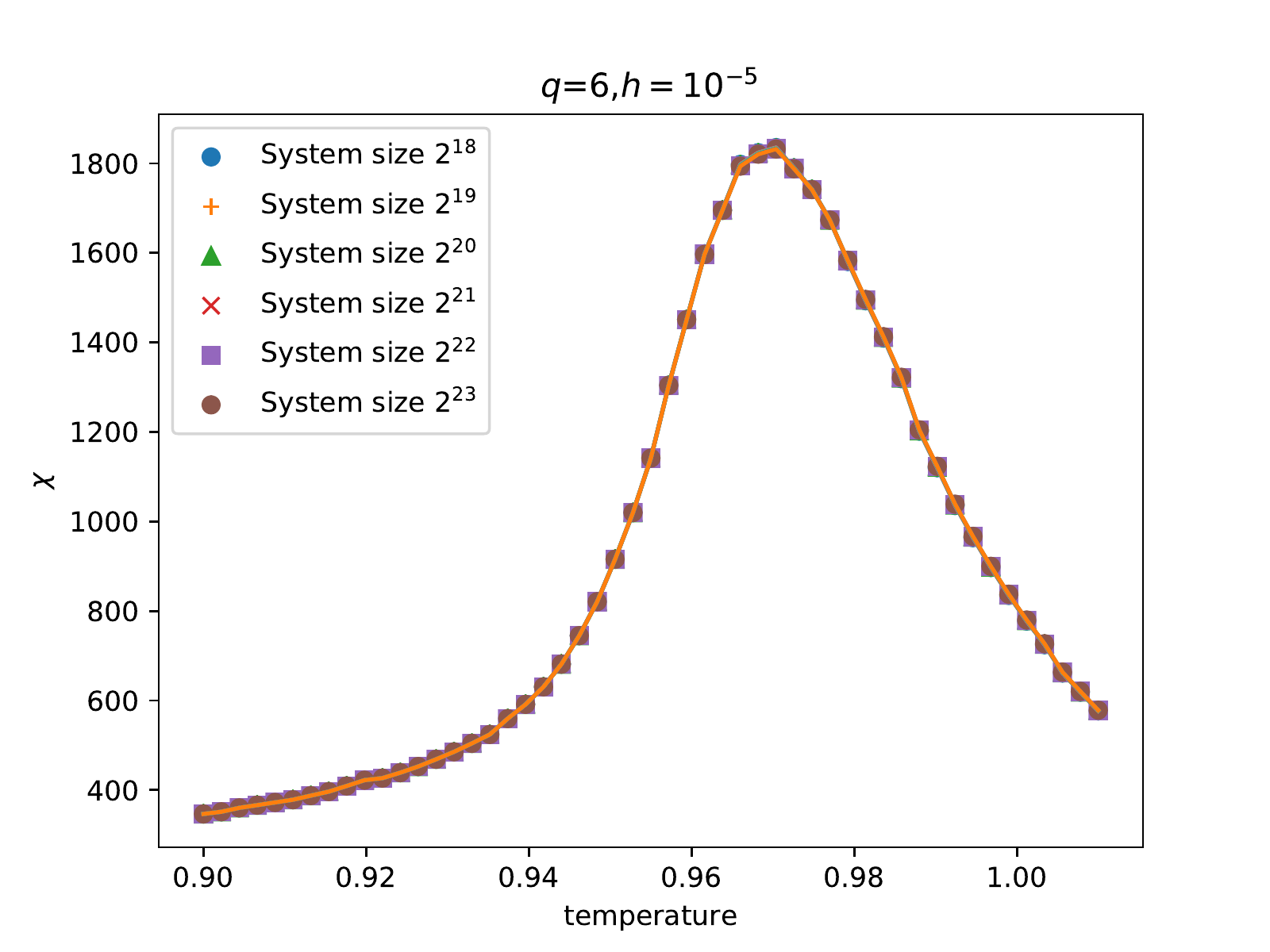}
\end{subfigure}
\caption{Susceptibility of $q=5$ and $q=6$ models with external field $h=10^{-5}$.}
\label{F9}
\end{figure}

\begin{figure}[h]
\includegraphics[width=8cm]{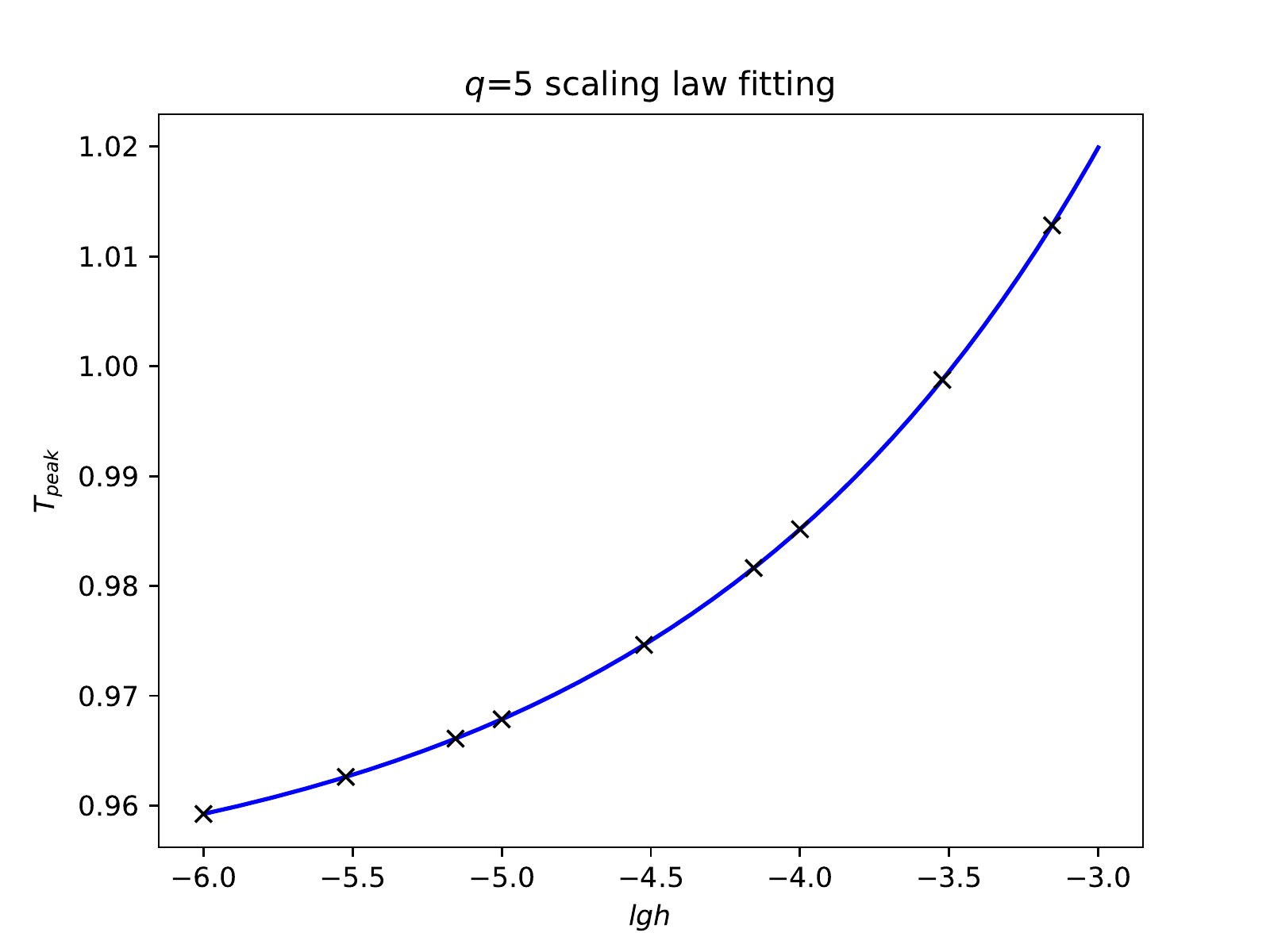}
\caption{Susceptibility peak temperature versus external field for $q=5$
model, from which we find that $T_{c}=0.9507(5)$, $a=0.5605$, $b=0.3028$.}
\label{F11}
\end{figure}

\begin{figure}[h]
\includegraphics[width=8cm]{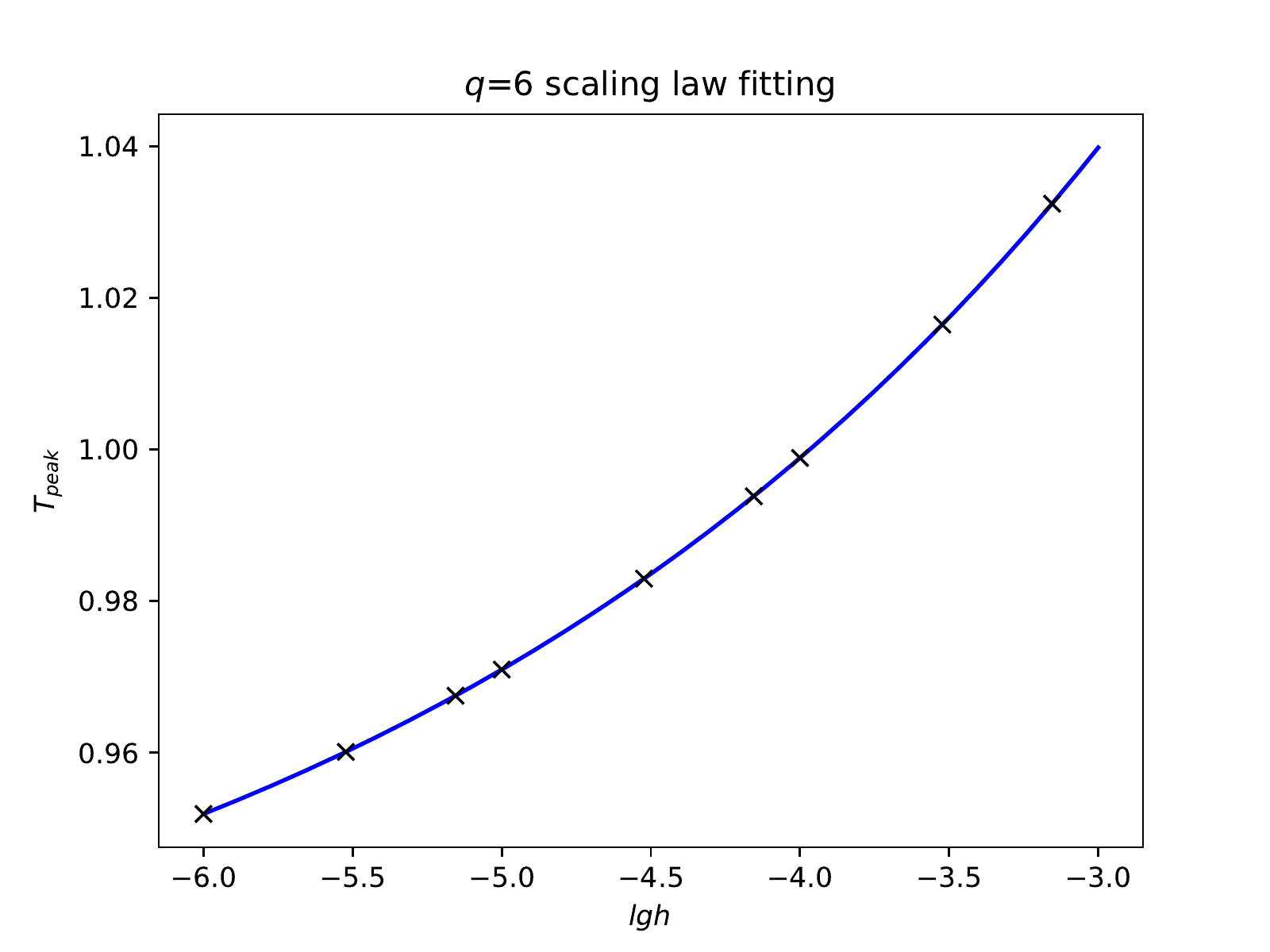}
\caption{Susceptibility peak temperature versus external field for $q=6$
model, from which we find that $T_{c}=0.9111(5)$, $a=0.4057$, $b=0.1662$.}
\label{F10}
\end{figure}

\begin{table}[h]
\centering
\begin{tabular}{ccc}
\hline\hline
& $T_{c1}$ & $T_{c2}$ \\ \hline
\multicolumn{3}{c}{$q=5$} \\ \hline
Ref.\cite{Toboc} & 0.8 & 1.1 \\
Ref.\cite{Bori1} & 0.905(1) & 0.951(1) \\
Ref.\cite{Kuman} & 0.908 & 0.944 \\
Ref.\cite{Chate} & 0.914(12) & 0.945(17) \\
Ref.\cite{Chatt} & 0.897(1) & - \\
Ref.\cite{ChenY} & 0.9029(1) & 0.9520(1) \\
Ref.\cite{Suru} & 0.911(5) & 0.940(5) \\
Ref.\cite{SHong} & 0.908 & 0.945 \\
Ref.\cite{Li} & 0.9059(2) & 0.9521(2) \\
our result & 0.908(2) & 0.9507(5) \\ \hline
\multicolumn{3}{c}{$q=6$} \\ \hline
Ref.\cite{Toboc} & 0.6 & 1.3 \\
Ref.\cite{Chall} & 68(2) & 0.92(1) \\
Ref.\cite{Yamag} & 0.68 & 0.90 \\
Ref.\cite{Tomit} & 0.7014(11) & 0.9008(6) \\
Ref.\cite{Hwang} & 0.632(2) & 0.997(2) \\
Ref.\cite{Brito} & 0.68(1) & 0.90(1) \\
Ref.\cite{Baek1} & - & 0.9020(5) \\
Ref.\cite{Kuman} & 0.700(4) & 0.904(5) \\
Ref.\cite{Krcma} & 0.70 & 0.88 \\
Ref.\cite{ChenJ} & 0.6658(5) & 0.8804(2) \\
Ref.\cite{Chatt} & 0.681(1) & - \\
Ref.\cite{Suru} & 0.701(5) & 0.898(5) \\
Ref.\cite{SHong} & 0.693 & 0.904 \\
Ref.\cite{Li} & 0.6901(4) & 0.9127(5) \\
our results & 0.696(2) & 0.9111(5) \\ \hline
\end{tabular}
\caption{A comparison of $T_{c1}$ and  $T_{c2}$ with previous results by using other methods.}
\label{compare}
\end{table}

Similar to the $q<5$ model, two transition points of $q=5$ model can be read out from the gauge invariant
quantity $\chi$. In Fig. \ref{F6}, we plot $\chi $ as a function of temperature near the
critical point. Very different from the $q<5$ model, there is no sharp change in $\chi$ near the two phase transition points. Similar to the $q<5$ model, in ordered phase, the tensor would
flow to the fixed point with $\chi=5$, while in disordered phase, the fixed
point tensor gives rise to $\chi=1$. However, in the middle phase, The structure of fixed point tensor is very complicated and we will discuss the details later. An interesting feature is that the gauge invariant quantity $\chi$ becomes size independent in the middle critical phase and this help us pin down the critical temperature for both high-temperature and low-temperature phase transitions.
As seen from Fig. \ref{F6}, we can read out that the low temperature
symmetry breaking transition point $T_{c1}$ is around 0.908(2) while the high-temperature BKT phase
transition point is around 0.952(2). Similar analysis can be applied to $q=6$ model as well, and we can read out from Fig. \ref{F20} that the low-temperature
critical point $T_{c1}$ is around 0.696(2), and high-temperature phase
transition point $T_{c2}$ is around 0.912(2). We note that in order to increase the accuracy, here and below we will use the $Z_q$ symmetric loop-TNR algorithm(see Appendix \ref{symmetry} for more details) with $D_{cut}=8q$ for simulating all $q$-state clock models.

Since the middle phase is
described by compactified boson model, we can further use the fixed point tensor to compute its central charge and scaling dimensions. As seen in Fig. \ref{F5}, we find $c=0.99987$ for $q=5$ model with $T=0.93k_{B}/J$, which is intrinsically close to the theoretical prediction with $c=1$.
It is well known that the
scaling dimensions of the primary fields of the compactified boson model can be expressed as:%
\begin{equation}
\Delta _{e,m}=\frac{m^{2}}{R^{2}}+\frac{e^{2}R^{2}}{4},\label{dimension}
\end{equation}%
where $R$ is the compactified radius and $m,e$ are integers which label the primary fields. In Fig. \ref{F5}, we also plot the scaling dimension for $q=5$ model with $T=0.93k_{B}/J$. We find that all the low scaling dimension can be fit quite well with $R=3.08607$.(We choose the scaling dimensions of RG steps from 15-20 to fit the compactification radius $R$). 
We note that the deviations for high scaling dimensions are due to the numerical error and we can further improve the accuracy by increasing $D_{cut}$ in the loop-TNR algorithm.


\begin{figure}[h]
\begin{subfigure}{.5\textwidth}
\centering
\includegraphics[width=8cm]{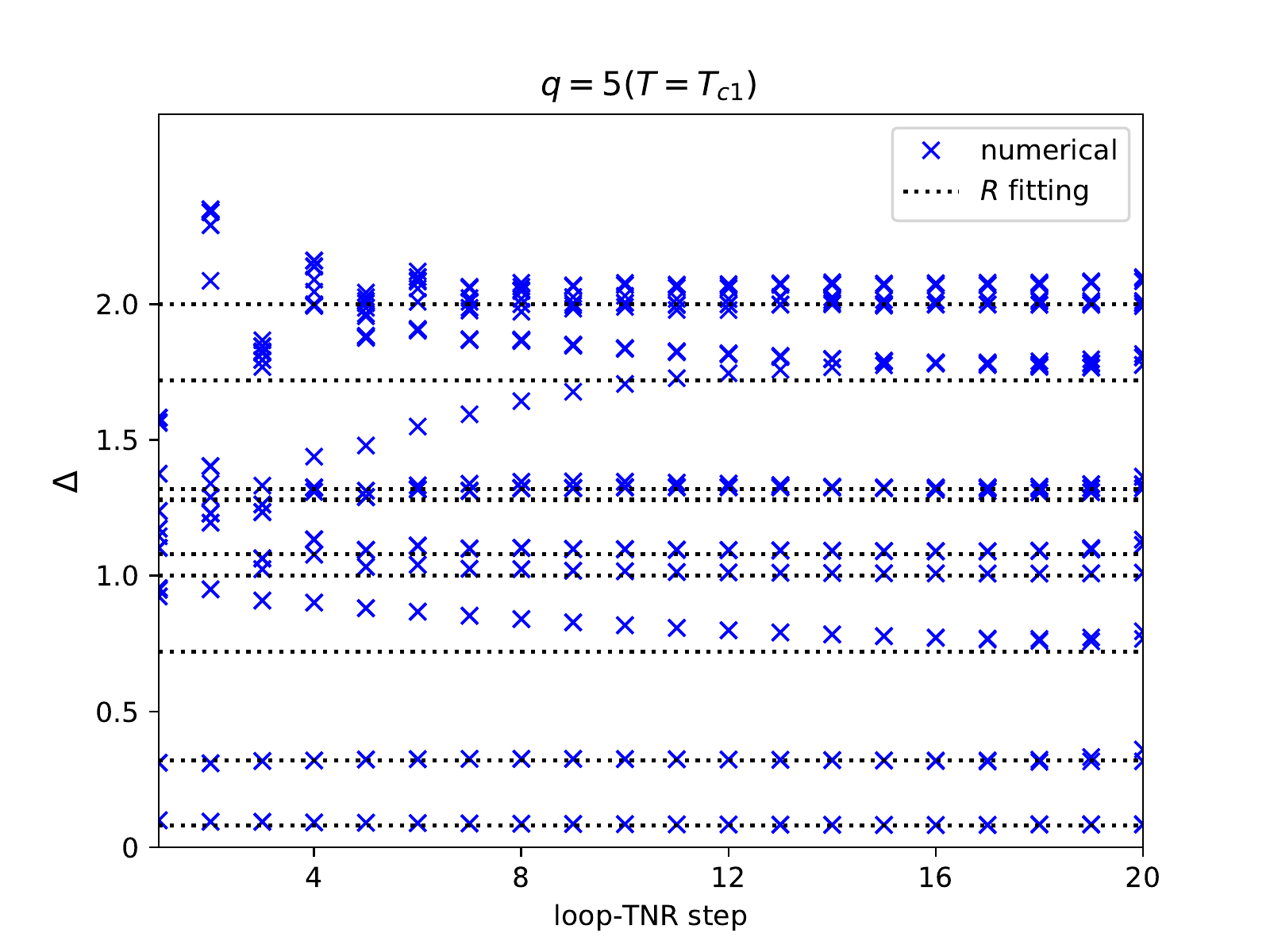}
\end{subfigure}
\par
\begin{subfigure}{.5\textwidth}
\includegraphics[width=8cm]{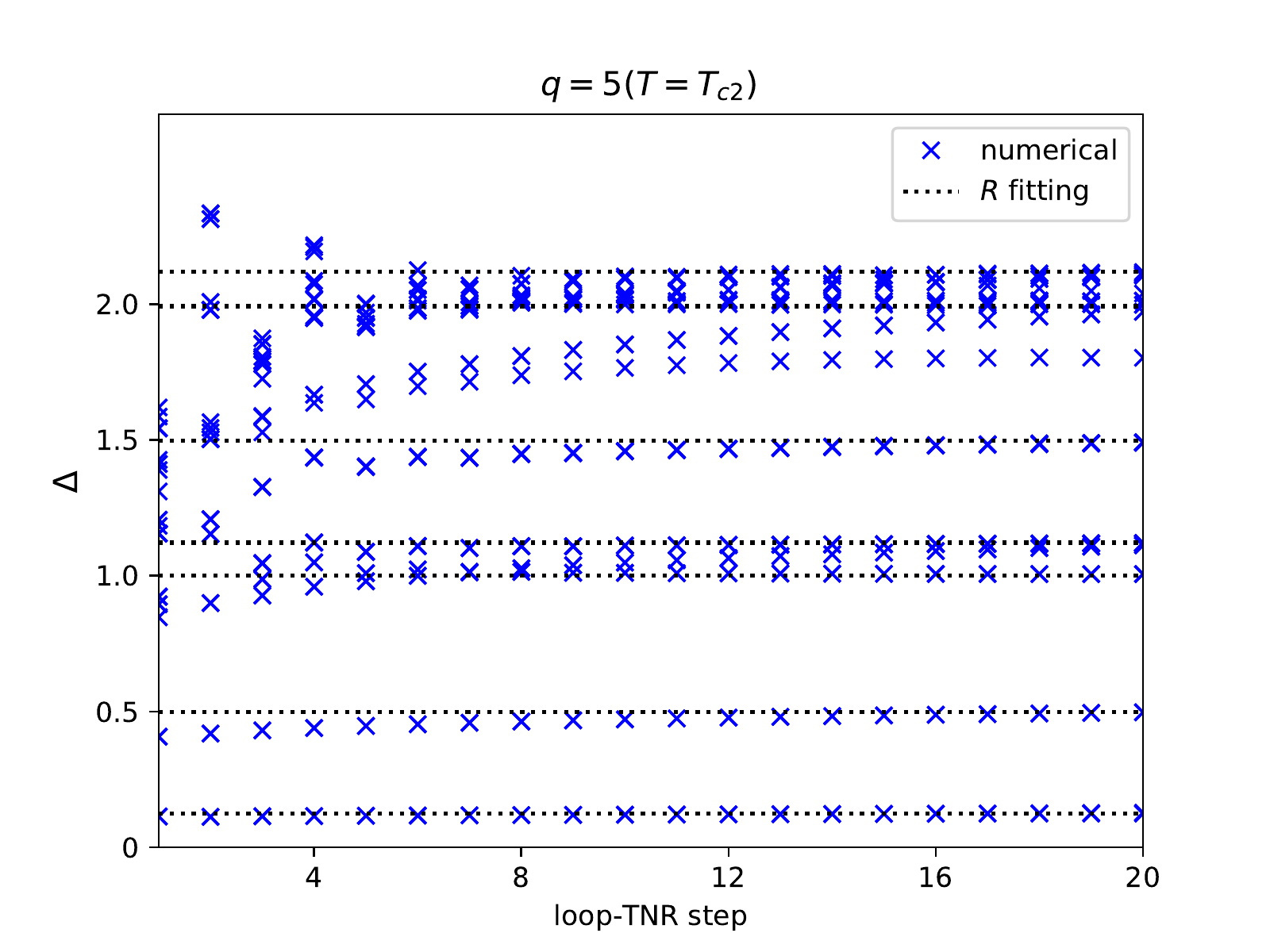}
\end{subfigure}
\caption{Scaling dimensions at the critical point $T_{c1}$ and $T_{c2}$
for $q=5$ model, from which we can fit the compactifiction radius $R$ of the compactified boson theory. We find that at $T_{c1}$, $R = 3.52954$, and at $T_{c2}$, $R = 2.83894$.}
\label{scaling5}
\end{figure}

\begin{figure}[h]
\begin{subfigure}{.5\textwidth}
\centering
\includegraphics[width=8cm]{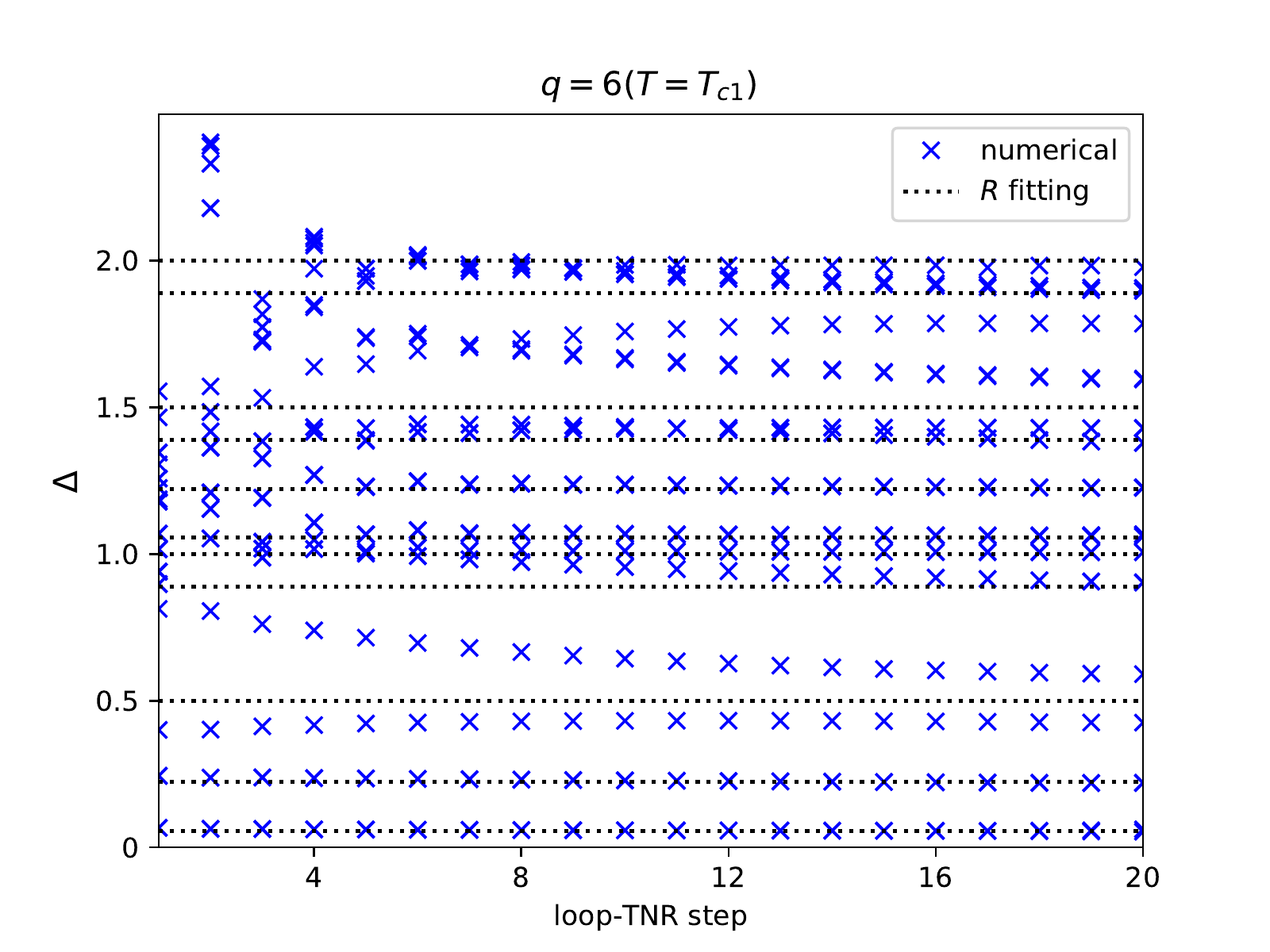}
\end{subfigure}
\par
\begin{subfigure}{.5\textwidth}
\includegraphics[width=8cm]{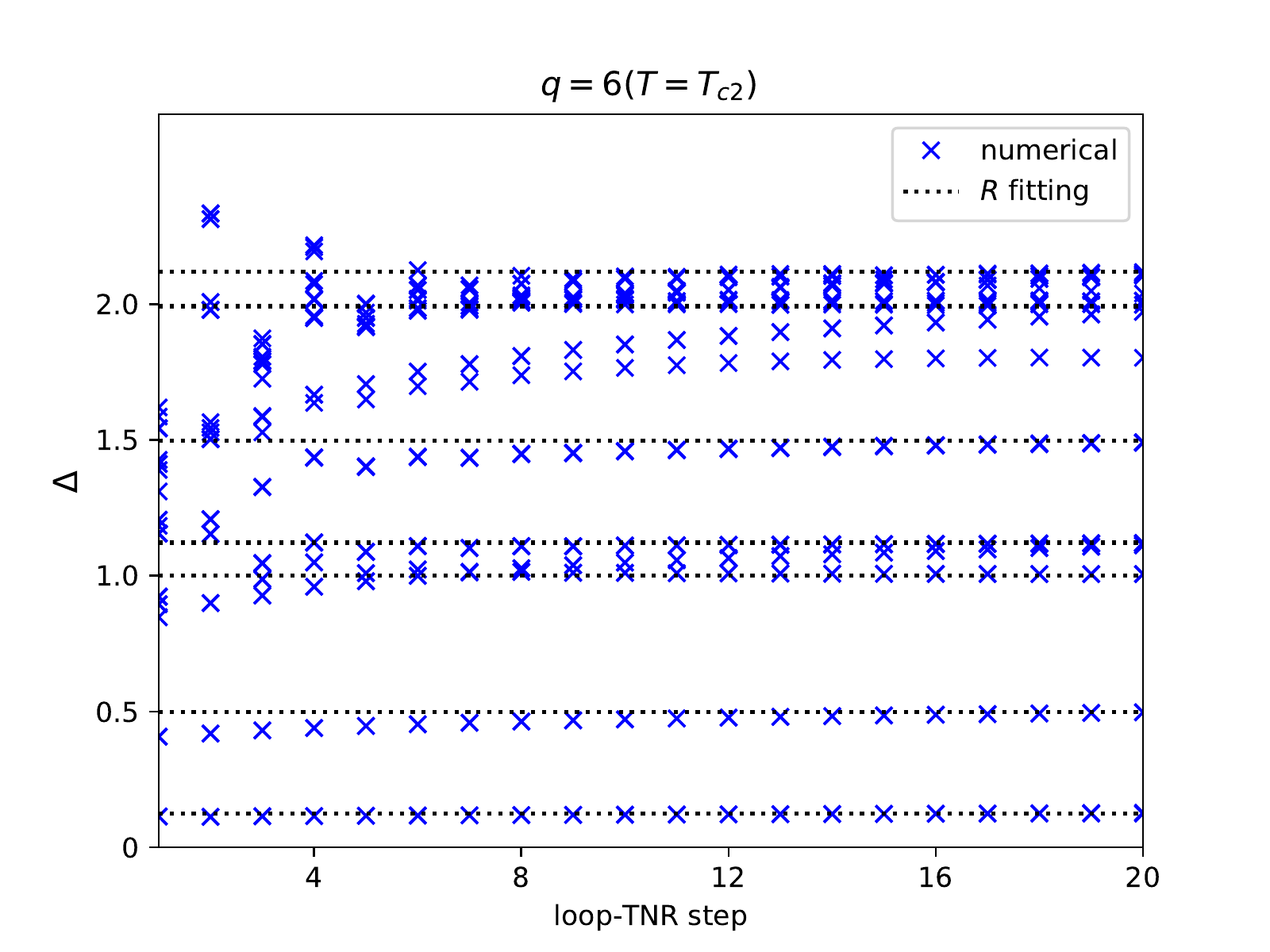}
\end{subfigure}
\caption{Scaling dimensions at the critical point $T_{c1}$ and $T_{c2}$
for $q=6$ model, from which we can fit the compactification radius $R$ of the compactified boson theory. We find that at $T_{c1}$, $R = 4.23870$, and at $T_{c2}$, $R = 2.82024$.}
\label{scaling6}
\end{figure}


The BKT transition point $T_{c2}$ can also be determined by the susceptibility peak method with extremely high accuracy.
First, by applying a very small external field, we can compute the susceptibility at different external field $h$ and temperature $T$\cite{Yu}:
\begin{equation}
\chi \left( h,T\right) =\left. \frac{\partial m}{\partial h}\right\vert _{T}.
\end{equation}%
For example, in Fig. \ref{F9}, we plot the susceptibility function at different system size for $q=5$ and $q=6$ models with a very small external field $h=10^{-5}$. We see that all the susceptibility functions collapse to a single curve, which implies the thermodynamic limit has already been achieved for physical quantities despite the fact that the gauge invariant quantity $\chi$ still has very strong size dependence near both critical temperatures.
By plotting the peak position of $\chi $ with different external fields, we can read
out $T_{c2}$ by using the following formula:
\begin{equation}
T_{peak}\left( h\right) =T_{c}+ah^{b}.
\label{peak}
\end{equation}%
We find that for
$q=5$ model, $T_{c2}=0.9507(5)$, $a=0.5605$, $b=0.3028$, and for $q=6$ model,
$T_{c2}=0.9111(5)$, $a=0.4057$, $b=0.1662$. Fig. \ref{F11} and Fig. \ref{F10}
show the susceptibility-peak fitting for $q=5$ and $q=6$ models, respectively. We see that the results
of $T_{c2}$ is comparable with what we get from the gauge
invariant quantity $\chi$.

In Table \ref{compare}, we compare our results with all previous known results for $T_{c1}$ and $T_{c2} $ using other methods. We see that our method gives much more accurate critical temperatures than HOTRG based method\cite{ChenJ,ChenY}, and the results are comparable with recent MPS based method\cite{Li} and large scale Monte-Carlo results\cite{Bori1,Suru}. We note that the small disagreement in the last digit might arise from the finite size effect in other methods.  Our loop-TNR method can handle system size up to $2^{23}$ with very high accuracy.


We further compute the scaling dimensions at
$T_{c1}$ and $T_{c2}$ for of both $q=5$ and $q=6$ models. From the results of scaling
dimension at each RG step, we can clearly observe the logarithmic flow of
some higher scaling dimensions, as seen in Fig. \ref{scaling5} and Fig. \ref{scaling6}. This implies  the existence of marginal irrelevant terms\cite{marginal}
for these transition points, and it explains why these transition points are very hard to be determined accurately in previous studies. From the scaling dimensions, we can fit the compactification radius $R$ by using Eq. (\ref{dimension}).  In Table \ref{chaR}, we list the compactification radius $R$ at both transition points and we find a perfect agreement with the field theory predictions. We stress that comparing with the very recent studies by using MPS based method\cite{Li}, our results give rise to much more accurate compactification radius $R$ at these phase transition points.

\begin{figure}[h]
\centering
\includegraphics[width=8cm]{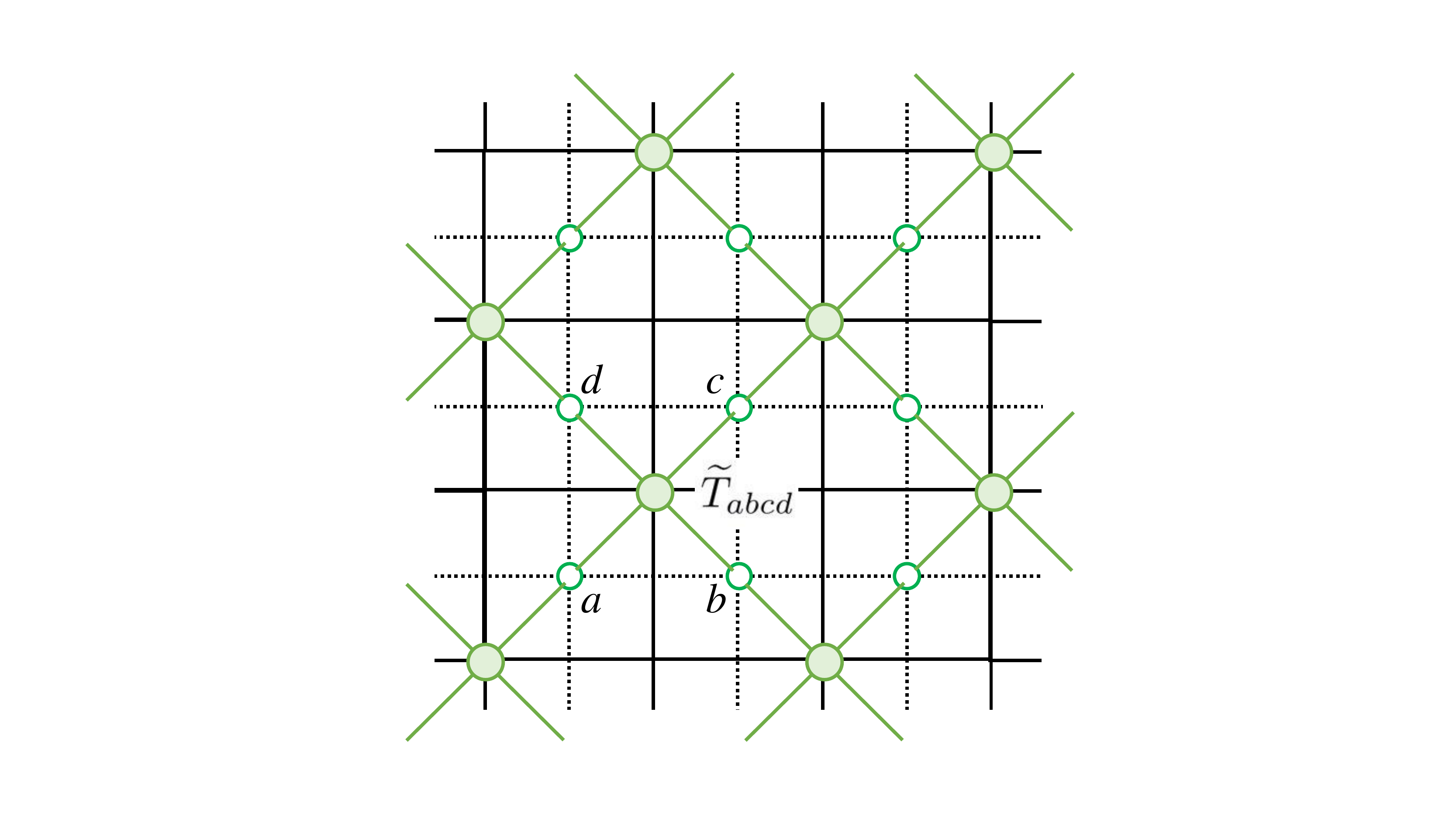}
\caption{Tensor network representation of dual model,
where the original lattice is shown by solid line and the dual lattice is shown
by dash line.}
\label{dual}
\end{figure}

\begin{figure}[h]
\begin{subfigure}{.5\textwidth}
\includegraphics[width=8cm]{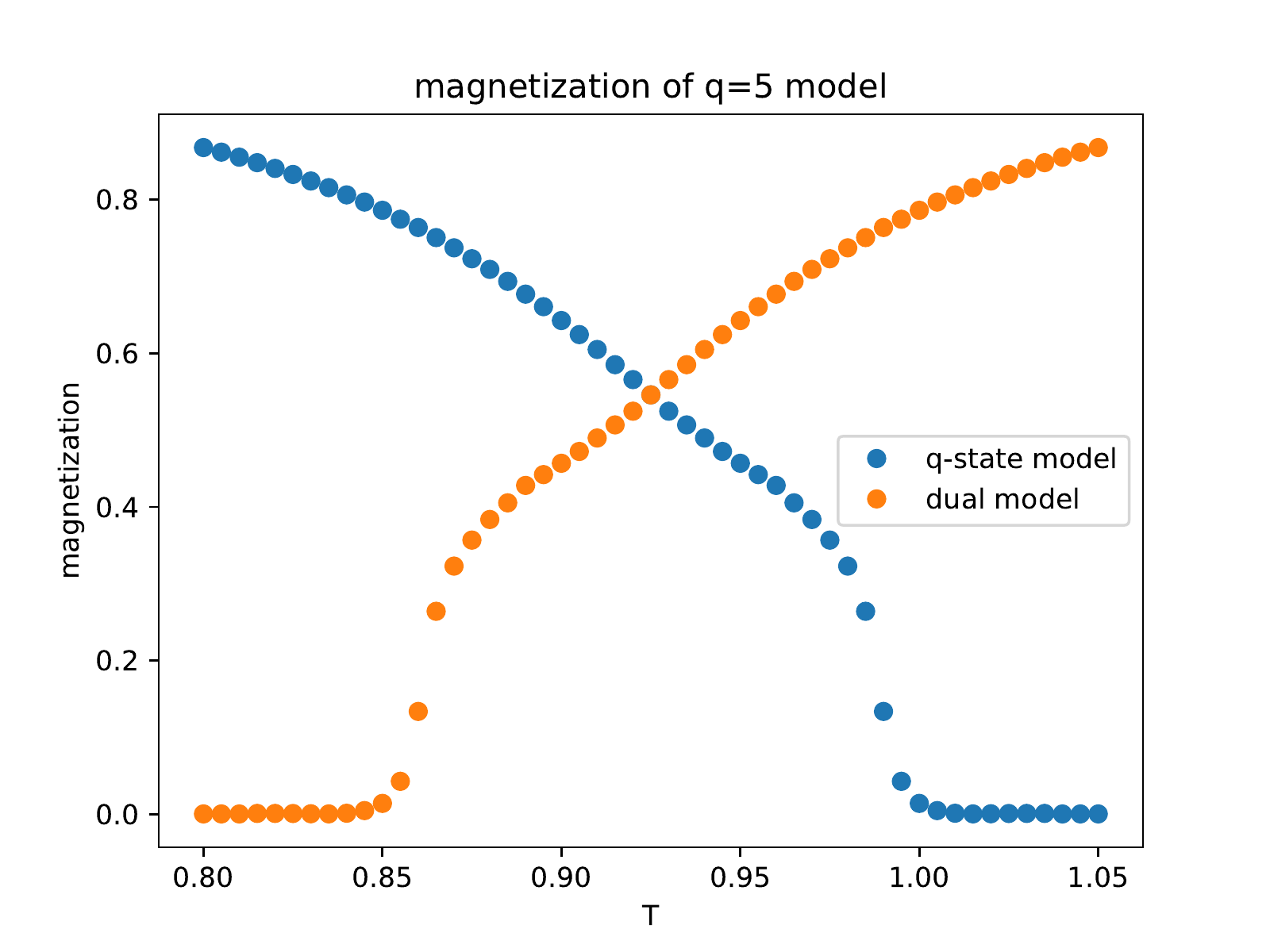}
\end{subfigure}
\begin{subfigure}{.5\textwidth}
\includegraphics[width=8cm]{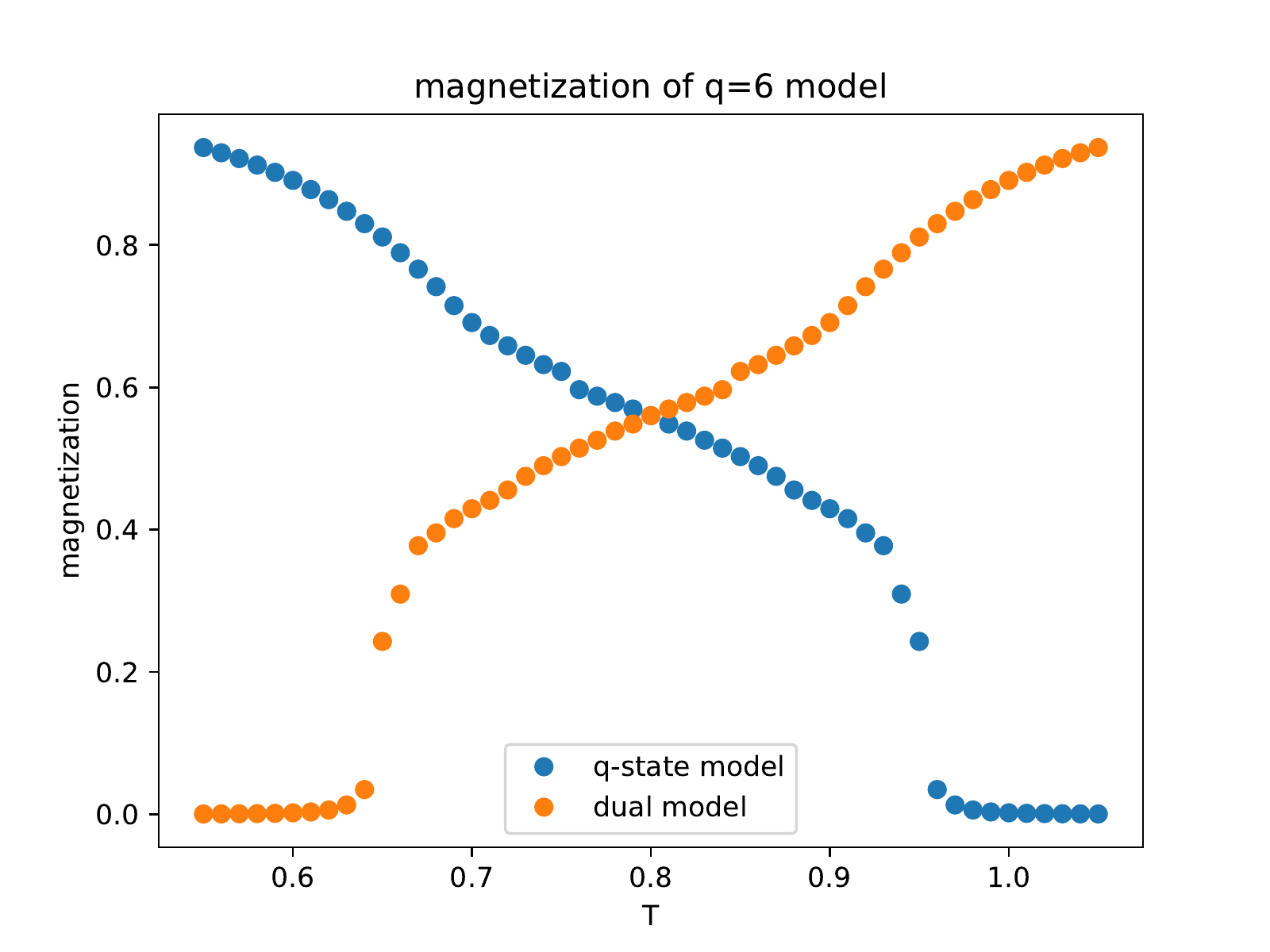}
\end{subfigure}
\caption{Magnetization of $q$-state clock model with $q=5,6$ and their corresponding dual model.}
\label{mag}
\end{figure}

\begin{figure}[h]
\begin{subfigure}{.5\textwidth}
\includegraphics[width=8cm]{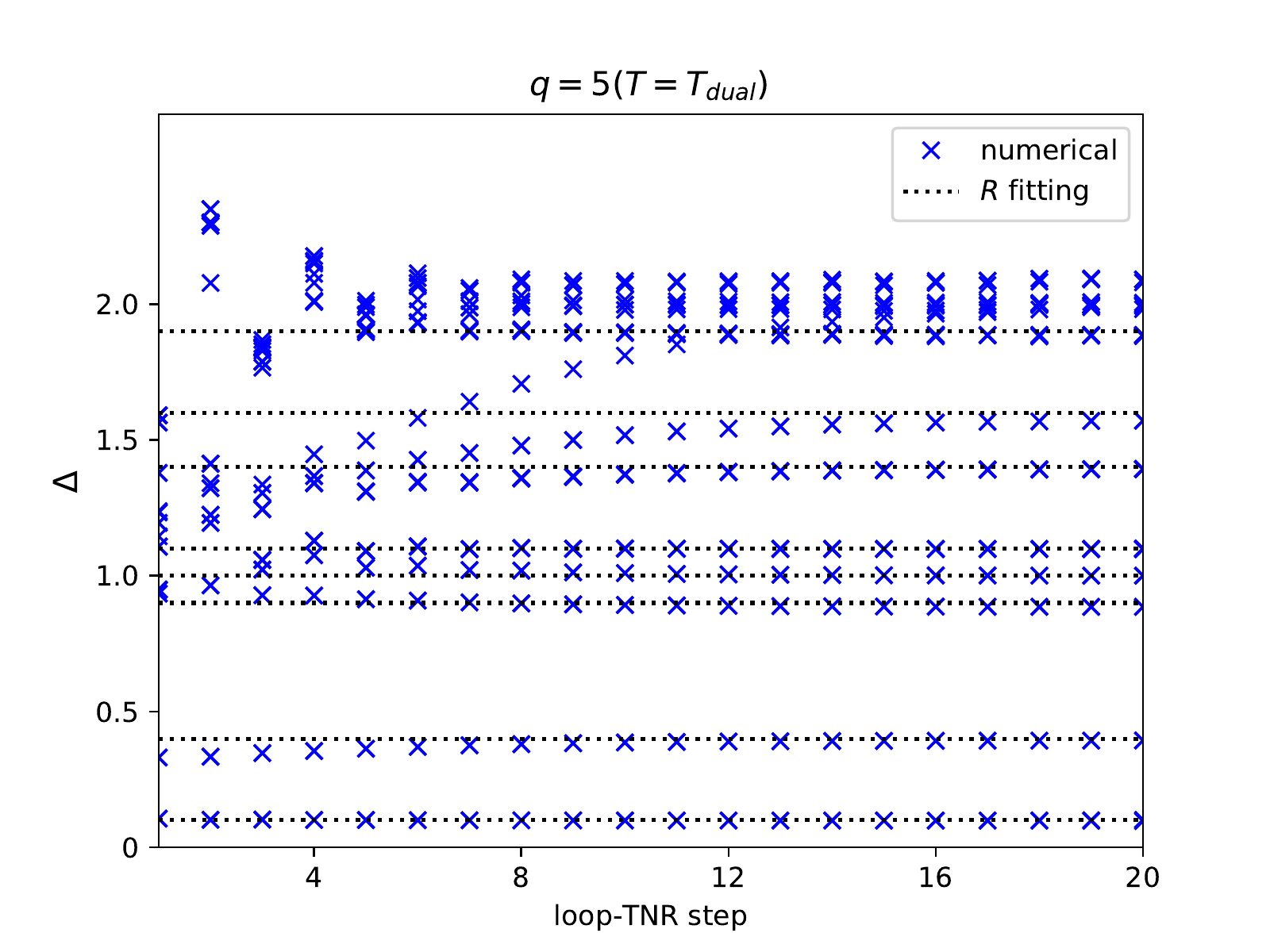}
\end{subfigure}
\par
\begin{subfigure}{.5\textwidth}
\includegraphics[width=8cm]{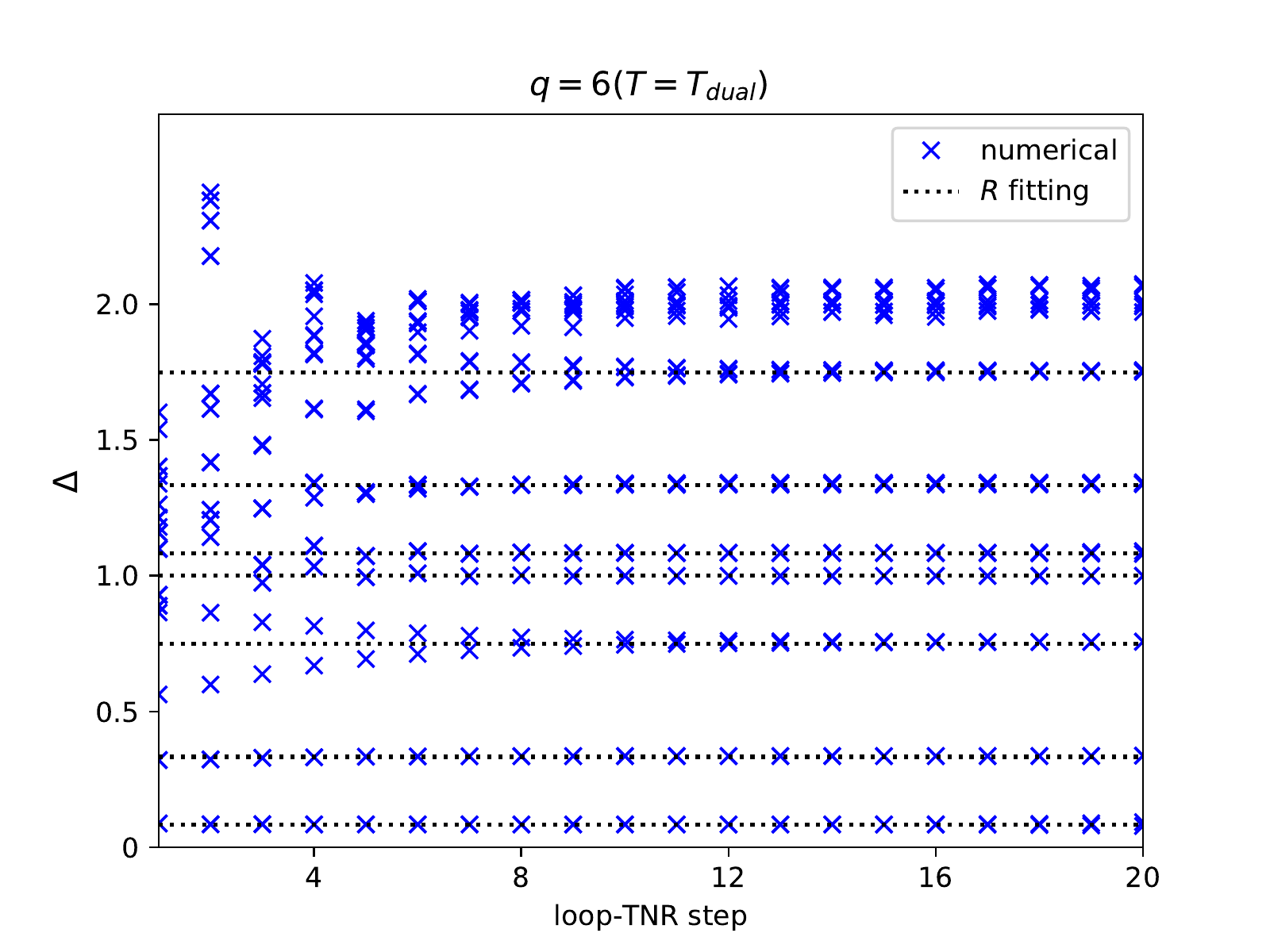}
\end{subfigure}
\caption{Scaling dimension on self-dual point for $q=5$ and $q=6$ models, from which we can fit the compactification radius $R$ of the compactified boson theory. We find that $R=3.17354$ for $q=5$ and $R=3.46002$ for $q=6$ model.}
\label{dualscaling}
\end{figure}

Finally, we investigate the scaling dimensions and compactification radius $R$ for the so-called self-dual point. The element tensor for dual model in
Fig. \ref{dual}, which is obtained by Kramers-Wannier transformation\cite{Kramers,Kramers2,Zamol},
could be expressed as:
\begin{eqnarray}
\widetilde{T}_{abcd}&=&\exp \frac{\beta}{2} \left( \cos \theta _{a}+\cos \theta _{b}+\cos \theta
_{c}+\cos \theta _{d}\right)\nonumber\\
&\times&\delta_{\rm{mod}(a+b+c+d,q),0}
\end{eqnarray}%
To determine the self-dual temperature, we compute the magnetization at different temperatures for both $q$-state model and its dual model. As seen in Fig. \ref{mag}, the crossing point corresponds to the dual temperature with $g_1=g_2$.
Again, we can use the loop-TNR algorithm to compute the scaling dimensions(see in Fig. \ref{dualscaling}) and from the scaling dimension data, we can further fit the compactification $R$.
In Table \ref{chaR}, we compare our results with the theoretical predictions. Again, we find a perfect agreement for both $q=5$ and $q=6$ models.

\begin{table}[tbp]
\centering
\begin{tabular}{ccccccc}
\hline\hline
$q$ & \multicolumn{2}{c}{$T_{c1}$} & \multicolumn{2}{c}{$T_{dual}$} & \multicolumn{2}{c}{$T_{c2}$} \\ \hline
& theory & numerical & theory & numerical & theory & numerical\\
5 & $\sqrt{25/2}$ & 3.52954 & $\sqrt{10}$ & 3.17354 & $2\sqrt{2}$ & 2.83894 \\
6 & $\sqrt{18}$ & 4.23870 & $\sqrt{12}$ & 3.46002 & $2\sqrt{2}$ & 2.82024 \\
\hline\hline
\end{tabular}%
\caption{Compactification radius $R$ on both critical points as well as self-dual point of $q$-state clock model with $q=5$ and $q=6$.}
\label{chaR}
\end{table}

\section{$q > 6$ models and fixed point tensor for BKT transition}
\subsection{Critical temperature and compactification radius}

By using the same methods for $q=5$ and $q=6$ models, we also studies
the phase diagram for $q>6$ models. By computing both the gauge invariant quantity $\chi$ and fitting the susceptibility peak position under different external field, we can determine both $T_{c1}$ and $T_{c2}$ accurately. In Table \ref{compare1}, we compare our results for $q=7,8,9$ models with previous studies using other methods. Remarkably, for models with big enough $q$, i.e., $q>6$, $T_{c2}$ becomes very close to the BKT transition value in classical XY model with $T_{c}=0.8929$.

Similar to $q=5$ and $q=6$ models, we can also use loop-TNR method to compute the scaling dimensions and fit the corresponding compactification radius $R$, see Appendix \ref{largeq} for more details. We find that the radius $R$ at $T_{c2}$ also saturates to a fixed value 2.81987 for big enough $q$, which is intrinsically close to the theoretical prediction $R=2\sqrt{2}$. We can also use the same method for $q=5$ and $q=6$ models to determine the self-dual point and fit the corresponding compactification radius $R$. In Table \ref{chaR1}, we also list the compactification radius $R$ for $T_{c1}$ and self-dual point $T_{dual}$. Again, we find a perfect agreement with theoretical predictions.

\begin{table}[tbp]
\centering
\begin{tabular}{ccc}
\hline\hline
& $T_{c1}$ & $T_{c2}$ \\ \hline
\multicolumn{3}{c}{$q=7$} \\ \hline
Ref.\cite{Bori2} & 0.533 & 0.900 \\
Ref.\cite{Chatt} & 0.531(6) & - \\
Ref.\cite{Li} & 0.5305(3) & 0.9071(5) \\
our results & 0.536(2) & 0.9065(5) \\ \hline
\multicolumn{3}{c}{$q=8$} \\ \hline
Ref.\cite{Tomit} & 0.4259(4) & 0.8936(7) \\
Ref.\cite{Baek} & 0.417(3) & 0.894(1) \\
Ref.\cite{Chatt} & 0.418(1) & - \\
Ref.\cite{Li} & 0.4172(3) & 0.9060(5) \\
our results & 0.4215(15) & 0.9051(5) \\ \hline
\multicolumn{3}{c}{$q=9$} \\ \hline
Ref.\cite{Chatt} & 0.334(1) & - \\
our results & 0.342(2) & 0.9051(5) \\ \hline\hline
\end{tabular}%
\caption{A comparison of $T_{c1}$ and  $T_{c2}$ with previous results by using other methods.}
\label{compare1}
\end{table}

\begin{table}[tbp]
\centering
\begin{tabular}{ccccccc}
\hline\hline
$q$ & \multicolumn{2}{c}{$T_{c1}$} & \multicolumn{2}{c}{$T_{dual}$} & \multicolumn{2}{c}{$T_{c2}$} \\ \hline
& theory & numerical & theory & numerical & theory & numerical\\
7 & $\sqrt{49/2}$ & 4.94072 & $\sqrt{14}$ & 3.75035 & $2\sqrt{2}$ & 2.83153 \\
8 & $\sqrt{32}$ & 5.67377 & $\sqrt{16}$ & 4.00726 & $2\sqrt{2}$ & 2.81987 \\
9 & $\sqrt{81/2}$ & 6.36759 & $\sqrt{18}$ & 4.23573 & $2\sqrt{2}$ & 2.81987 \\ \hline\hline
\end{tabular}%
\caption{Compactification radius $R$ on critical points and self-dual point for $q=7,8,9$ models.}
\label{chaR1}
\end{table}


\subsection{Fixed point tensor for BKT transition}
Since the $T_{c2}$ for $q>6$ models is already very close to the BKT transition in classical XY model, and the compactification radius $R$ is also approaching the expected value for BKT transition, it is natural to ask whether the corresponding fixed point tensors in these models also converge to the same one(up to numerical errors) which contains the complete information for BKT transitions. Below we will study the structure of fixed point tensor for $q>6$ models at BKT transition and try to read out the OPE coefficient directly for the corresponding compactified boson CFT.

\subsubsection{The gauge choice of the fixed point tensor}

It is well known that there exists a gauge degree of freedom for the fixed point tensor in any TNR scheme and it is actually the major difficulty for us to understand the full structure of fixed point tensors for critical systems.

We will begin with some general discussion for the nature of such a gauge degree of freedom and explain why it can be fixed by introducing enough symmetry conditions. Apparently, if we apply some invertible matrices on every legs of a tensor, the transformed
tensor actually forms the same tensor network as before:
\begin{equation}
T_{ijkl}^{\prime }=\underset{i^{\prime }j^{\prime }k^{\prime }l^{\prime }}{%
\sum }T_{i^{\prime }j^{\prime }k^{\prime }l^{\prime }}U_{i^{\prime
}i}V_{j^{\prime }j}\left[ U^{-1}\right] _{kk^{\prime }}\left[ V^{-1}\right]
_{ll^{\prime }}\label{gauge}
\end{equation}%

This gives rise to great difficulty to analyze the properties of the tensor
components of the fixed point tensor, since they could be randomly affected by the gauge choice in numerical calculations. To
get a proper gauge fixing, we have the following considerations:

$\bullet$ The fixed point tensor(defined on the 2 by 2 plaquette composed by $T_{A}$ and $T_{B}$ tensors, as shown in Fig. \ref{F42}) should preserve the $C_{4}$ lattice symmetry during the loop-TNR process(see Appendix \ref{symmetry} for more details). Preserving $C_{4}$ symmetry will reduce the gauge freedom of the fixed
point tensor. The gauge transformation in Eq. (\ref{gauge}) can be simplified as:
\begin{equation}
T_{ijkl}^{\prime }=\underset{i^{\prime }j^{\prime }k^{\prime }l^{\prime }}{%
\sum }T_{i^{\prime }j^{\prime }k^{\prime }l^{\prime }}O_{i^{\prime
}i}O_{j^{\prime }j}O_{k^\prime k }O
_{l^{\prime }l}\label{gauge1}
\end{equation}%
where $O$ is an orthogonal matrix.(We assume all the tensors are real.)

$\bullet$ Since the $q$-state clock model has a $Z_q$ internal symmetry, we should also keep such an internal symmetry during the whole loop-TNR process(see Appendix \ref{symmetry} for more details). By keeping the
$Z_q$ symmetry, we can further reduce the gauge degrees of freedom.   In fact this is a crucial step to obtain the right fusion rule for fixed point tensor. It is well known that the fusion rule of compactifiled boson theory has a $U(1)$ symmetry which can be realized explicitly on $XY$ model. However, if we only focus on the leading components of primary fields and descendant fields, $Z_q$ symmetry is a very good approximation for $U(1)$.

\begin{figure}[h]
\centering
\includegraphics[width=9cm]{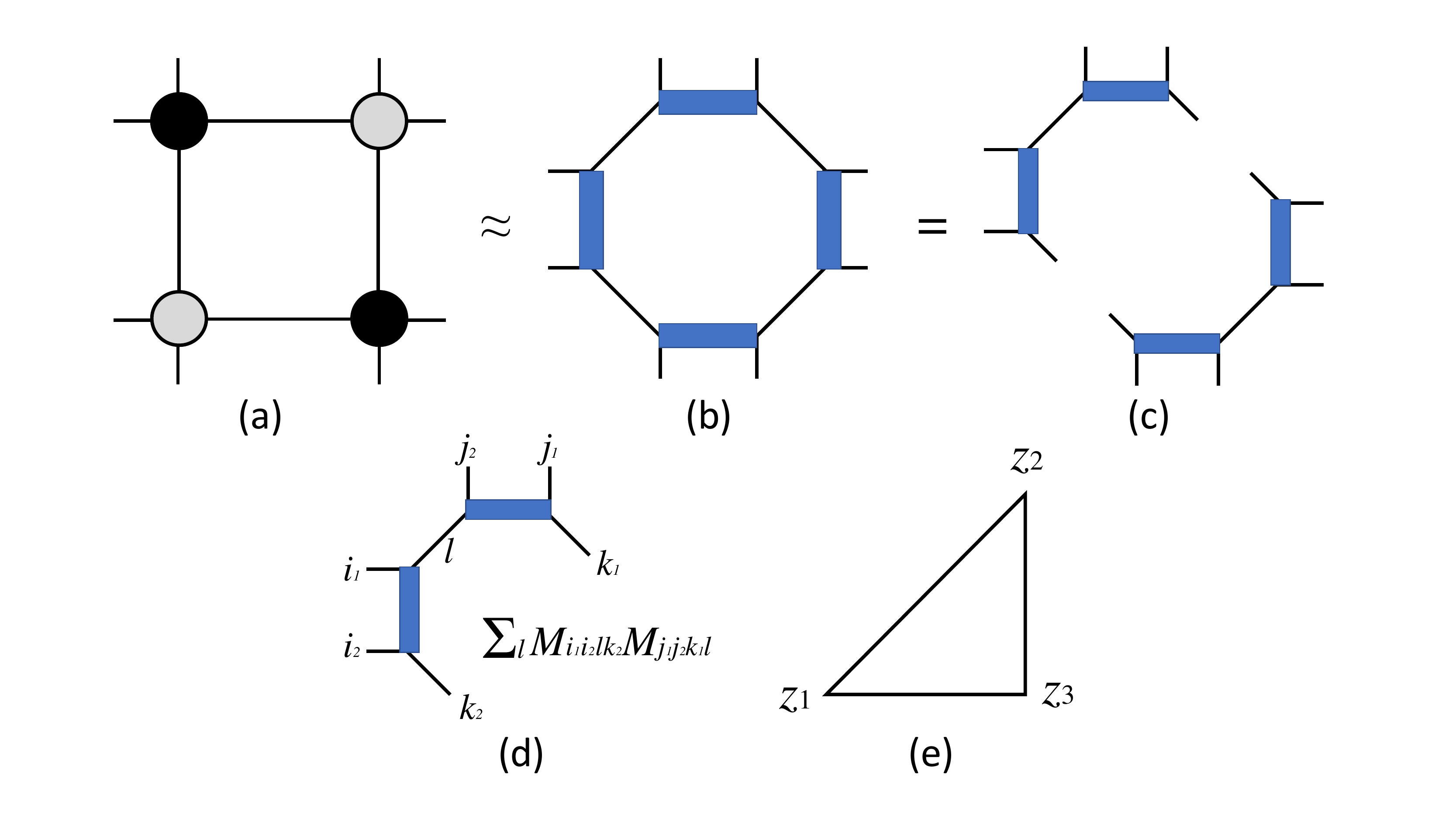}
\caption{From loop-TNR algorithm, a square fixed point tensor (a) could be approximately represented
by MPS on the octagon lattice (b). Then, we decompose octagon MPS from (b)
to (c). The rank-3 tensor in (d) is the fixed point tensor we will study here. (e) The geometry of the corresponding 3-point correlation functions.}
\label{fixten2}
\end{figure}
\begin{figure}[h]
\centering
\includegraphics[width=9cm]{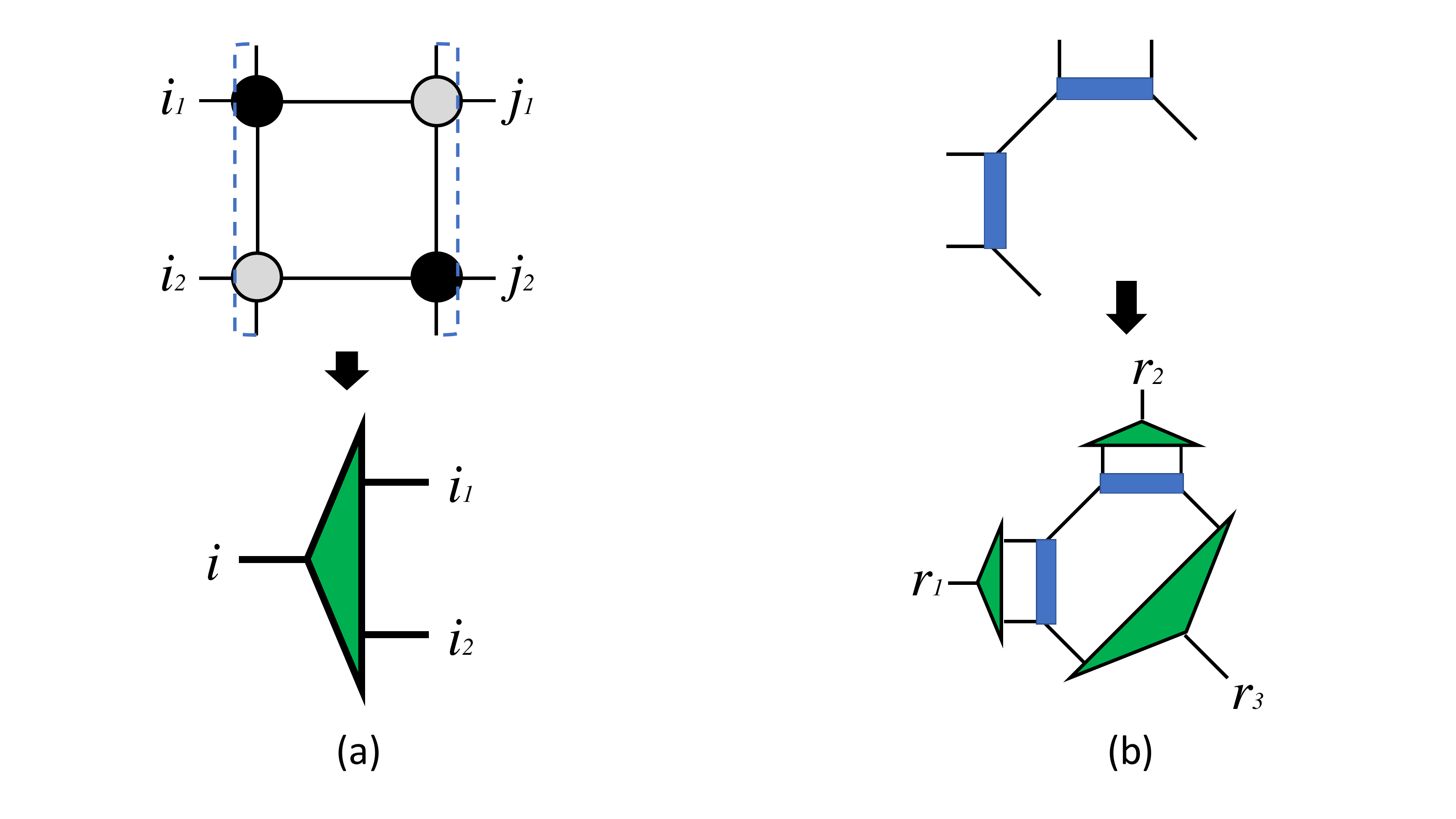}
\caption{(a) We choose the eigenstates of the transfer matrix in as our basis. (b) We then
project the rank-3 fixed point tensors on to these basis.}
\label{fixten3}
\end{figure}

$\bullet$ If we want the indices of the fixed point tensor to represent the
primary fields and their descendants for the corresponding compactified boson theory, we need to choose a proper basis. The eigenstate of the transfer matrix is a good choice.
As shown in Fig. \ref{fixten2} (d), we construct a rank-3 tensor with the building block tensor $M_{ijkl}$ in
$C_{4}$-loop-TNR algorithm(see Appendix \ref{symmetry} for more details). This is because in usual CFT, the 3-point correlation function is more fundamental and has a much simpler
form than the 4-point correlation function. In fact, the basic renormalzation step in loop-TNR is similar to the crossing symmetry for 4-point correlation function. Thus, we conjecture that the rank-3 tensor constructed here could be regarded as a 3-point correlation function(at least for primary fields).
As illustrated in Fig. \ref{fixten3}, we construct the $2\times 2$ transfer matrix as shown in Fig \ref{fixten3} (a), and apply the eigenvalue
decomposition:
\begin{equation}
M_{(i_{1}i_{2})\left( j_{1}j_{2}\right) }=\underset{k}{\sum }%
U_{(i_{1}i_{2})k}\lambda _{k}\left[ U^{-1}\right] _{k(i_{1}i_{2})}.
\end{equation}%
We use
eigenvectors $U_{(i_{1}i_{2})i}$ as the basis for the fixed point tensor, as
shown in Fig. \ref{fixten3} (b). As a result, the fixed point tensor is projected onto the basis representing primary fields and their descendants.

\subsubsection{Operator product expansion(OPE) coefficient from the fixed point tensor}

In Table \ref{fixdata}, we list the leading non-zero
components of the fixed point tensors of different $q$-state clock models at BKT critical point. Here we
normalize the largest component $T_{III}=1$. We use
$I$, $\alpha$, $\beta$, $\gamma$, $\delta$, $\lambda$ and $\eta$ to represent the leading primary fields $(0,0)$, $(1,0)$,
$(-1,0)$, $(2,0)$, $(-2,0)$, $(3,0)$ and $(-3,0)$.

\begin{table}[tbp]
\centering
\begin{tabular}{ccccccc}
\hline\hline
$T^{q=7}_{r_{1}r_{2}r_{3}}$ & $T^{q=8}_{r_{1}r_{2}r_{3}}$ & $%
T^{q=9}_{r_{1}r_{2}r_{3}}$ & $T^{q=10}_{r_{1}r_{2}r_{3}}$ & $%
r_{1}$ & $r_{2}$ & $r_{3}$ \\ \hline
1.00000 & 1.00000 & 1.00000 & 1.00000 & $I$ & $I$ & $I$ \\
0.81215 & 0.81594 & 0.81555 & 0.81725 & $I$ & $\alpha $ & $\beta $ \\
0.81215 & 0.81594 & 0.81555 & 0.81725 & $I$ & $\beta $ & $\alpha $ \\
0.44178 & 0.44747 & 0.45242 & 0.45253 & $I$ & $\gamma $ & $\delta $ \\
0.44178 & 0.44747 & 0.45242 & 0.45253 & $I$ & $\delta $ & $\gamma $ \\
0.81215 & 0.81594 & 0.81555 & 0.81725 & $\alpha $ & $I$ & $\beta $ \\
0.59058 & 0.59609 & 0.59756 & 0.59821 & $\alpha $ & $\alpha $ & $\delta $ \\
0.74242 & 0.74698 & 0.74671 & 0.74901 & $\alpha $ & $\beta $ & $I$ \\
0.45419 & 0.46021 & 0.49495 & 0.46479 & $\alpha $ & $\delta $ & $\alpha $ \\
0.81215 & 0.81594 & 0.81555 & 0.81725 & $\beta $ & $I$ & $\alpha $ \\
0.74242 & 0.74698 & 0.74671 & 0.74901 & $\beta $ & $\alpha $ & $I$ \\
0.59058 & 0.59609 & 0.59756 & 0.59821 & $\beta $ & $\beta $ & $\gamma $ \\
0.45419 & 0.46021 & 0.46369 & 0.46479 & $\beta $ & $\gamma $ & $\beta $ \\
0.44178 & 0.44747 & 0.45242 & 0.45253 & $\gamma $ & $I$ & $\delta $ \\
0.45419 & 0.46021 & 0.46369 & 0.46479 & $\gamma $ & $\beta $ & $\beta $ \\
0.30919 & 0.31491 & 0.32114 & 0.32120 & $\gamma $ & $\delta $ & $I$ \\
0.44178 & 0.44747 & 0.45242 & 0.45253 & $\delta $ & $I$ & $\gamma $ \\
0.45419 & 0.46021 & 0.46369 & 0.46479 & $\delta $ & $\alpha $ & $\alpha $ \\
0.30919 & 0.31491 & 0.32114 & 0.32120 & $\delta $ & $\gamma $ & $I$ \\
\hline\hline
\end{tabular}%
\caption{A comparison of non-zero leading components of the fixed point tensor of q-state clock models with $q=7,8,9,10$ at BKT critical point.}
\label{fixdata}
\end{table}

It is well known that the fusion
rule of the primary fields in compactified boson theory satisfies:
\begin{equation}
\left[ \phi _{e_{1},m_{1}}\right] \times \left[ \phi _{e_{2},m_{2}}\right] =%
\left[ \phi _{e_{1}+e_{2},m_{1}+m_{2}}\right] ,
\end{equation}%
where $\left[ \phi _{e,m}\right] $ is a conformal family generated by
primary field $\phi _{e,m}$ with conformal dimension $\left( \left(
e/R+mR/2\right) ^{2}/2,\left( e/R-mR/2\right) ^{2}/2\right) $. In
particular, the primary field with $m=0$ 
is just the vertex operator and it can be written as:
\begin{equation}
\phi _{e,0}\left( z,\overline{z}\right) =e^{ie\varphi \left( z,\overline{z}%
\right) /R},
\end{equation}%
with $\varphi \left( z,\overline{z}\right) $ the free boson field. The 3-point function has a
pretty simple form:
\begin{eqnarray}
&&\langle \phi _{e_{1},0}\left( z_{1},\overline{z}_{1}\right) \phi
_{e_{2},0}\left( z_{2},\overline{z}_{2}\right) \phi _{e_{3},0}\left( z_{3},
\overline{z}_{3}\right) \rangle  \nonumber \\
&=& \frac{C_{123}}{\left\vert z_{12}\right\vert ^{\Delta _{1}+\Delta _{2}-\Delta
_{3}}\left\vert z_{23}\right\vert ^{\Delta _{2}+\Delta _{3}-\Delta
_{1}}\left\vert z_{31}\right\vert ^{\Delta _{3}+\Delta _{1}-\Delta _{2}}},
\label{3point}
\end{eqnarray}%
where $C_{123}$ is the OPE coefficient, which equals 1 for $e_{1}+e_{2}+e_{3}\neq 0$ and vanishes for $e_{1}+e_{2}+e_{3}=0$.
$\left\vert z_{12}\right\vert \equiv \left\vert z_{1}-z_{2}\right\vert$,
and the scaling dimension $\Delta_i=\frac{e_{i}^{2}}{R^{2}}$. We note that in general only leading primary fields in our numerical fixed point tensor can satisfy the fusion rule since we use the $Z_q$ symmetry to approximate the $U(1)$ symmetry in the gauge fixing procedure, and with increasing $q$, more and more primary fields with correct fusion rules can be resolved numerically. (Although we believe that the emergent $U(1)$ must be present for all finite $q$ with $q>4$, it is in general very hard to find the proper gauge choice for small $q$, especially for $q=5$ and $q=6$.)

Next, we can try to fit our numerical fixed point tensor by using the 3-point correlation function Eq.  (\ref{3point}). Let $z_{13}=\lambda _{1}x$, $%
z_{23}=\lambda _{2}x$, $z_{12}=\lambda _{3}x$. We can rewrite the right
hand side of Eq. (\ref{3point}) as:
\begin{eqnarray}
&&\frac{C_{123}}{\left( \lambda _{1}x\right) ^{\Delta _{1}+\Delta
_{3}-\Delta _{2}}\left( \lambda _{2}x\right) ^{\Delta _{2}+\Delta
_{3}-\Delta _{1}}\left( \lambda _{3}x\right) ^{\Delta _{1}+\Delta
_{2}-\Delta _{3}}}  \nonumber \\
&=&C_{123}\left( \frac{\lambda _{2}}{\lambda _{1}\lambda _{3}x}\right)
^{\Delta _{1}}\left( \frac{\lambda _{1}}{\lambda _{2}\lambda _{3}x}\right)
^{\Delta _{2}}\left( \frac{\lambda _{3}}{\lambda _{1}\lambda _{2}x}\right)
^{\Delta _{3}}  \nonumber \\
&\equiv &C_{123}l_{1}^{\Delta _{1}}l_{2}^{\Delta _{2}}l_{3}^{\Delta _{3}},
\label{fit1}
\end{eqnarray}%
with $l_{1}=\frac{\lambda _{2}}{\lambda _{1}\lambda _{3}x}$, $l_{2}=\frac{\lambda _{1}}{\lambda _{2}\lambda _{3}x}$ and $l_{3}=\frac{\lambda _{3}}{\lambda _{1}\lambda _{2}x}$, respectively. From the
geometry of the square lattice, we conjecture that our rank-3 fixed point tensor
can be regarded as 3-point correlation(at least for primary fields) function on the vertex of an isosceles right triangle on the complex plane, as seen in Fig. \ref{fixten2} (e). Thus we can choose $\lambda _{1}=\lambda _{2}=\lambda _{3}/\sqrt{2}=1$ and Eq. (\ref{fit1}) can be simplified as:
\begin{equation}
C_{123}l_{1}^{\Delta _{1}}l_{2}^{\Delta _{2}}l_{3}^{\Delta _{3}}\equiv
C_{123}l^{\Delta _{1}}l^{\Delta _{2}}\left( 2l\right) ^{\Delta _{3}}.
\label{fit2}
\end{equation}%
where $l=\frac{1}{\sqrt{2}x}$ is a fundamental inverse length scale.
For $q=10$ model at the temperature $T_{c2}$, the
non-zero leading components of fixed point tensor are given by Table  \ref{fixdata2}. If
we fit our data with Eq. (\ref{fit2}), we find:
\begin{eqnarray}
\Delta _{\left( \pm 1,0\right) } &=&0.12684  \nonumber \\
\Delta _{\left( \pm 2,0\right) } &=&0.50869  \nonumber \\
\Delta _{\left( \pm 3,0\right) } &=&1.17789,
\end{eqnarray}%
which match well with the results from our previous transfer matrix calculation, with $%
\Delta _{\left( \pm 1,0\right) }=0.12539$, $\Delta _{\left( \pm 2,0\right)
}=0.50158$ and $\Delta _{\left( \pm 3,0\right) }=1.12851$(the corresponding compactification radius $R=2.82402$). The fundamental length scale can also be fitted as $%
x=2.23035$, and the corresponding OPE coefficients are listed in Table \ref{OPE1}. The relative error
of our fitting is estimated as:
\begin{equation}
\sqrt{\frac{\underset{N}{\sum }\left[ \left\vert \left\vert
T_{r_{1}r_{2}r_{3}}\right\vert -C_{123}l^{\Delta _{1}}l^{\Delta _{2}}\left(
2l\right) ^{\Delta _{3}}\right\vert /\left\vert
T_{r_{1}r_{2}r_{3}}\right\vert \right] ^{2}}{N}},
\end{equation}%
where $N$ is the total number of components in our consideration. We find the
fitting error is around $4.0\times 10^{-3}$. Thus, we conclude that the fixed point tensor can be well described by the 3-point function(at least for primary fields) and the OPE coeeficient can be read out directly.

\begin{table}[tbp]
\centering
\begin{tabular}{cccccccc}
\hline\hline
$T_{r_{1}r_{2}r_{3}}$ & $r_{1}$ & $r_{2}$ & $r_{3}$ & $T_{r_{1}r_{2}r_{3}}$ & $r_{1}$ & $r_{2}$ & $r_{3}$ \\
\hline
1.00000 & $I$ & $I$ & $I$ & 0.45253 & $\gamma$ & $I$ & $\delta$ \\
0.81725 & $I$ & $\alpha$ & $\beta$ & 0.30155 & $\gamma$ & $\alpha$ & $\eta$ \\
0.81725 & $I$ & $\beta$ & $\alpha$ & 0.46479 & $\gamma$ & $\beta$ & $\beta$ \\
0.45253 & $I$ & $\gamma$ & $\delta$ & 0.32120 & $\gamma$ & $\delta$ & $I$ \\
0.45253 & $I$ & $\delta$ & $\gamma$ & 0.15522 & $\gamma$ & $\eta$ & $\alpha$ \\
0.17675 & $I$ & $\lambda$ & $\eta$ & 0.45253 & $\delta$ & $I$ & $\gamma$ \\
0.17675 & $I$ & $\eta$ & $\lambda$ & 0.46479 & $\delta$ & $\alpha$ & $\alpha$ \\
0.81725 & $\alpha$ & $I$ & $\beta$ & 0.30155 & $\delta$ & $\beta$ & $\lambda$ \\
0.59820 & $\alpha$ & $\alpha$ & $\delta$ & 0.32120 & $\delta$ & $\gamma$ & $I$ \\
0.74901 & $\alpha$ & $\beta$ & $I$ & 0.15522 & $\delta$ & $\lambda$ & $\beta$ \\
0.30155 & $\alpha$ & $\gamma$ & $\eta$ & 0.17675 & $\lambda$ & $I$ & $\eta$ \\
0.46479 & $\alpha$ & $\delta$ & $\alpha$ & 0.20004 & $\lambda$ & $\beta$ & $\delta$ \\
0.20004 & $\alpha$ & $\eta$ & $\gamma$ & 0.15522 & $\lambda$ & $\delta$ & $\beta$ \\
0.81725 & $\beta$ & $I$ & $\alpha$ & 0.08449 & $\lambda$ & $\eta$ & $I$ \\
0.74901 & $\beta$ & $\alpha$ & $I$ & 0.17675 & $\eta$ & $I$ & $\lambda$ \\
0.59820 & $\beta$ & $\beta$ & $\gamma$ & 0.20004 & $\eta$ & $\alpha$ & $\gamma$ \\
0.46479 & $\beta$ & $\gamma$ & $\beta$ & 0.15522 & $\eta$ & $\gamma$ & $\alpha$ \\
0.30155 & $\beta$ & $\delta$ & $\lambda$ & 0.08449 & $\eta$ & $\lambda$ & $I$ \\
0.20004 & $\beta$ & $\lambda$ & $\delta$ & & & &  \\
\hline\hline
\end{tabular}%
\caption{Leading non zero components of the fixed point tensor of $q=10$ model
at BKT critical point $T_{c2}$.}
\label{fixdata2}
\end{table}

\begin{table}[tbp]
\centering
\begin{tabular}{cccccccc}
\hline\hline
$C_{r_{1}r_{2}r_{3}}$ & $r_{1}$ & $r_{2}$ & $r_{3}$ & $C_{r_{1}r_{2}r_{3}}$ & $r_{1}$ & $r_{2}$ & $r_{3}$ \\
\hline
1.00000 & $I$ & $I$ & $I$ & 1.00000 & $\gamma$ & $I$ & $\delta$ \\
1.00000 & $I$ & $\alpha$ & $\beta$ & 1.01105 & $\gamma$ & $\alpha$ & $\eta$ \\
1.00000 & $I$ & $\beta$ & $\alpha$ & 1.00350 & $\gamma$ & $\beta$ & $\beta$ \\
1.00000 & $I$ & $\gamma$ & $\delta$ & 1.00000 & $\gamma$ & $\delta$ & $I$ \\
1.00000 & $I$ & $\delta$ & $\gamma$ & 0.99779 & $\gamma$ & $\eta$ & $\alpha$ \\
1.00000 & $I$ & $\lambda$ & $\eta$ & 1.00000 & $\delta$ & $I$ & $\gamma$ \\
1.00000 & $I$ & $\eta$ & $\lambda$ & 1.00350 & $\delta$ & $\alpha$ & $\alpha$ \\
1.00000 & $\alpha$ & $I$ & $\beta$ & 1.01105 & $\delta$ & $\beta$ & $\lambda$ \\
1.00023 & $\alpha$ & $\alpha$ & $\delta$ & 1.00000 & $\delta$ & $\gamma$ & $I$ \\
1.00000 & $\alpha$ & $\beta$ & $I$ & 0.99779 & $\delta$ & $\lambda$ & $\beta$ \\
1.01105 & $\alpha$ & $\gamma$ & $\eta$ & 1.00000 & $\lambda$ & $I$ & $\eta$ \\
1.00350 & $\alpha$ & $\delta$ & $\alpha$ & 0.99590 & $\lambda$ & $\beta$ & $\delta$ \\
1.01105 & $\alpha$ & $\eta$ & $\gamma$ & 0.99779 & $\lambda$ & $\delta$ & $\beta$ \\
1.00000 & $\beta$ & $I$ & $\alpha$ & 1.00000 & $\lambda$ & $\eta$ & $I$ \\
1.00000 & $\beta$ & $\alpha$ & $I$ & 1.00000 & $\eta$ & $I$ & $\lambda$ \\
1.00023 & $\beta$ & $\beta$ & $\gamma$ & 0.99590 & $\eta$ & $\alpha$ & $\gamma$ \\
1.00350 & $\beta$ & $\gamma$ & $\beta$ & 0.99779 & $\eta$ & $\gamma$ & $\alpha$ \\
1.01105 & $\beta$ & $\delta$ & $\lambda$ & 1.00000 & $\eta$ & $\lambda$ & $I$ \\
0.99590 & $\beta$ & $\lambda$ & $\delta$ & & & &  \\
\hline\hline
\end{tabular}%
\caption{Fitting OPE coefficients in Eq. (\ref{3point}) of $q=10$ model at temperature $T=T_{c2}$,
we see that they approach the expected value 1.}
\label{OPE1}
\end{table}

\subsection{Fixed point tensor for general cases}
In fact, the above structure of fixed point tensor holds for the whole critical phase between $T_{c1}$ and $T_{c2}$.
In the following, we further study the fixed point tensor for the $q=10$ case at different temperatures.
Table \ref{OPEcoe} shows that all the OPE coefficients are very close to 1, as expected from the compactified boson theory.
Table \ref{scadim} shows the comparison between the scaling dimensions read from the fixed point
tensor and from the direct calculation of transfer matrix. We see that they also match very well.

\begin{table}[tbp]
\centering
\begin{tabular}{cccccc}
\hline\hline
$C^{t=0.70}_{r_{1}r_{2}r_{3}}$ & $C^{t=0.76}_{r_{1}r_{2}r_{3}}$ & $%
C^{t=0.80}_{r_{1}r_{2}r_{3}}$ & $r_{1}$ & $r_{2}$ & $r_{3}$ \\ \hline
1.00000 & 1.00000 & 1.00000 & $I$ & $I$ & $I$ \\
1.00000 & 1.00000 & 1.00000 & $I$ & $\alpha $ & $\beta $ \\
1.00000 & 1.00000 & 1.00000 & $I$ & $\beta $ & $\alpha $ \\
1.00000 & 1.00000 & 1.00000 & $I$ & $\gamma $ & $\delta $ \\
1.00000 & 1.00000 & 1.00000 & $I$ & $\delta $ & $\gamma $ \\
1.00000 & 1.00000 & 1.00000 & $\alpha $ & $I$ & $\beta $ \\
0.99967 & 0.99961 & 0.99955 & $\alpha $ & $\alpha $ & $\delta $ \\
1.00000 & 1.00000 & 1.00000 & $\alpha $ & $\beta $ & $I$ \\
1.00016 & 1.00015 & 1.00010 & $\alpha $ & $\delta $ & $\alpha $ \\
1.00000 & 1.00000 & 1.00000 & $\beta $ & $I$ & $\alpha $ \\
1.00000 & 1.00000 & 1.00000 & $\beta $ & $\alpha $ & $I$ \\
0.99967 & 0.99961 & 0.99955 & $\beta $ & $\beta $ & $\gamma $ \\
1.00016 & 1.00015 & 1.00010 & $\beta $ & $\gamma $ & $\beta $ \\
1.00000 & 1.00000 & 1.00000 & $\gamma $ & $I$ & $\delta $ \\
1.00016 & 1.00015 & 1.00010 & $\gamma $ & $\beta $ & $\beta $ \\
1.00000 & 1.00000 & 1.00000 & $\gamma $ & $\delta $ & $I$ \\
1.00000 & 1.00000 & 1.00000 & $\delta $ & $I$ & $\gamma $ \\
1.00016 & 1.00015 & 1.00010 & $\delta $ & $\alpha $ & $\alpha $ \\
1.00000 & 1.00000 & 1.00000 & $\delta $ & $\gamma $ & $I$ \\
\hline\hline
\end{tabular}%
\caption{OPE coefficient in Eq. (\ref{3point}) fitting from the data of model $q=10$
on different tempreature}
\label{OPEcoe}
\end{table}

\begin{table}[tbp]
\centering
\begin{tabular}{ccccc}
\hline\hline
Temperature & $\Delta_{1} $ & $\Delta_{2}$ & $\Delta_{3}$ & Fitting radius $R$ \\
\hline
\multicolumn{5}{c}{From fixed point tensor} \\
\hline
0.70 & 0.06941 & 0.27824 & 0.62659 & 3.79267 \\
0.72 & 0.07228 & 0.28971 & 0.65169 & 3.72312 \\
0.74 & 0.07532 & 0.30182 & 0.67813 & 3.64573 \\
0.76 & 0.07851 & 0.31456 & 0.70589 & 3.57106 \\
0.78 & 0.08191 & 0.32813 & 0.73536 & 3.49684 \\
0.80 & 0.08566 & 0.34298 & 0.76683 & 3.41513 \\
\hline
\multicolumn{5}{c}{From transfer matrix} \\
\hline
0.70 & 0.06940 & 0.27795 & 0.62510 & 3.79498 \\
0.72 & 0.07217 & 0.28845 & 0.65024 & 3.72290 \\
0.74 & 0.07508 & 0.30057 & 0.67854 & 3.64905 \\
0.76 & 0.07830 & 0.31386 & 0.70314 & 3.57325 \\
0.78 & 0.08178 & 0.32827 & 0.73366 & 3.49727 \\
0.80 & 0.08571 & 0.34675 & 0.76734 & 3.41494 \\
\hline\hline
\end{tabular}%
\caption{Scaling dimension of the first 3 levels reads by fitting fixed point tensor with 3-point function of CFT
and from the calculation of transfer matrix.}
\label{scadim}
\end{table}

Therefore, we find very strong evidence that the fixed point tensor can be described by three-point correlation
function for primary fields. Such a structure also explains why loop-TNR is a very accurate algorithm for critical systems since primary fields with higher scaling dimensions will lead to a rapid decay for the corresponding tensor components. We also find that the components for descendant fields are always smaller than the corresponding primary field in the fixed point tensor. We believe this is also because descendant fields will have bigger scaling dimensions. However, the explicit fixed point tensor structure for descendant fields is rather complicated and we will leave this problem in our future study.



\section{Conclusions and discussions}

In summary, we use loop-TNR algorithm to study the phase transition properties
of the $q$-state clock model. For $q<5$ models, we compute the central charge and scaling dimensions at  the self-dual critical points and find perfect agreement with previous CFT predictions. For $q>5$ models, we accurately determine the critical temperatures $T_{c1}$ and $T_{c2}$ for both phase transitions. By further computing the central charge and scaling dimensions at $T_{c1}$ and $T_{c2}$, we can further obtain the compactification radius $R$ which also perfectly agrees with the $Z_q$ deformed sine-Gordon theory predictions. Interestingly, for big enough $q$, we find that the fixed point tensor at $T_{c2}$ converges to the same one(up to numerical errors) that describes the well known BKT transitions, and the corresponding OPE coefficient can also be read out directly.

For our future work, it will be of great interest to investigate the explicit expression of the infinite dimensional fixed point tensor description for the compactified boson theory as well as general CFT. In fact, the fixed point tensor provides us a purely algebraic way to describe CFT which origins from a geometric perspective. Very recently, it has been shown that the p-adic CFT\cite{p-adic} admits an explicit finite dimensional tensor network representation. It is somewhat not quite surprising since p-adic CFT has no descendant fields. Since descendant fields might tell us how geometry emerges from basic algebraic data, it would be very important to understand the explicit form of fixed point tensor descriptions for descendant fields in usual CFT.


\begin{acknowledgments}
We are grateful to Ling-Yan Hung and Gong Cheng for very enlightening discussions for the structure of fixed point tensors at critical points. This work is supported by funding from Hong Kong’s Research Grants Council (GRF no.14301219) and Direct Grant no. 4053409 from The Chinese University of Hong Kong.
\end{acknowledgments}

\ \ \

\appendix

\section{Transition temperatures and compactification radius $R$ for $q>6$ models}
\label{largeq}
\begin{figure}[h]
\begin{subfigure}{.5\textwidth}
\includegraphics[width=8cm]{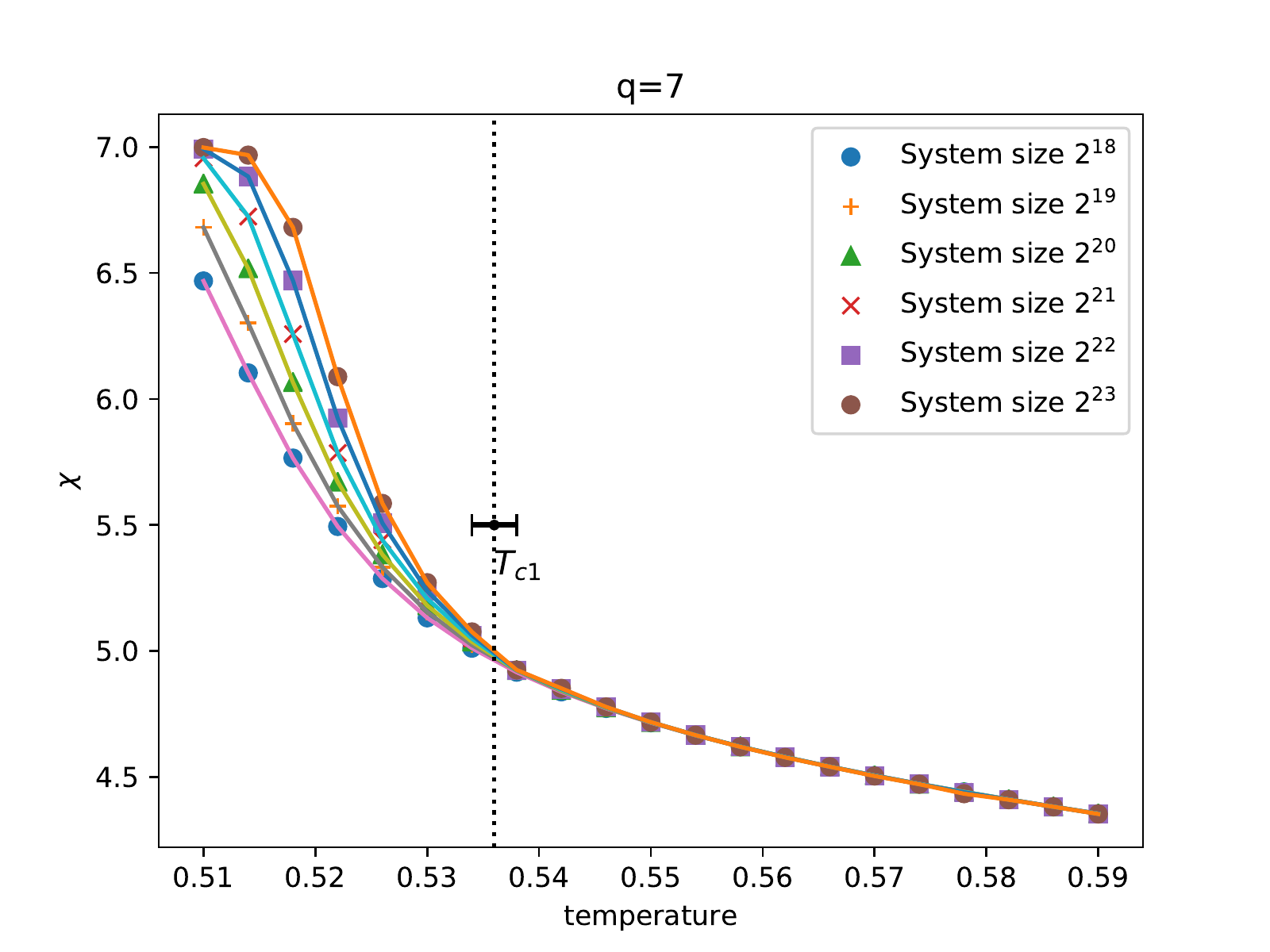}
\end{subfigure}
\par
\begin{subfigure}{.5\textwidth}
\includegraphics[width=8cm]{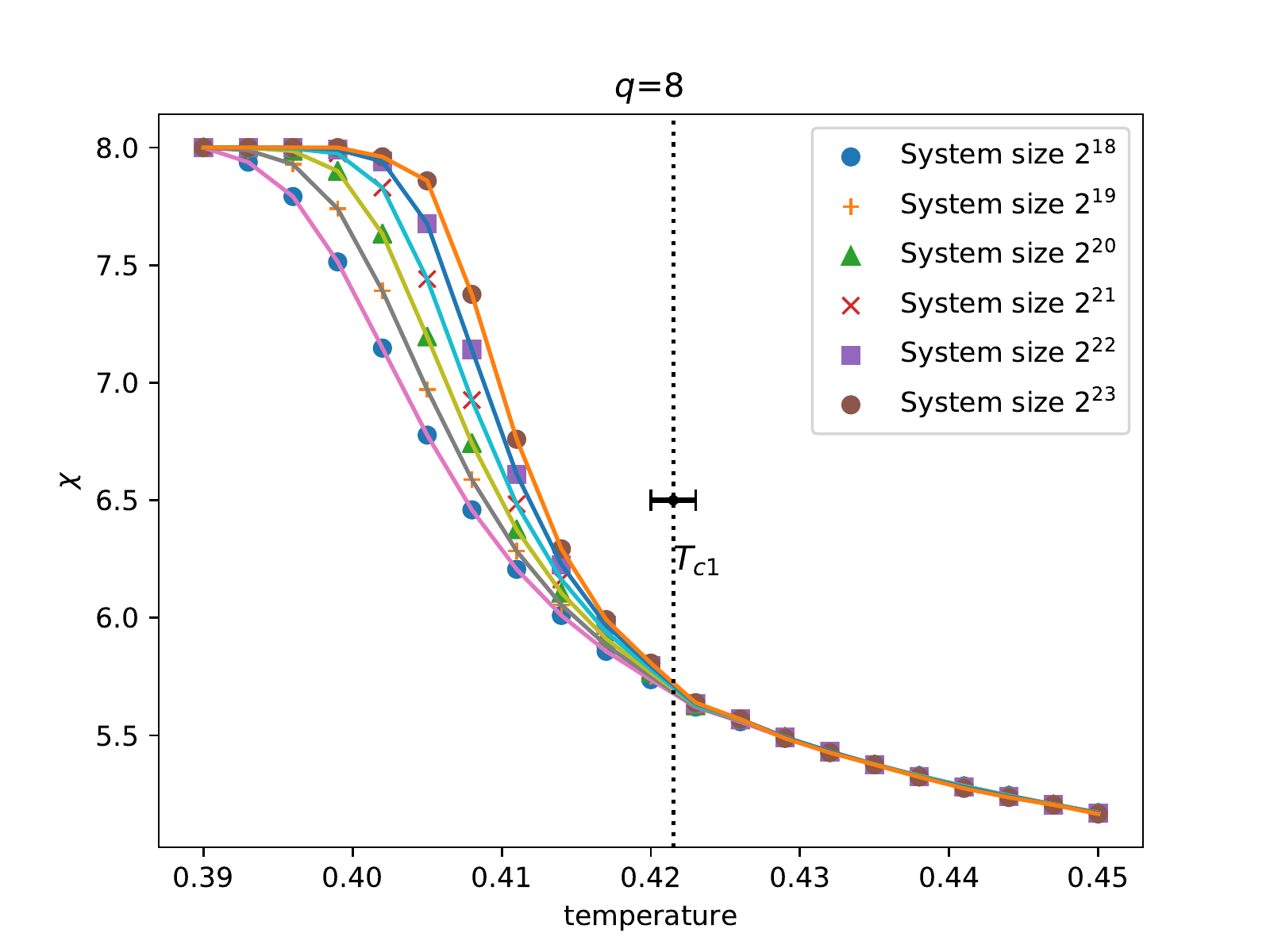}
\end{subfigure}
\par
\begin{subfigure}{.5\textwidth}
\includegraphics[width=8cm]{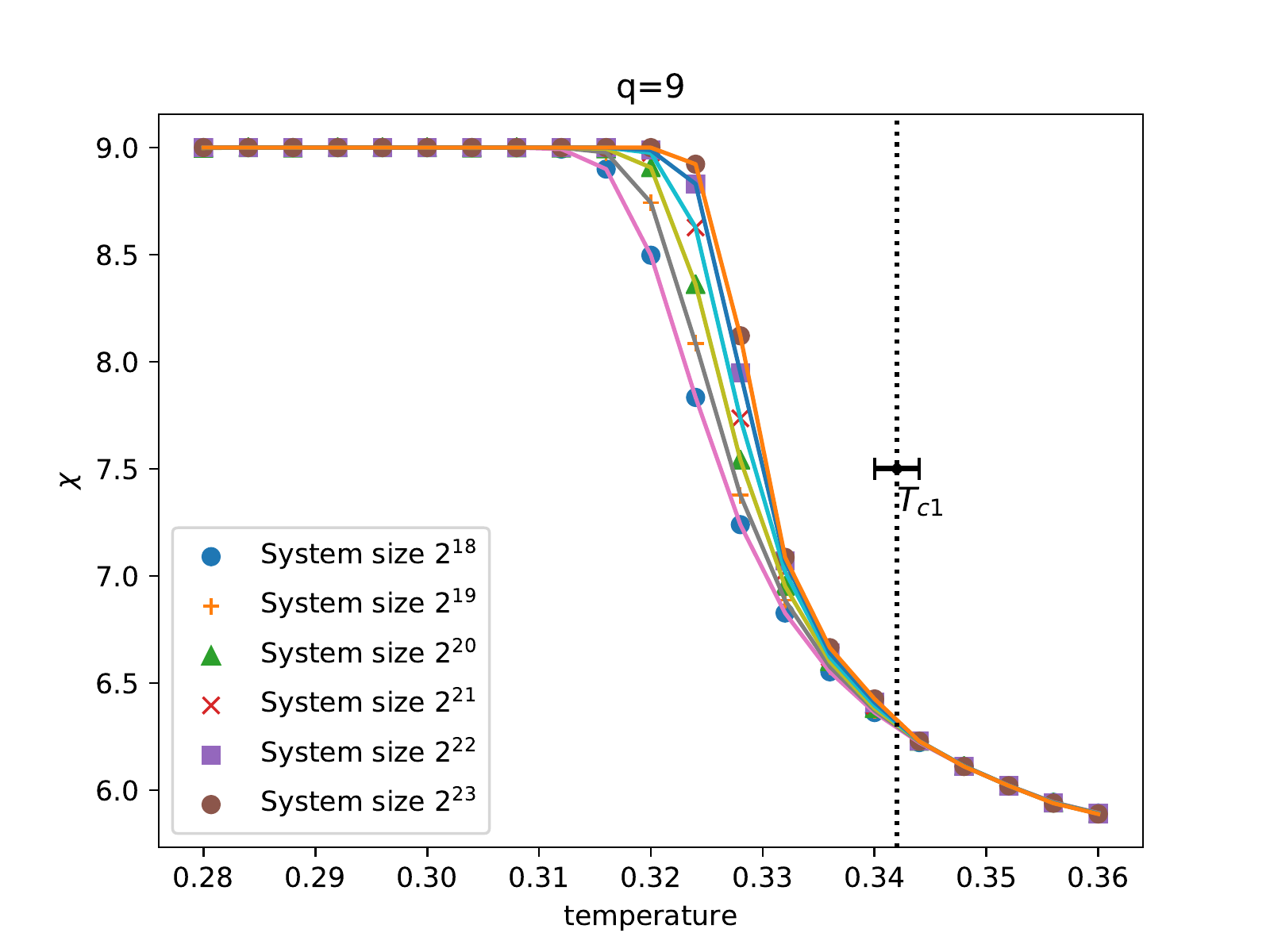}
\end{subfigure}
\caption{Invariant quantity of for q-state clock model with $q=7,8,9$ around
$T_{c1}$}
\label{invariant1}
\end{figure}

\begin{figure}[h]
\begin{subfigure}{.5\textwidth}
\includegraphics[width=8cm]{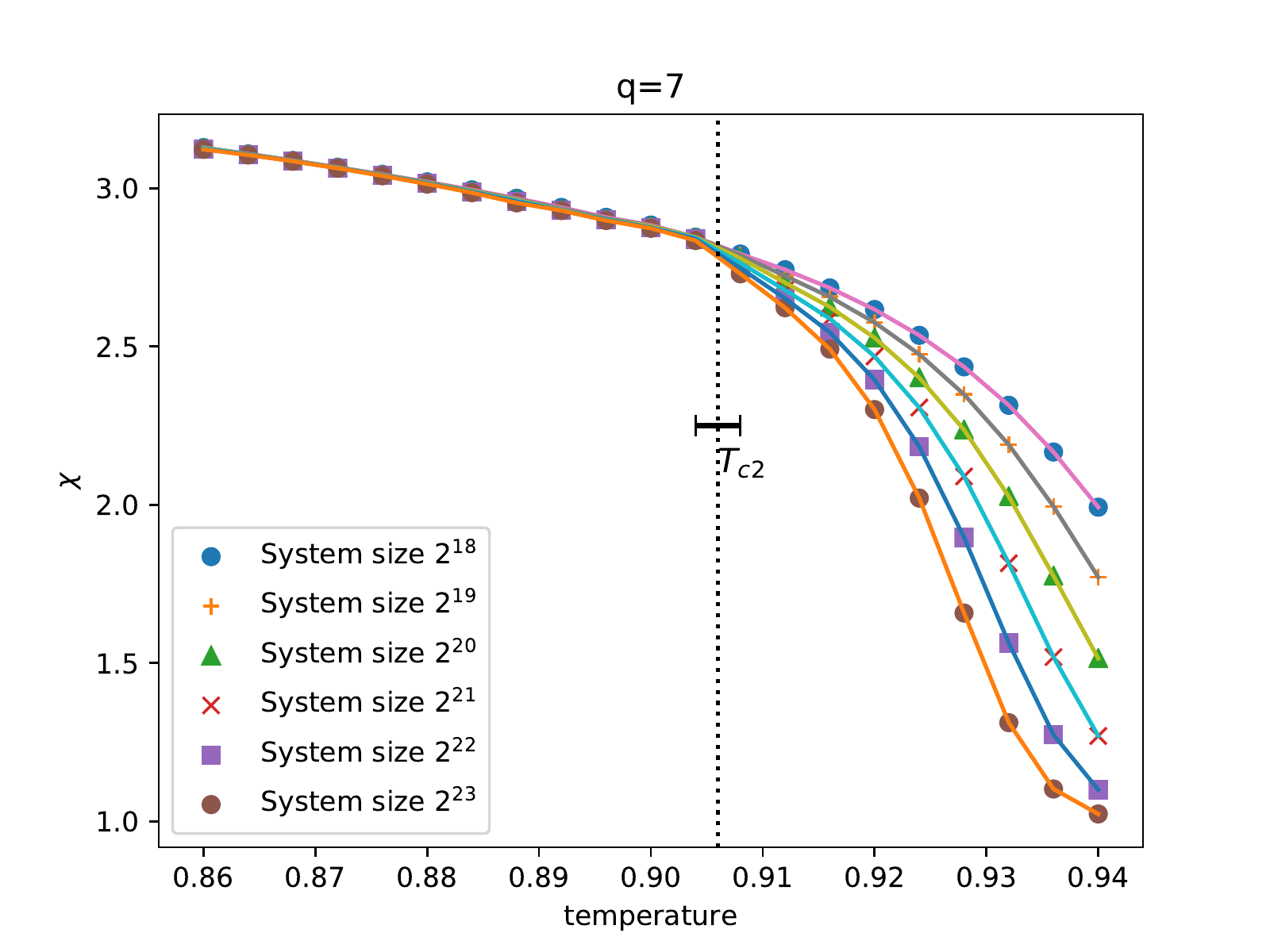}
\end{subfigure}
\par
\begin{subfigure}{.5\textwidth}
\includegraphics[width=8cm]{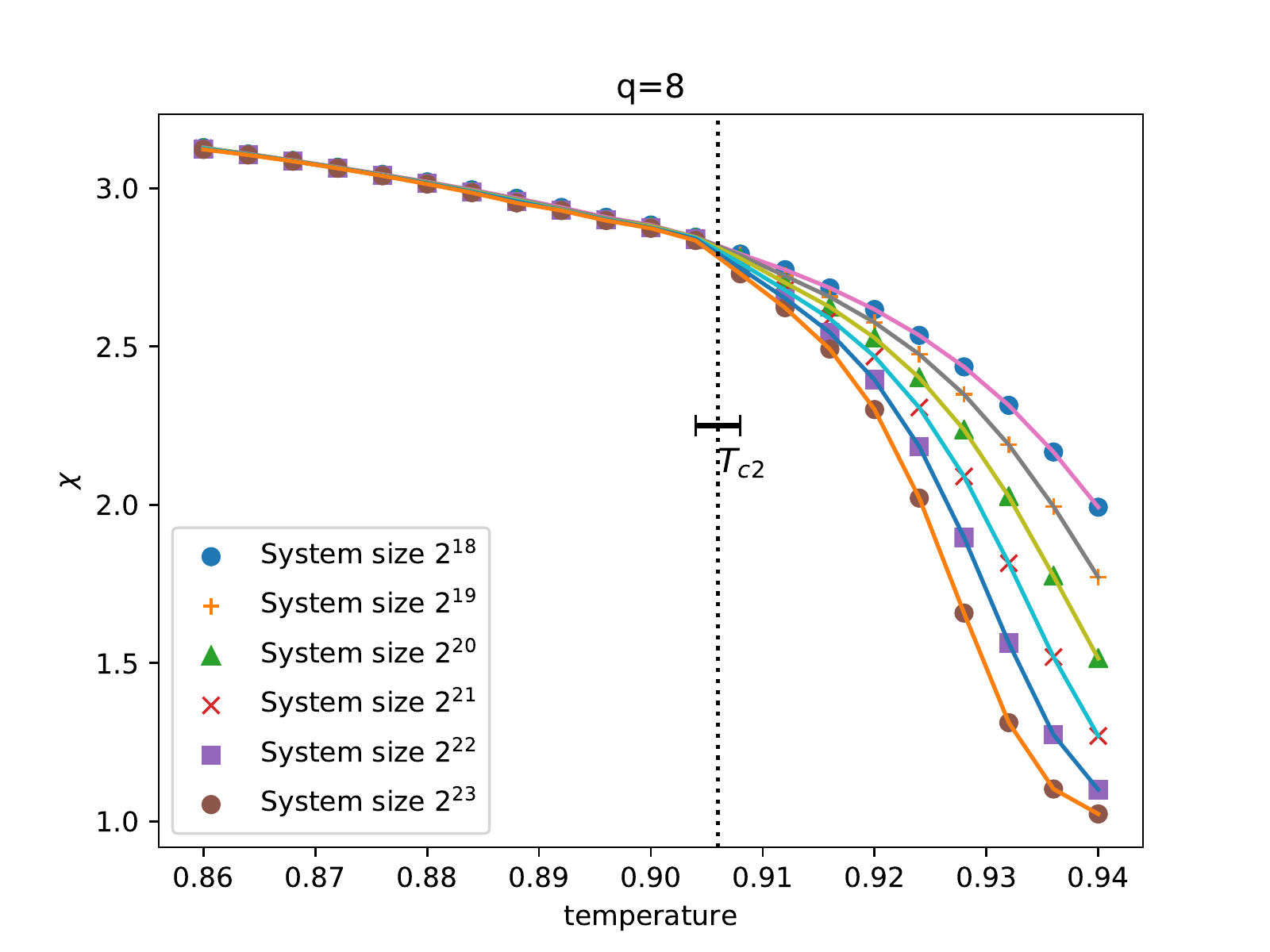}
\end{subfigure}
\par
\begin{subfigure}{.5\textwidth}
\includegraphics[width=8cm]{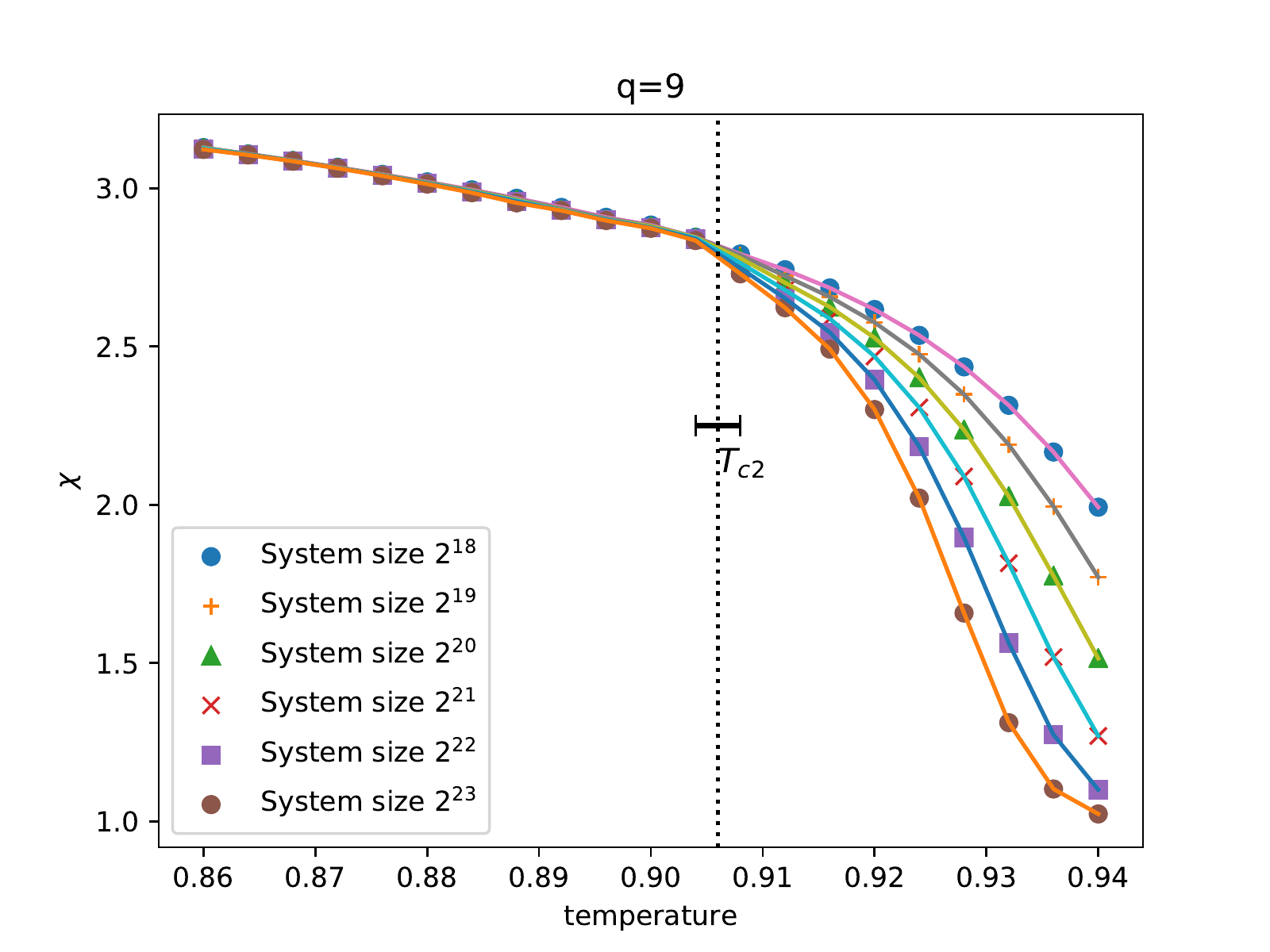}
\end{subfigure}
\caption{Invariant quantity of for q-state clock model with $q=7,8,9$ around
$T_{c2}$}
\label{invariant2}
\end{figure}

\begin{figure}[h]
\includegraphics[width=8cm]{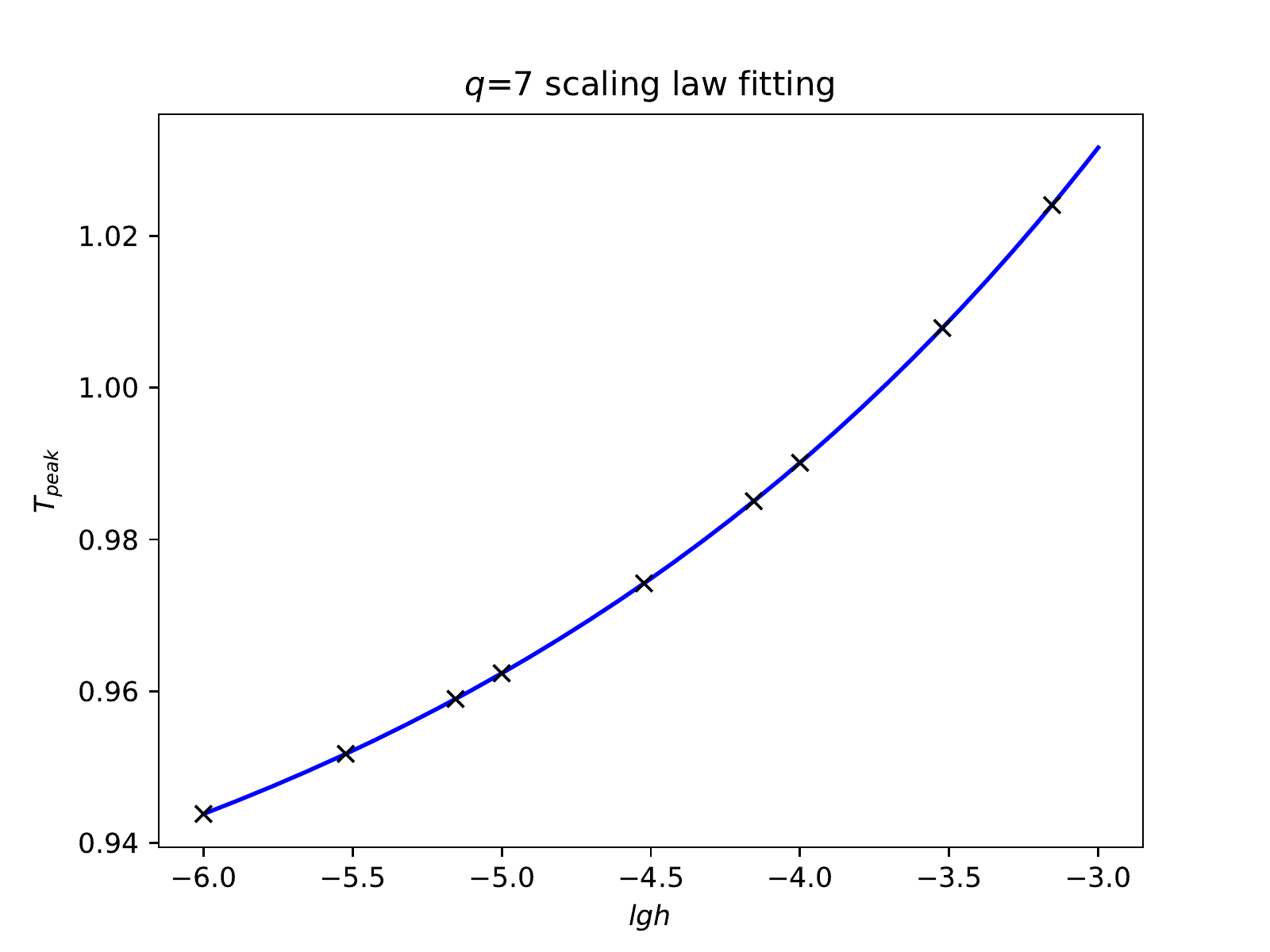}
\caption{Susceptibility peak temperature versus external field. For $q=7$
model, we find that $T_{c}=0.9065(5),$ $a=0.4198,$ $b=0.1752.$.}
\label{F13}
\end{figure}

\begin{figure}[h]
\includegraphics[width=8cm]{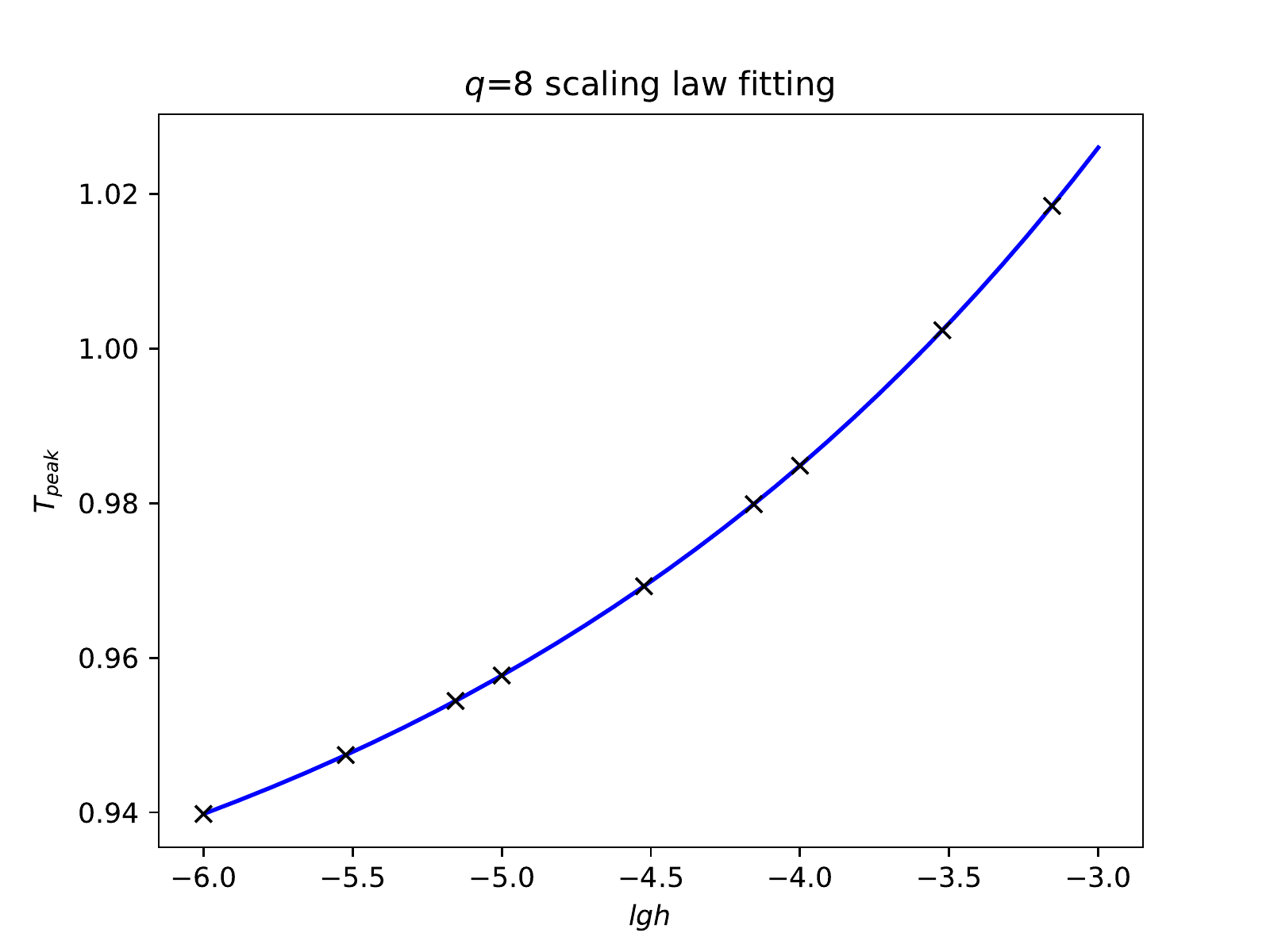}
\caption{Susceptibility peak temperature versus external field. For $q=8$
model, we find that $T_{c}=0.9051(5),$ $a=0.4213$, $b=0.1807$.}
\label{F14}
\end{figure}

\begin{figure}[h]
\includegraphics[width=8cm]{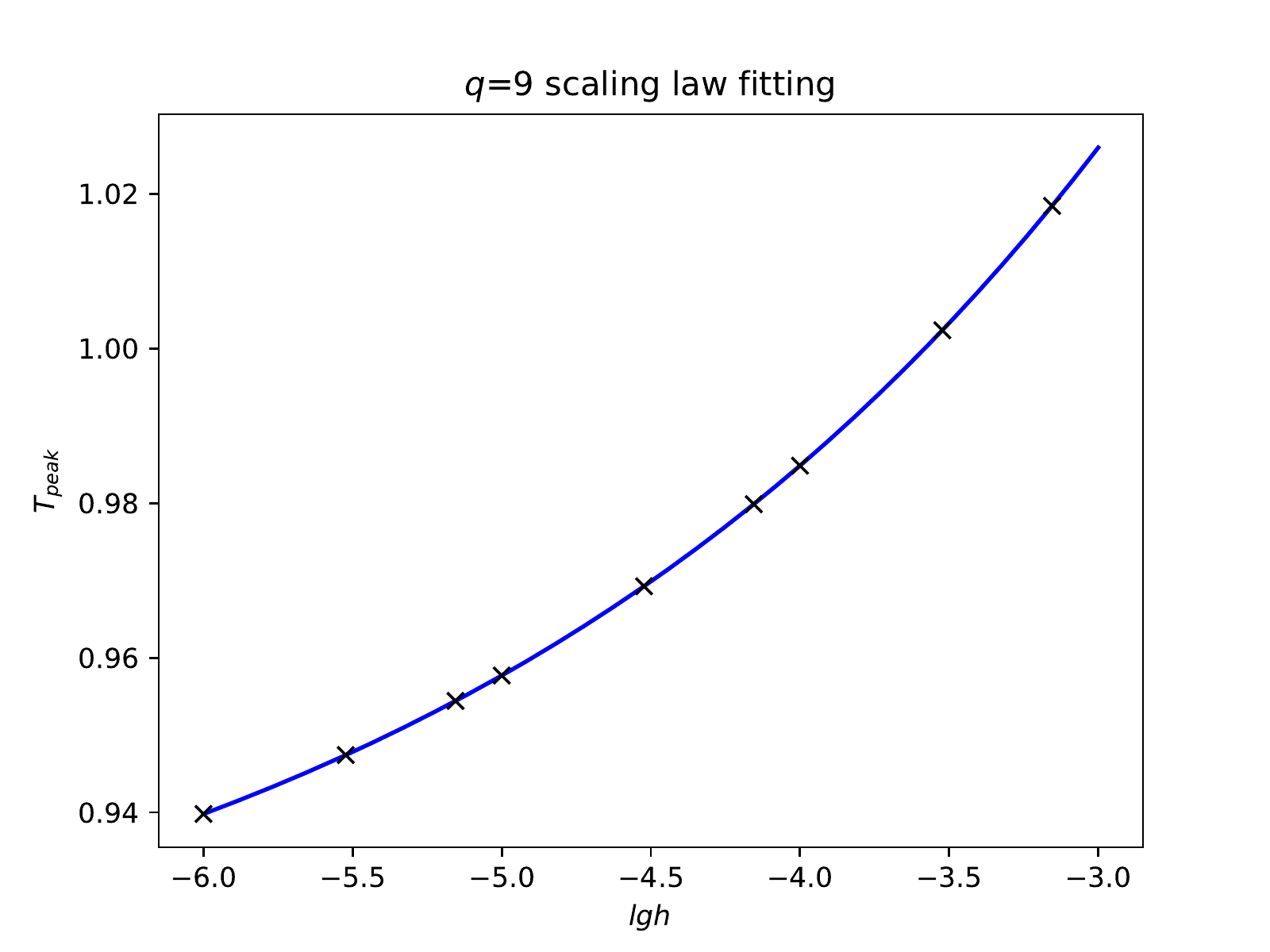}
\caption{Susceptibility peak temperature versus external field. For $q=9$
model, we find that $T_{c}=0.9051(5),$ $a=0.4213$, $b=0.1807$.}
\label{F15}
\end{figure}

\begin{figure}[h]
\begin{subfigure}{.5\textwidth}
\centering
\includegraphics[width=8cm]{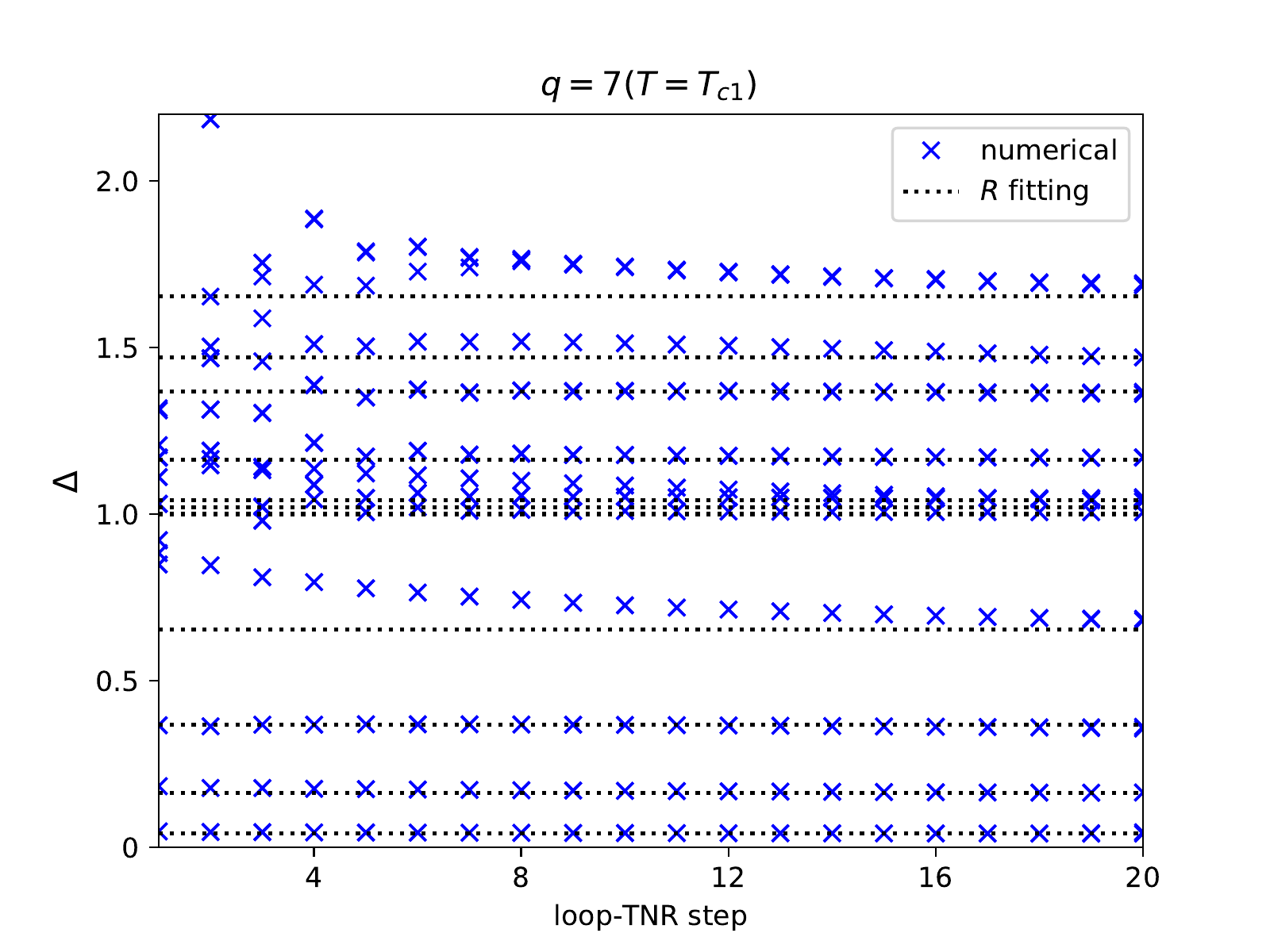}
\end{subfigure}
\par
\begin{subfigure}{.5\textwidth}
\centering
\includegraphics[width=8cm]{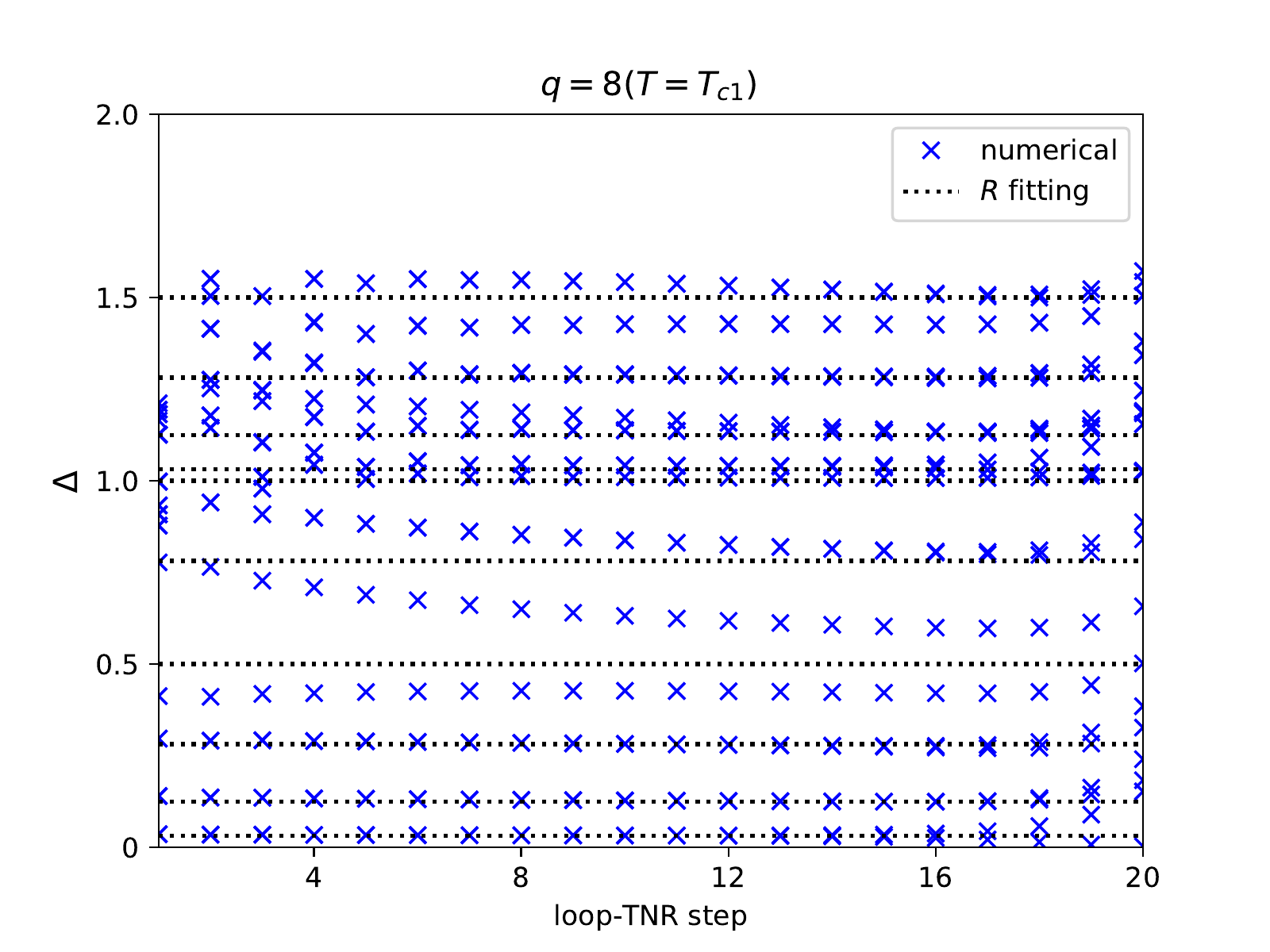}
\end{subfigure}
\par
\begin{subfigure}{.5\textwidth}
\includegraphics[width=8cm]{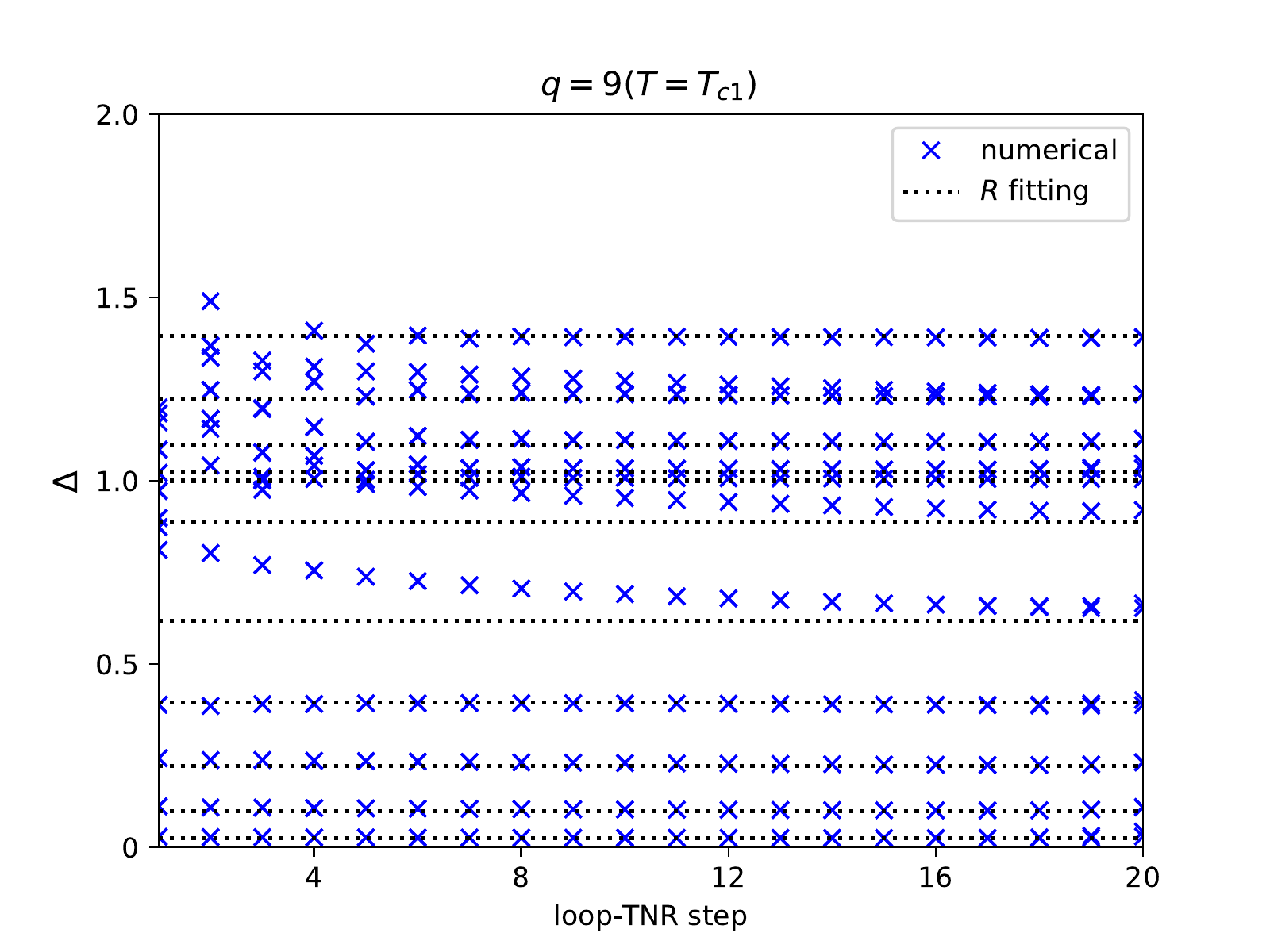}
\end{subfigure}
\caption{Fitting of scaling dimensions at the critical point $T_{c1}$ for $%
q=7,8,9$ model, from which we can read the character radius of the
compactified boson}
\label{scaling1}
\end{figure}

\begin{figure}[h]
\begin{subfigure}{.5\textwidth}
\includegraphics[width=8cm]{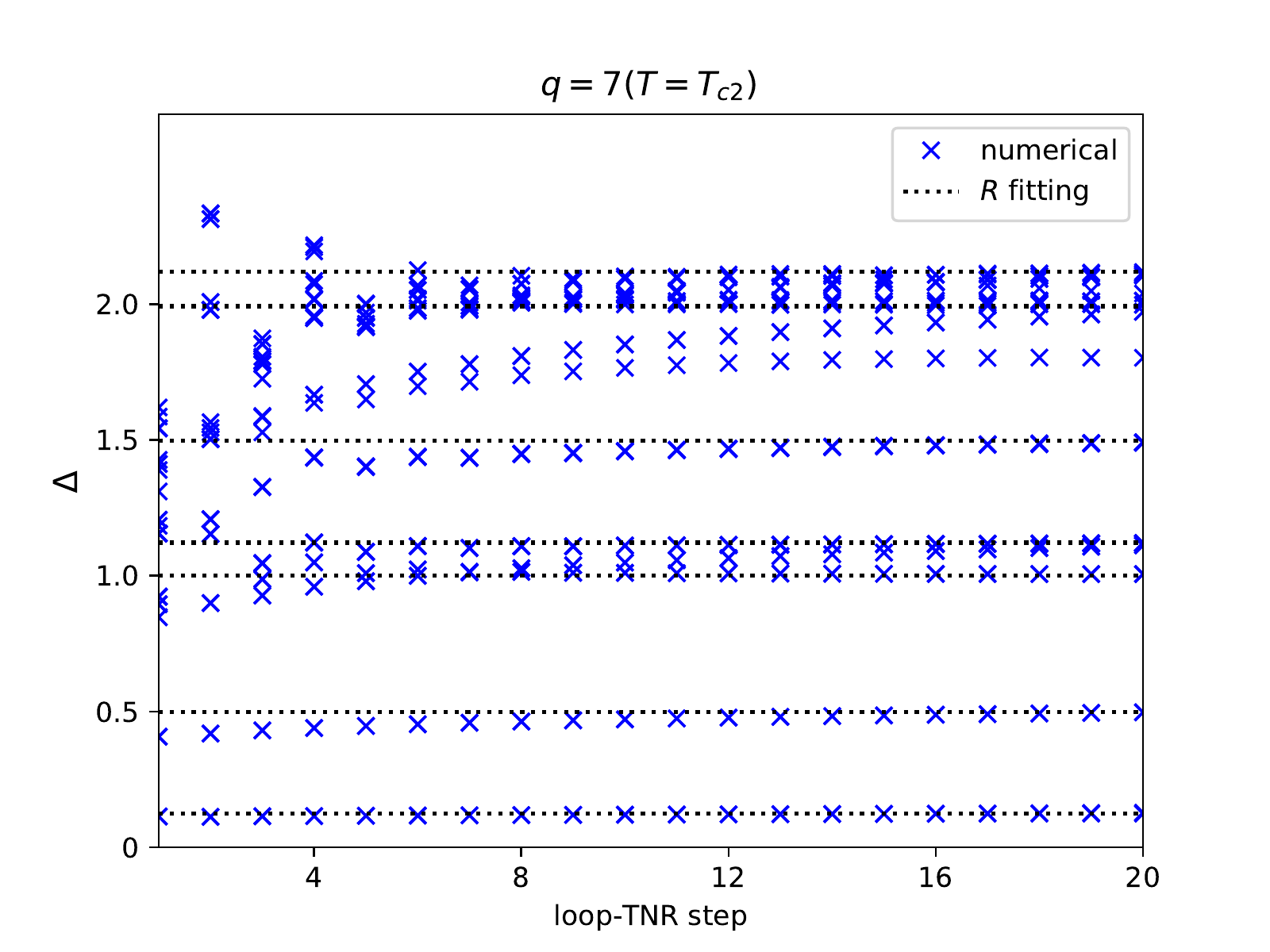}
\label{F35}
\end{subfigure}
\par
\begin{subfigure}{.5\textwidth}
\includegraphics[width=8cm]{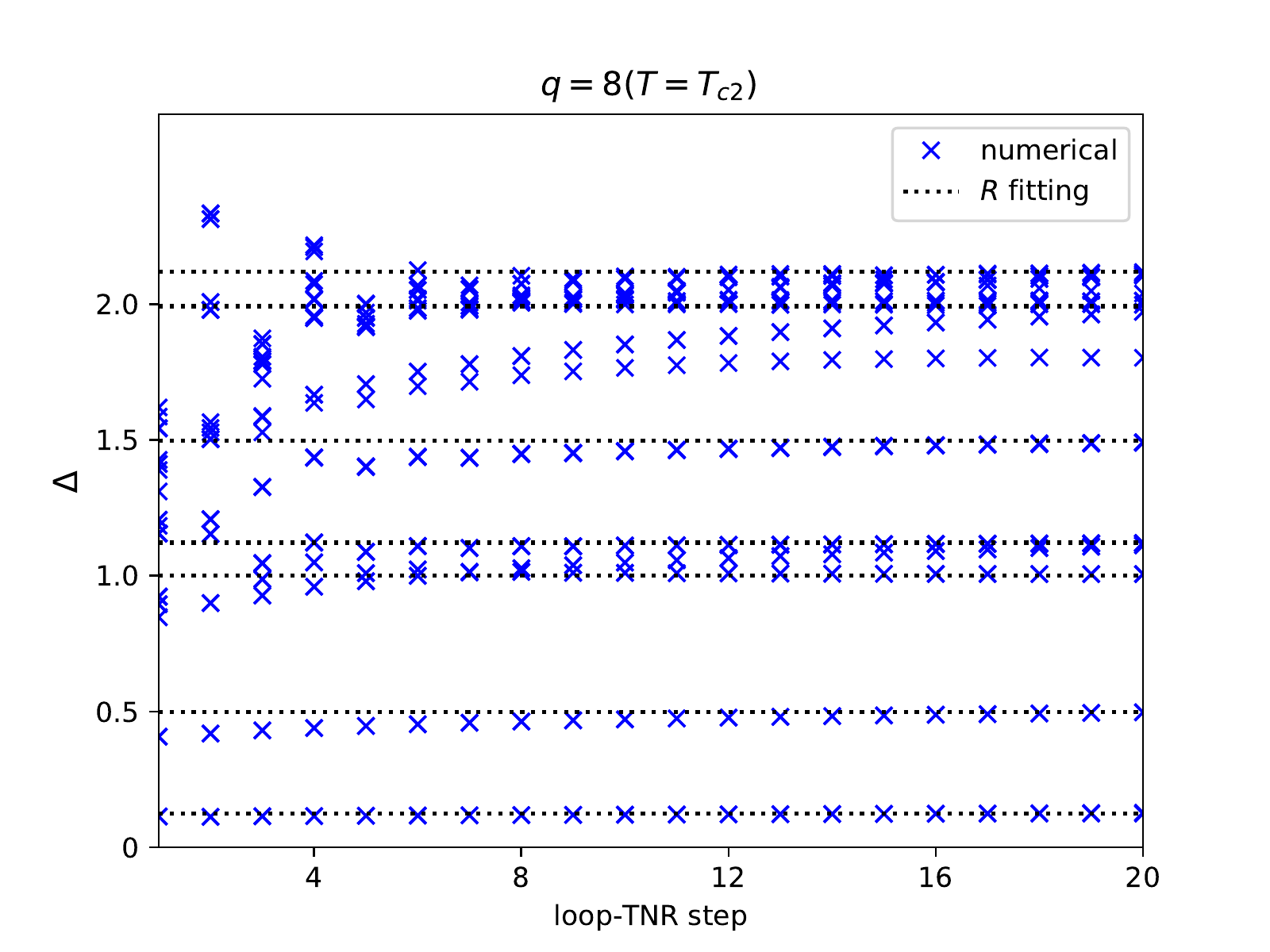}
\label{F36}
\end{subfigure}
\par
\begin{subfigure}{.5\textwidth}
\includegraphics[width=8cm]{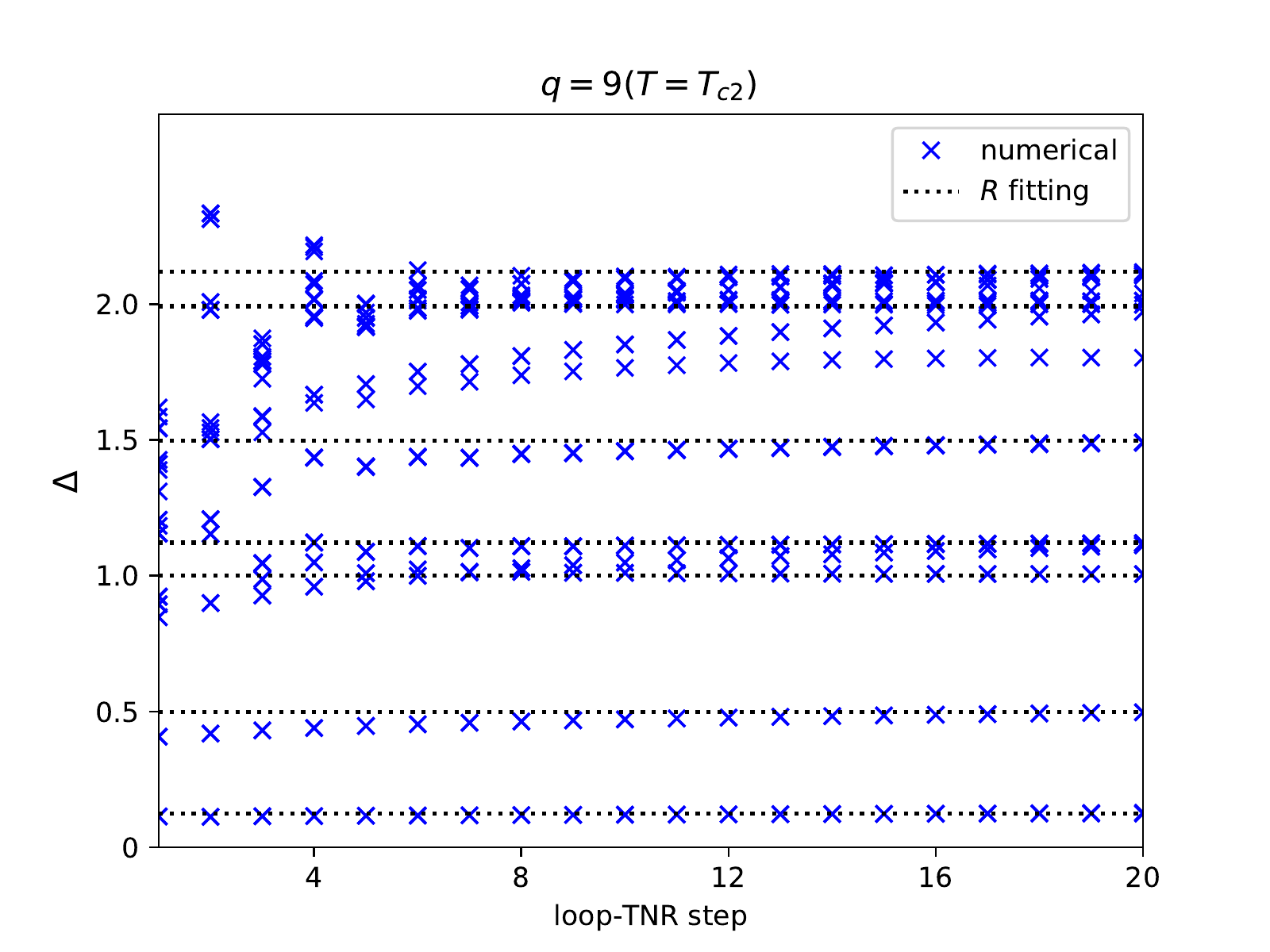}
\label{F37}
\end{subfigure}
\caption{Fitting of scaling dimensions at the critical point $T_{c2}$ for
q-state clock models with $q=7,8,9$ from which we can read the character
radius of the compactified boson}
\label{scaling2}
\end{figure}

\begin{figure}[h]
\begin{subfigure}{.5\textwidth}
\includegraphics[width=8cm]{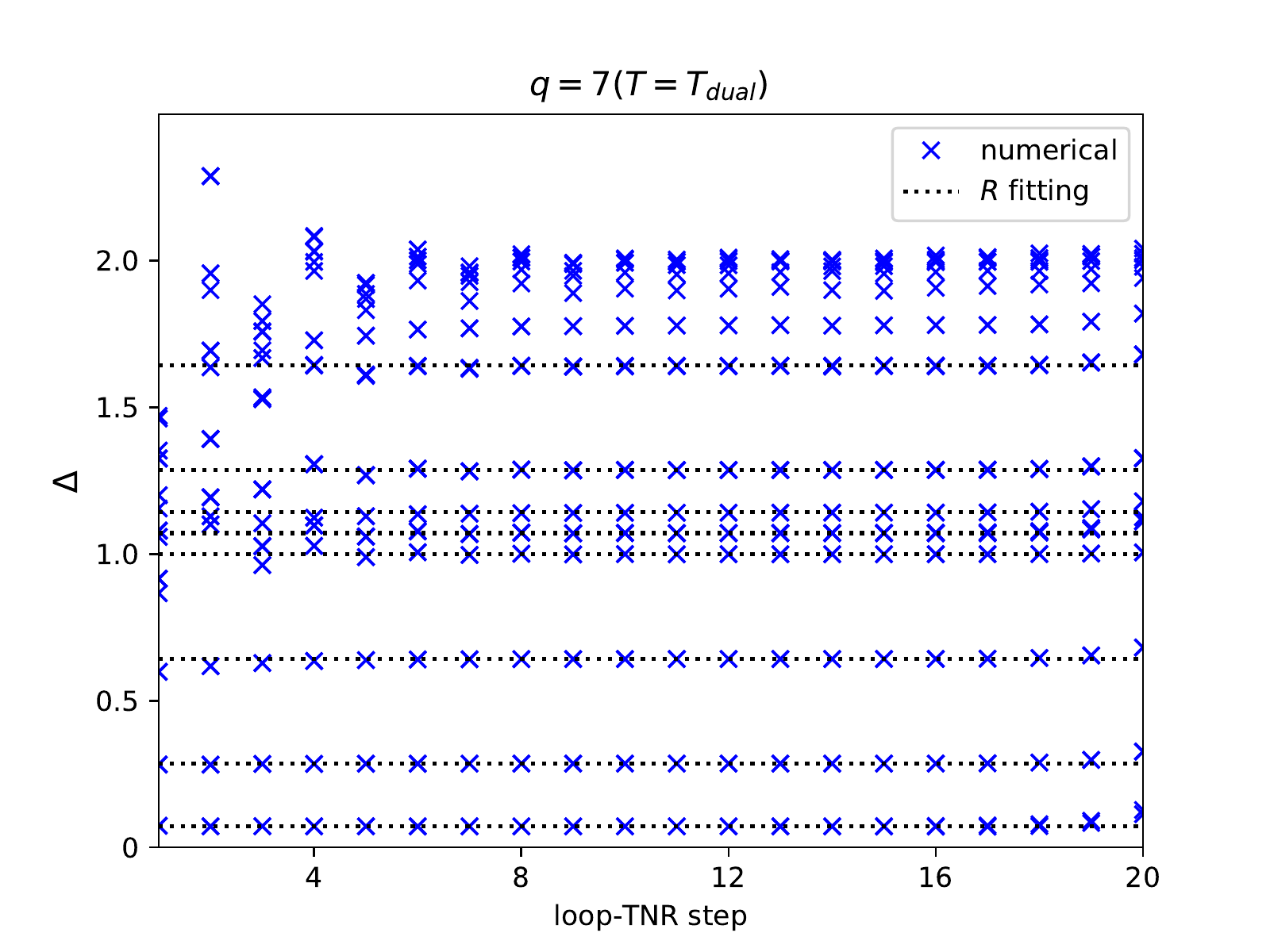}
\label{F38}
\end{subfigure}
\par
\begin{subfigure}{.5\textwidth}
\includegraphics[width=8cm]{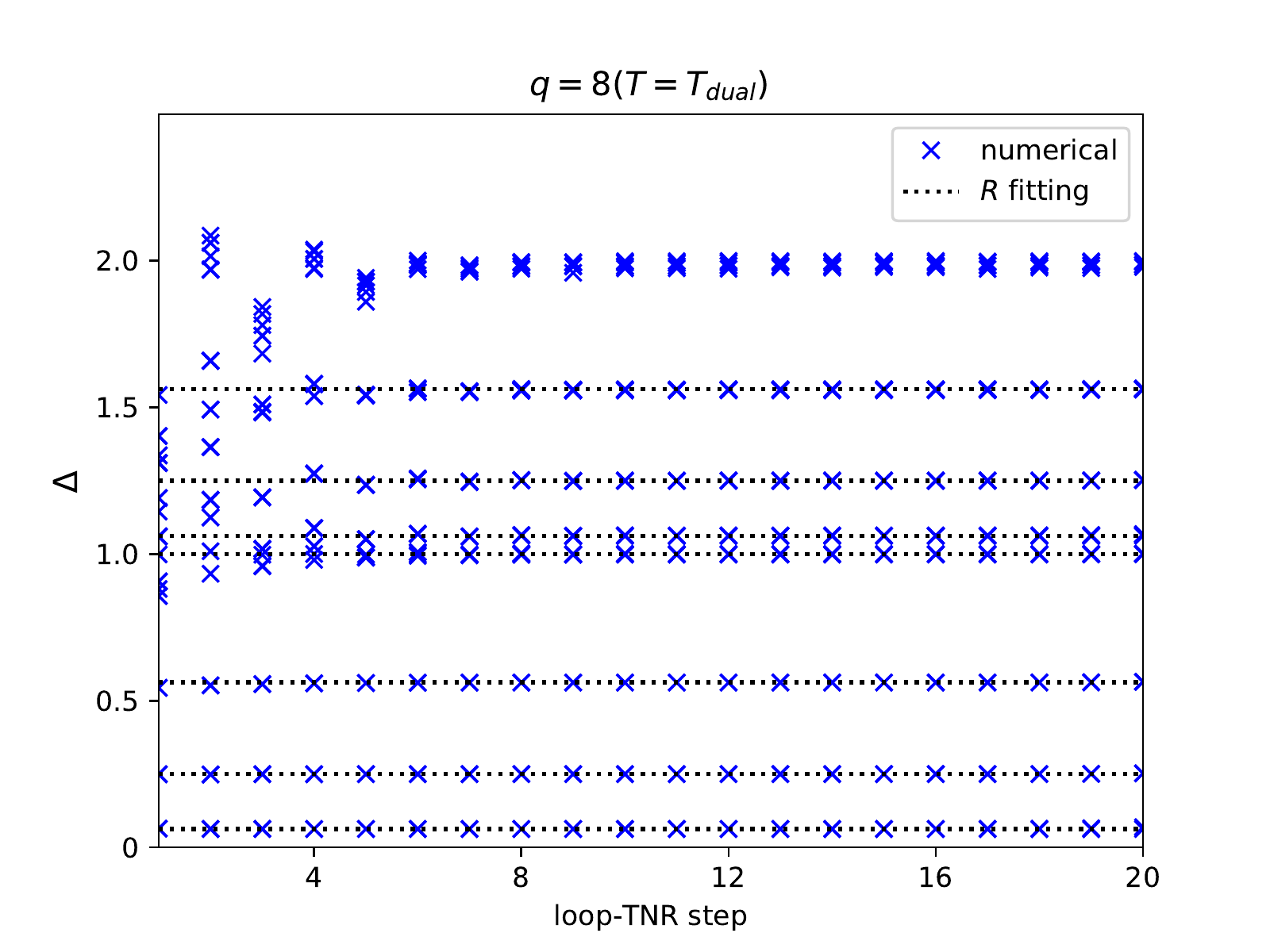}
\label{F39}
\end{subfigure}
\par
\begin{subfigure}{.5\textwidth}
\includegraphics[width=8cm]{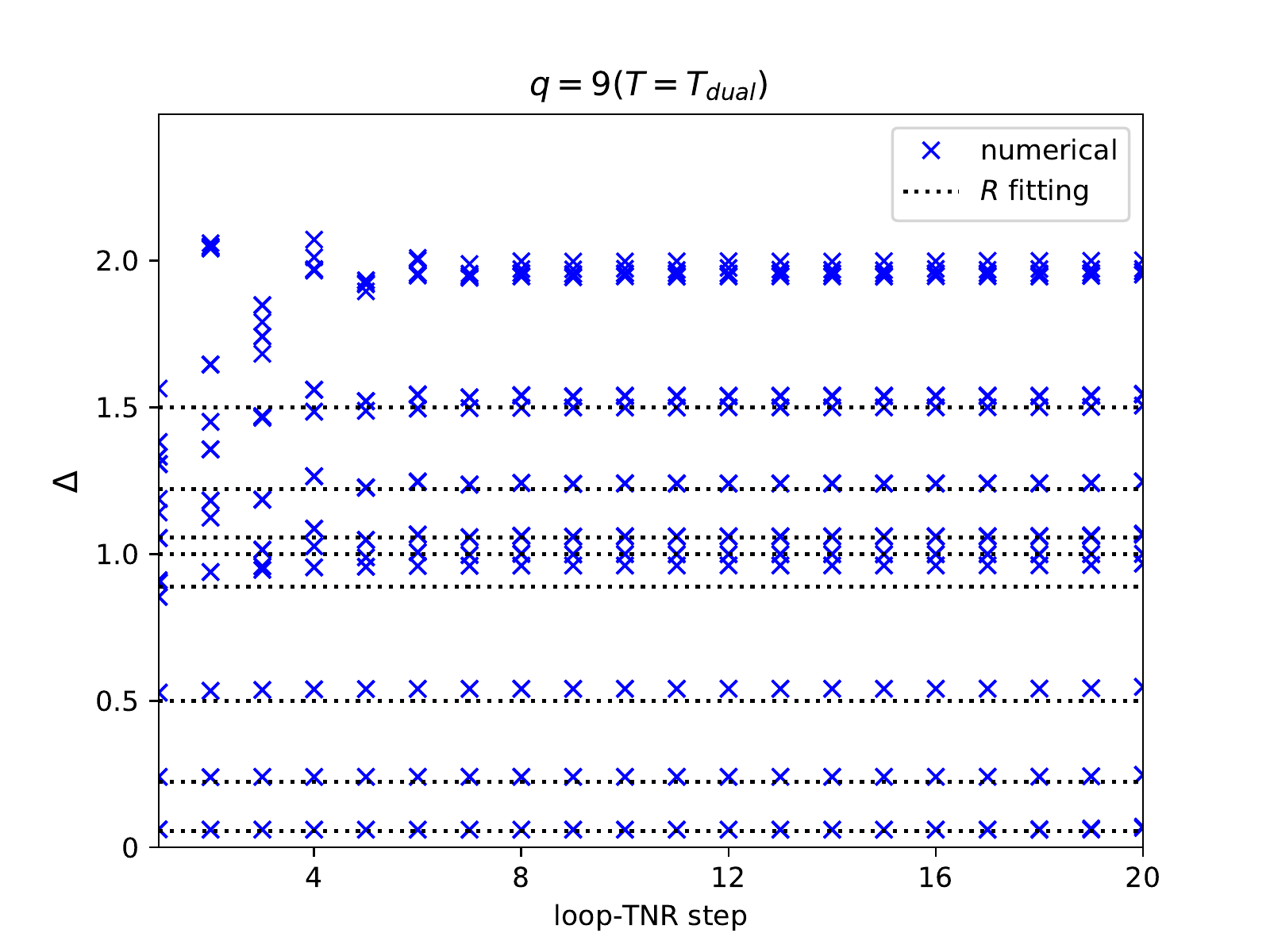}
\label{F40}
\end{subfigure}
\caption{Fitting of scaling dimensions on self-dual point for q-state clock models with $%
q=7,8,9$}
\label{scaling3}
\end{figure}

For models with $q>6$, e.g. $q=7,8,9$ we can also use the invariant quantity $\chi$ to
determine the transition temperature for $T_{c1}$ and $T_{c2}$, as seen in Fig. \ref{invariant1} and Fig. \ref{invariant2}. In Fig. \ref{F13}, Fig. \ref{F14} and Fig. \ref{F15}, we also use the susceptibility peak method Eq. (\ref{peak}) to determine the BTK transition temperature $T_{c2}$ with very high accuracy. Remarkably, we find that for $q>6$, the fitting paramters $a$ and $b$ are already very close to those obtained from classical XY model\cite{Yu}. Finally, we use the loop-TNR algorithm to compute the scaling dimensions at both high-temperature and low temperature critical points as well as the self-dual point, as seen in Fig. \ref{scaling1}, Fig. \ref{scaling2} and Fig. \ref{scaling3}. The corresponding compactification radius $R$ can also be fitted by using Eq. (\ref{dimension}).

\section{Imposing $C_4$ rotational symmetry and $Z_q$ internal symmetry in loop-TNR algorithm}
\label{symmetry}

\begin{figure}[h]
\centering
\includegraphics[width=8cm]{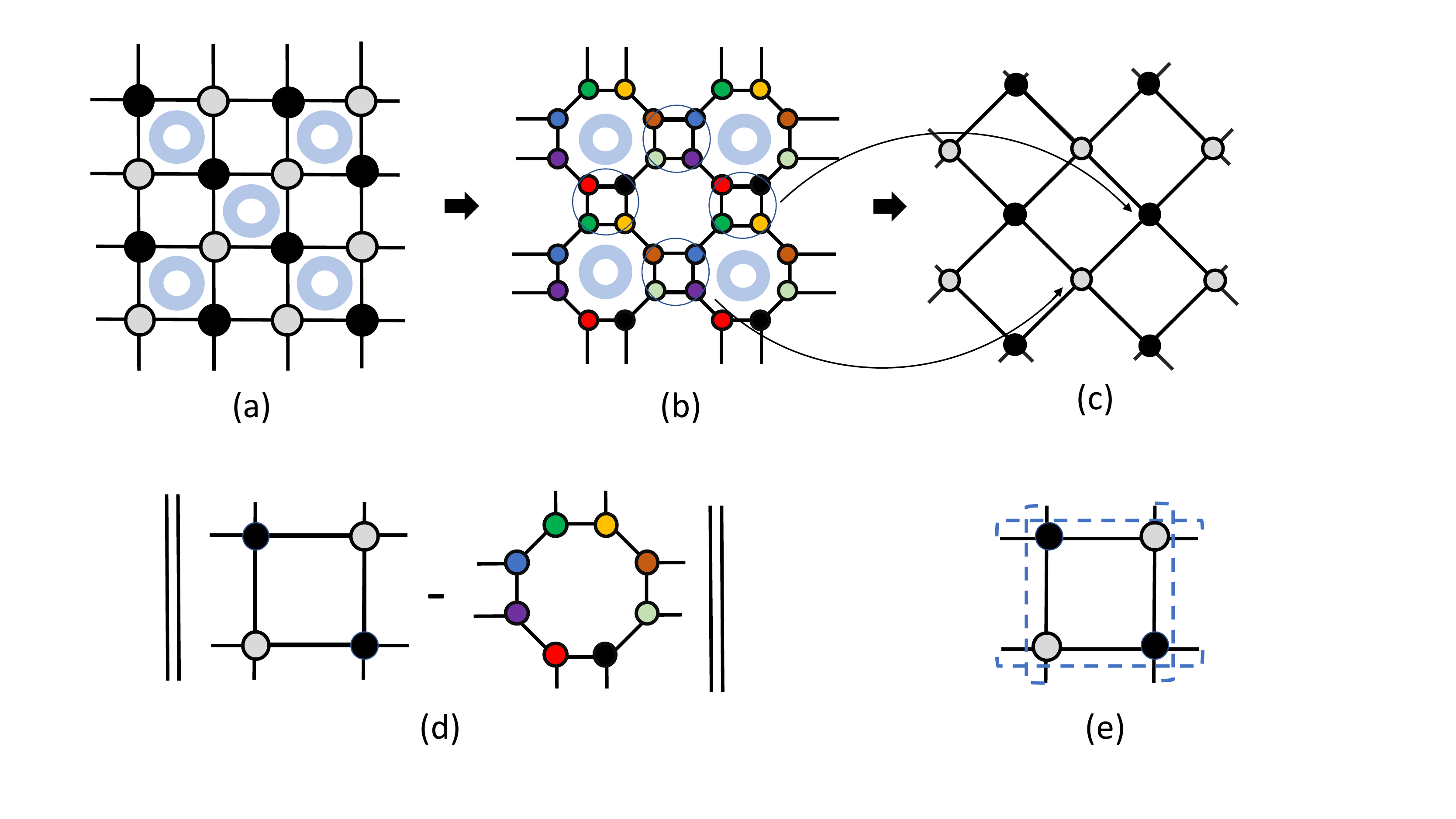}
\caption{Loop optimization procedure, in step (a), we apply entanglement
filtering, and in step (b) we find the optimal $S_{i}$ to minimize cost
function as shown (d). Then we trace the indices on the small square marked by the
circle. (e) is gauge invariant quantity, which will be used as the overall normalization factor.}
\label{loopTNR}
\end{figure}

In this appendix we first give a short review for the loop-TNR algorithm \cite{Yang}. Then we will discuss how to implement the $C_4$ lattice symmetry and the internal $Z_q$ symmetry. Loop-TNR method mainly contains the following steps, as shown in Fig. \ref{loopTNR}.
In general, there will be two types of tensors $T_{A}$
and $T_{B}$ on sublattices A and B during the renormaliation process.

$\bullet $ In step (a), we apply entanglement filtering to remove the corner
double line(CDL) tensor. The CDL tensor only contains local entanglement and
cannot be the fixed point tensor describing critical systems.
Ref. \cite{Gu,Yang} gives very clear explanation on how to remove such
short range entanglement.

\begin{figure}[h]
\centering
\includegraphics[width=8cm]{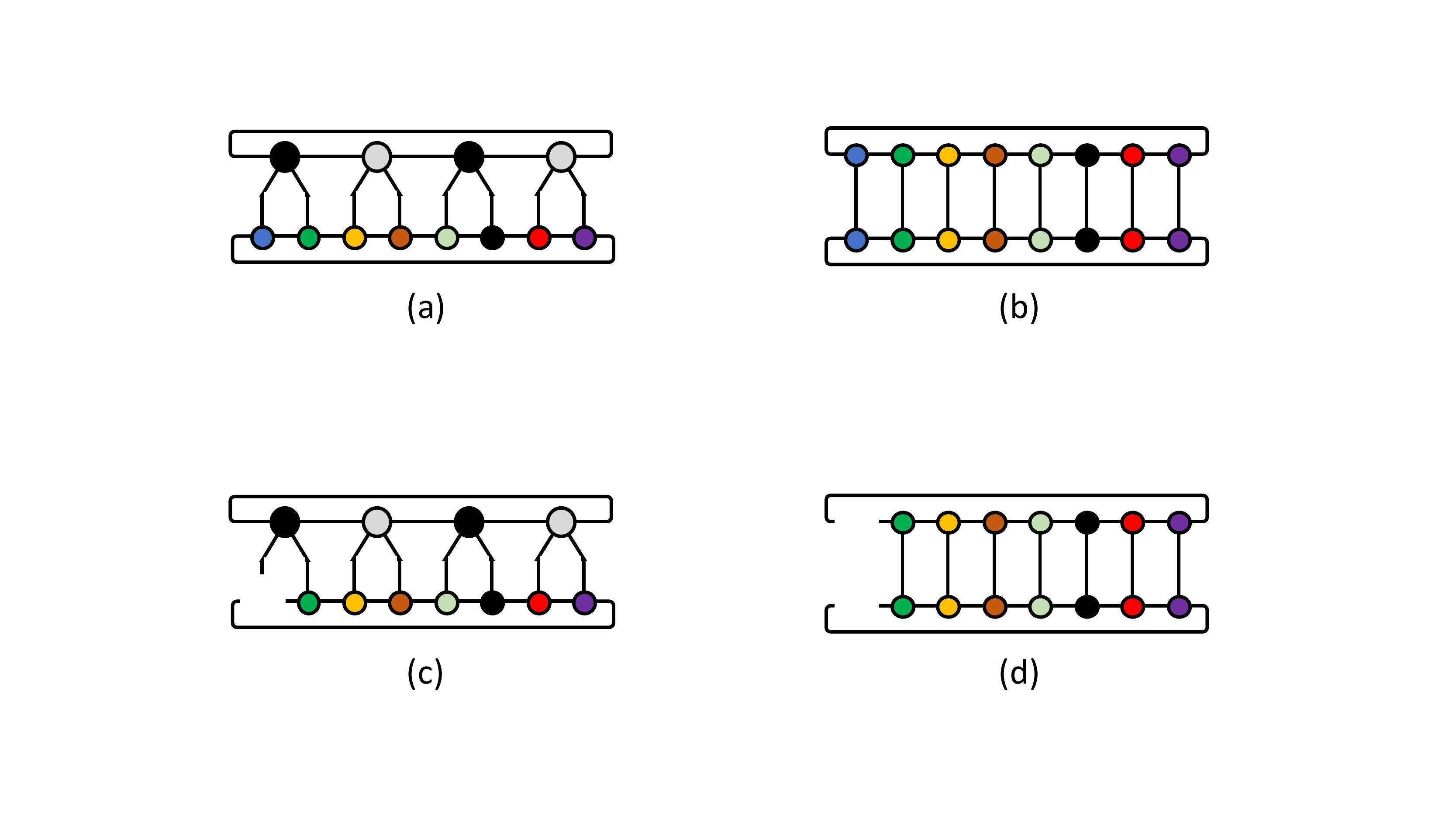}
\caption{Components of the cost function and its derivative.
(a) is $\left\langle \Psi _{A}|\Psi _{B}\right\rangle $. (b) is $%
\left\langle \Psi _{B}|\Psi _{B}\right\rangle $. (c) and (d) are the
quantity $W_{1}$ and $N_{1}$ in (\protect\ref{OPTfun}), respectively.}
\label{cost}
\end{figure}

$\bullet $ In step (b), we find 8 rank-3 tensor to form a octagon matrix
product state(MPS) to approximate the square MPS, as shown in Fig. \ref%
{loopTNR} (d). We're aiming to find the optimal choice of those 8 rank-3
tensors $S_{1},S_{2},...S_{8}$ to minimize the cost function in Fig.\ref%
{loopTNR} (d), which can be expressed as%
\begin{equation}
C\left( S_{1},S_{2},...,S_{8}\right) =\left\Vert T_{A}\cdot T_{B}\cdot
T_{A}\cdot T_{B}-S_{1}\cdot S_{2}\cdot ...\cdot S_{8}\right\Vert ^{2}.
\end{equation}%
Since $S_{1},S_{2},...S_{8}$ are independent variables, we can minimize this
cost function with variation method. We denote the two MPS's in above cost
function as:
\begin{eqnarray}
\left\vert \Psi _{A}\right\rangle &=&T_{A}\cdot T_{B}\cdot T_{A}\cdot T_{B}
\notag \\
\left\vert \Psi _{B}\right\rangle &=&S_{1}\cdot S_{2}\cdot ...\cdot S_{8}
\end{eqnarray}%
Then, the cost function could be write down as%
\begin{equation}
C\left( S_{1},S_{2},...,S_{8}\right) =\left\langle \Psi _{A}|\Psi
_{A}\right\rangle +\left\langle \Psi _{B}|\Psi _{B}\right\rangle
-\left\langle \Psi _{A}|\Psi _{B}\right\rangle -\left\langle \Psi _{B}|\Psi
_{A}\right\rangle .
\end{equation}%
Taking variation on $S_{1},$ we get%
\begin{eqnarray}
\left. \frac{\delta C}{\delta S_{1}^{\dagger }}\right\vert
_{S_{2},S_{3},..S_{8}} &=&\left\langle \left. \frac{\delta \Psi _{B}}{\delta
S_{1}^{\dagger }}\right\vert \Psi _{B}\right\rangle -\left\langle \left.
\frac{\delta \Psi _{B}}{\delta S_{1}^{\dagger }}\right\vert \Psi
_{A}\right\rangle  \notag \\
&\equiv &\left[ N_{1}\cdot S_{1}-W_{1}\right] .
\end{eqnarray}%
The minimum of $C\left( S_{1}\right) $ is given by the solution of the linear equation:
\begin{equation}
N_{1}\cdot S_{1}=W_{1}.  \label{OPTfun}
\end{equation}%
The cost function (\ref{OPTfun}) and $N_{1}$, $W_{1}$ are
illustrated in Fig. \ref{cost}. After optimizing $S_{1}$, we can go on to the
next site, and if we finish the optimization from $S_{1}$ to $S_{8}$, we
finish one circle. By repeating this variation optimization, we can minimize the cost function.

$\bullet $ After minimizing the cost function, we trace the inner indices in
the small circles, as shown in Fig. \ref{loopTNR} (b), and get the
coarse-grained tensor $T_{A}^{\prime }$ and $T_{B}^{\prime },$ as in Fig. \ref%
{loopTNR} (c). Compared with the original tensor network, we find the tensor
network composed of the renormalized tensor elements $T_{A}^{\prime }$ and $%
T_{B}^{\prime }$ (a) rotates an angle of $\pi /4$ and (b) the system size of
the new network reduced to be half of the original. Then, we can start the new
RG step from this tensor network

$\bullet $ We will normalize
the tensor $T_{A}$ and $T_{B}$ in every RG step with the normalization factor as shown in Fig. \ref{loopTNR} (e).

\subsection{loop-TNR with $C_{4}$ lattice symmetry}

\begin{figure}[h]
\includegraphics[width=8cm]{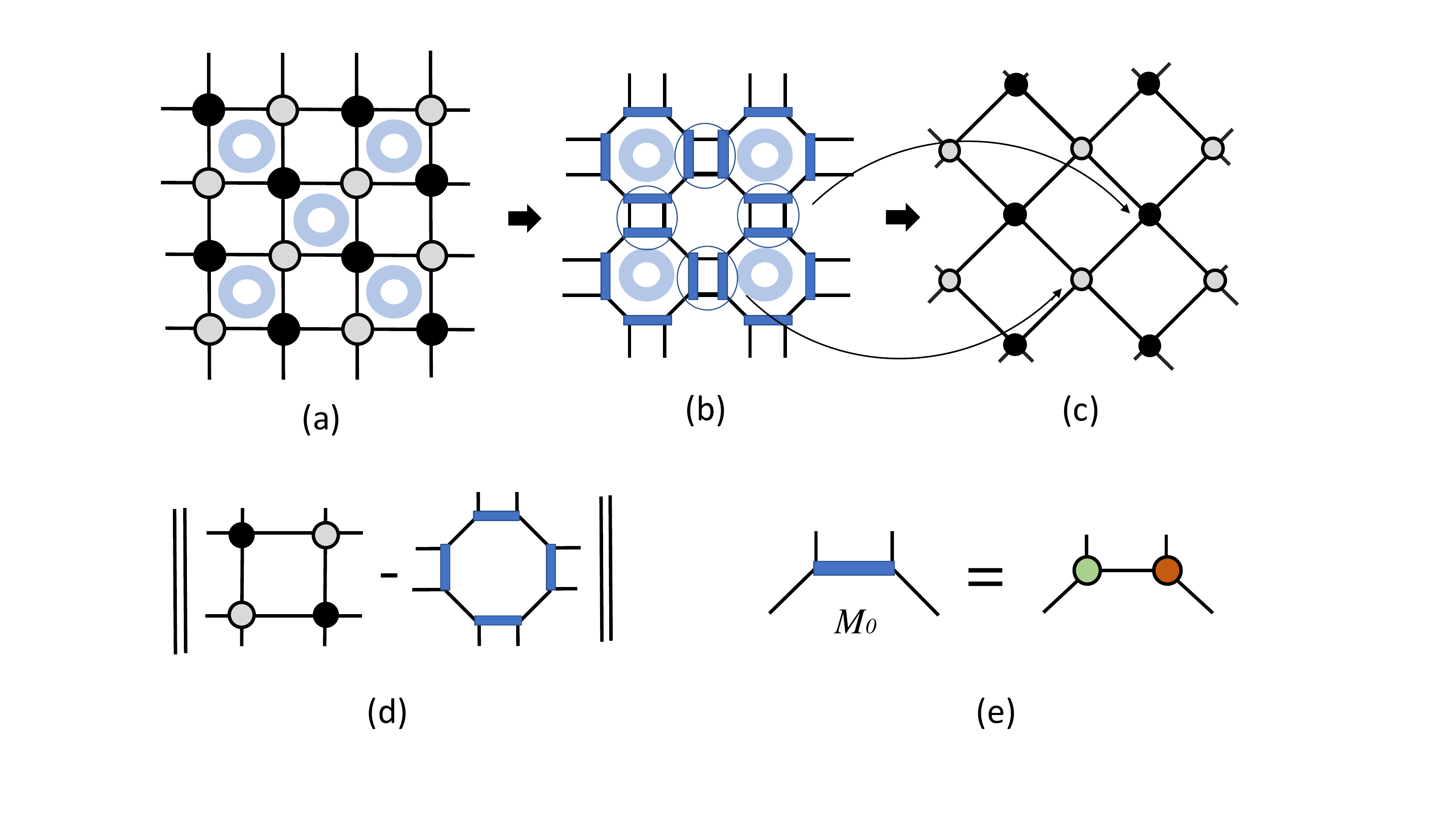}
\caption{Loop-TNR algorithm with $C_4$ lattice symmetry is similar with usual
loop-TNR. Notice that the cost function in this case is nonlinear, so we
need to use nonlinear optimization algorithm, such as conjugate gradient
method.}
\label{C4lpTNR}
\end{figure}


To keep the lattice symmetry in the renormalization process, we
need to find a octagon MPS with $C_{4}$ symmetry when minimizing the cost
function in Fig. \ref{C4lpTNR} (d). We can construct this octagon MPS
with the rank-4 block tensor $M_{ijkl}$, as shown in Fig. \ref{C4lpTNR} (b).
Then we can use the conjugate gradient method to minimize the
cost function:
\begin{equation}
C=\left\Vert T_{A}\cdot T_{B}\cdot T_{A}\cdot T_{B}-M\cdot M\cdot M\cdot
M\right\Vert ^{2}.
\end{equation}%
After the optimization, we can use tensor $M$ to build the renormalized
tensor $T_{A}^{\prime },T_{B}^{\prime }$, as shown in Fig.\ref{C4lpTNR}
(c)%
\begin{eqnarray}
T_{ruld}^{A\prime } &=&\underset{ij}{\sum }M_{ijrd}M_{jilu}  \notag \\
T_{ruld}^{B\prime } &=&\underset{ij}{\sum }M_{ijld}M_{jiru}.
\end{eqnarray}%
Since the octagon network has $C_{4}$ symmetry, the coarse-grained tensor
network on the square lattice marked by blue circle has the same $C_4$
symmetries.

The initial value of the tensor $M$ is very important for the numerical
accuracy. We can decompose tensor $T_{A}$ and $T_{B}$ by SVD method%
\begin{eqnarray}
T_{ruld}^{A} &\approx &\underset{x}{\sum }S_{ldx}^{1}S_{rux}^{2}=\underset{x}%
{\sum }S_{rux}^{1}S_{ldx}^{2}  \notag \\
T_{ruld}^{B} &\approx &\underset{x}{\sum }S_{ulx}^{1}S_{drx}^{2}=\underset{x}%
{\sum }S_{drx}^{1}S_{ulx}^{2}.
\end{eqnarray}%
Thus, the initial $M$ is could be constructed as Fig. \ref{C4lpTNR} (e), with
\begin{equation}
M_{ijkl}^{0}=\underset{x}{\sum }S_{ixk}^{1}S_{xjl}^{2}.
\end{equation}%
By keeping $C_4$ lattice symmetries in each iteration step, we have partially
fixed the gauge of the building block $M$, which would be very important for
studying the structure of the fixed point tensor.

\subsection{loop-TNR with $Z(q)$ symmetry in Hamiltonian}
As the original tensor element of $q$-state
model $T_{ijkl}$ contains $Z\left( q\right) $ symmetry, we can keep such a symmetry for every step in the loop-TNR algorithm. As $Z\left( q\right) $ is a cyclic
group, which contains group elements $\left\{ I,g,g^{2},...g^{q-1}\right\} $%
, and the generator $g$ has the $q$-dimension faithful representation%
\begin{equation}
G_{q}=\left(
\begin{array}{cccccc}
0 & 0 & 0 & ... & 0 & 1 \\
1 & 0 & 0 & ... & 0 & 0 \\
0 & 1 & 0 & ... & 0 & 0 \\
0 & 0 & 1 & ... & 0 & 0 \\
... & ... & ... & ... & ... & ... \\
0 & 0 & 0 & ... & 1 & 0%
\end{array}%
\right) .
\end{equation}%
It is easy to check that:
\begin{eqnarray}
T_{ruld}^{\prime } &=&\underset{r^{\prime }u^{\prime }l^{\prime }d^{\prime }}%
{\sum }\left[ G_{q}\right] _{rr^{\prime }}\left[ G_{q}\right] _{uu^{\prime }}%
\left[ G_{q}\right] _{ll^{\prime }}\left[ G_{q}\right] _{dd^{\prime
}}T_{r^{\prime }u^{\prime }l^{\prime }d^{\prime }}  \notag \\
&=&T_{ruld}.
\end{eqnarray}%
In order to find out all the irreducible representation of the $Z_q$ symmetry, we can just do
eigenvalue decomposition for $G_{q}$,%
\begin{equation}
G_{q}=V\Lambda V^{-1},
\end{equation}%
with eigenvalues $\Lambda _{nn}=\lambda _{n}=e^{2\pi in/q},n\in
\{0,1,2,...,q-1\}$, and the components of the matrix $V$ is given by:
\begin{equation}
V_{mn}=\frac{e^{2\pi imn/q}}{\sqrt{q}},m,n\in \{0,1,2,...,q-1\}.
\end{equation}%
Such that:
\begin{equation}
G_{q}=V^{-1}\Lambda ^{-1}V.
\end{equation}%
Then, we define two tensors:
\begin{eqnarray}
T^A_{ruld} &=&\underset{r^{\prime }u^{\prime }l^{\prime }d^{\prime }}{\sum }%
\left[ V^{-1}\right] _{rr^{\prime }}\left[ V^{-1}\right] _{uu^{\prime }}%
\left[ V^{-1}\right] _{ll^{\prime }}\left[ V^{-1}\right] _{dd^{\prime
}}T_{r^{\prime }u^{\prime }l^{\prime }d^{\prime }}  \notag \\
T^B_{ruld} &=&\underset{r^{\prime }u^{\prime }l^{\prime }d^{\prime }}{\sum }%
V_{rr^{\prime }}V_{uu^{\prime }}V_{ll^{\prime }}V_{dd^{\prime }}T_{r^{\prime
}u^{\prime }l^{\prime }d^{\prime }}.
\end{eqnarray}%
Obviously, tensor $T^A$ and $T^B$ form the same tensor network with $T$. In the new basis tensors $T^A$ and $T^B$ satisfy:
\begin{eqnarray}
T^A_{ruld} &=&\lambda _{r}\lambda _{u}\lambda _{l}\lambda _{d}T^A_{ruld}  \notag
\\
T^B_{ruld} &=&\lambda _{r}^{-1}\lambda _{u}^{-1}\lambda _{l}^{-1}\lambda
_{d}^{-1}T^B_{ruld},
\end{eqnarray}%
which implies that $T^A_{ruld}$ and $T^B_{ruld}$ only have non-zero components when $%
r+u+l+d\equiv 0\left( \text{mod } q\right)$. Thus, tensors $T^A$ and $T^B$ are
block diagonalized. It turns out that if we keep such block diagonalized
property during RG process, i.e., in every RG step, we keep $r+u+l+d\equiv
0\left( \text{mod } q\right)$, $Z\left( q\right) $ symmetry is preserved during loop-TNR process.
In particular,
we can keep $D_{cut}=nq$(where $n$ is an arbitrary integer), such that we can always construct a dimension $nq$ by $nq$ block diagonalized matrix $\Lambda ^{\prime }$
\begin{equation}
\Lambda ^{\prime }=\left(
\begin{array}{cccc}
\Lambda  & \mathbf{0} & ... & \mathbf{0} \\
\mathbf{0} & \Lambda  & ... & \mathbf{0} \\
... & ... & ... & ... \\
\mathbf{0} & \mathbf{0} & ... & \Lambda
\end{array}%
\right) .
\end{equation}%
Obviously, $\Lambda ^{\prime }$ is a representation of $Z\left( q\right) $ symmetry. So that $Z\left( q\right) $ symmetry is kept.


\bibliography{Ref}

\end{document}